\documentclass[useAMS,usenatbib]{mn2e}
\usepackage{graphicx}
\usepackage{txfonts}
\usepackage{wrapfig}
\usepackage{longtable}
\usepackage[usenames]{color}
\def\aj{AJ}%
\def\araa{ARA\&A}%
\def\apj{ApJ}%
\def\apjs{ApJS}%
\def\aap{A\&A}%
\def\mnras{MNRAS}%
\begin{document}
\title[mom0]{The Bluedisks project, a study of unusually HI-rich galaxies: I. HI Sizes and Morphology }

\author[Jing ~Wang et al.]
{Jing Wang$^{1}$\thanks{Email: wangj@mpa-garching.mpg.de}, Guinevere Kauffmann$^{1}$, Gyula I. G. J\'ozsa$^{2,3}$, Paolo Serra$^{2}$, Thijs van der Hulst$^{4}$,  \\
      \newauthor   Frank Bigiel$^{5}$, Jarle Brinchmann$^{6}$, M.A.W. Verheijen$^{4}$, Tom Oosterloo$^{2,4}$,\\
     \newauthor   Enci Wang$^{7}$, Cheng Li$^{7}$, Milan den Heijer$^{3}$\thanks{Member of the International Max
Planck Research School (IMPRS) for Astronomy and Astrophysics at the
Universities of Bonn and Cologne}, J\"urgen Kerp$^{3}$\\
 $^1$Max--Planck--Institut f\"ur Astrophysik,
        Karl--Schwarzschild--Str. 1, D-85741 Garching, Germany        \\
 $^2$Netherlands Institute for Radio Astronomy (ASTRON), Postbus 2, 7990 AA Dwingeloo, The Netherlands \\
 $^3$Argelander-Institut f\"ur Astronomie, Auf dem Hügel 71, D-53121 Bonn, Germany\\
 $^4$University of Groningen,  Kapteyn Astronomical Institute, Landleven 12,  9747 AD, Groningen, The Netherlands\\
 $^5$Institut f\"ur theoretische Astrophysik, Zentrum für Astronomie der Universität Heidelberg, Albert-Ueberle Str. 2, D-69120 Heidelberg\\
 $^6$Leiden Observatory, Leiden University, PO Box 9513, 2300 RA Leiden, The Netherlands \\
 $^7$Partner Group of the Max Planck Institute for Astrophysics and Key Laboratory for Research in Galaxies and Cosmology of Chinese Academy of Sciences, \\
Shanghai Astronomical Observatory, Nandan Road 80, Shanghai 200030, China  
}
\date{Accepted 2011 ???? ??
      Received 2011 ???? ??;
      in original form 2012 November}

\pubyear{2012}
\maketitle
\begin{abstract}
We introduce the ``Bluedisk'' project, a large program at the Westerbork Synthesis Radio Telescope (WSRT) 
that has mapped the HI in a sample of 23 nearby galaxies with unusually high HI mass fractions,
along with a similar-sized sample of control galaxies matched in stellar mass, size, inclination and redshift. 
This paper presents the sample selection, observational set-up,  data
reduction strategy, and a first analysis of the sizes and structural properties of the HI disks.
We find that the HI-rich galaxies lie on the same HI mass versus HI size relation as normal 
spiral galaxies, extending it to total  HI masses of $2 \times 10^{10} M_{\odot}$ and radii R1
of $\sim 100$ kpc  (where R1 is defined as the radius where
the HI column density reaches 1 $M_{\odot}$ pc$^{-2}$).
HI-rich galaxies  have significantly larger values of HI-to-optical size ratio at  fixed stellar mass,
concentration index, stellar and star formation rate surface density compared to
the control sample. The disks of HI-rich galaxies are
also significantly more clumpy (i.e. have higher HI Gini and $\Delta$Area coefficient) than those
of normal spirals. There is no evidence that the disks of HI-rich galaxies are more disturbed:
HI-rich galaxies exhibit no difference with respect to control samples in their
distributions of HI asymmetry indices or optical/HI disk position angle differences.
In fact, the center of the HI distribution corresponds more closely with the center
of the optical light in the HI-rich galaxies than in the controls. 
All these results argue against a scenario in which new gas has been brought in by mergers.
It is possible that they may be more
consistent with cooling from a surrounding quasi-static halo of warm/hot gas. 

\end{abstract}

\begin{keywords}
disk galaxies; atomic gas; synthesis radio mapping
\end{keywords}

\section{introduction}
The mechanisms by which galaxies acquire their gas remain one of the
key unsolved problems in galaxy formation. Only $\sim$ 20 percent  of the available
baryons in  dark matter halos surrounding present-day $L_*$ spiral
galaxies  have cooled and been transformed into stars. 
The star formation rates in these galaxies imply
that their atomic and molecular gas reservoirs will be used up on
timescales of a few Gyr \citep{Kennicutt83,Fraternali12}. In order to maintain star formation at its observed level over
long timescales, it is often assumed that gas must accrete from the
external environment. 
Additional indirect evidence for gas infall comes 
from the fact that the metallicity distribution of main sequence stars
in the solar neighbourhood is incompatible with a closed-box scenario
\citep[``The G-dwarf problem",][]{vandenBergh62,Pagel75},
implying early inflow of metal-poor gas onto the Milky Way.

Estimates of the total mass in cold gas clouds infalling onto the Milky Way
yield $\sim 10^9 M_{\odot}$ within 150 kpc and $\sim 5 \times 10^8
M_{\odot}$ within 60 kpc \citep{Putman06}. These estimates assume that the
fraction of a cloud that is detectable as neutral hydrogen is around 10\%
of the total mass of the cloud. Nevertheless, these numbers are an order of
magnitude smaller than predicted by simple models that attempt to calculate
the rate at which gas would be cooling and fragmenting into clouds in a
typical dark matter halo in a $\Lambda$CDM universe \citep[e.g.][]{Maller04}.  An independent analysis by \citet{Richter12} yields an estimate of the
neutral gas accretion rate onto M31/Milky Way type galaxies of 0.7
M$_{\odot}$ yr$^{-1}$, a factor of $\sim 3-5$ less than the observed star
formation rates these systems. 

In external galaxies,  HI-rich companions, warped or
lopsided HI disks are often cited as evidence of ongoing cold gas
accretion in local spiral galaxies. In addition, there is evidence for 
large quantities of extra-planar gas around a few nearby spirals  \citep{Fraternali02, Oosterloo07}. The kinematical structure 
of this gas indicates their (partially) extragalactic origin \citep[e.g.,][]{Benjamin00,Collins02,Fraternali06,Heald07,Kamphuis07,Fraternali08,Marinacci11}. 

The other way in which galaxies may be ``refuelled'' is via cooling from a 
hot halo of gas that is in virial equilibrium with the
dark matter \citep{White78}.
Studies of the hot gas around galaxies have been hampered by the fact that
X-ray halos that are not directly associated with galactic winds from an ongoing
starburst, are only detected individually around the very most luminous
spirals \citep{Anderson11,Dai12}.
Recent attempts to get around this problem by analyzing the X-ray emission
around stacked samples of nearby spirals have clarified that hot gas 
not associated with present-day galactic winds is
present around these systems, but estimates of  cooling rates
are still a factor $\sim 2$  too low to explain the observed
star formation in these systems \citep{Anderson13}. On the other hand, \citet{Marinacci12} estimated a higher ``fountain'' driven accretion rate from the galactic halo arround the Milky Way \citep[see also][]{Fraternali13}.

At this point it is still difficult to assess what the dominant gas fuelling
mechanism is in the local Universe.  
Deep HI observations of nearby galaxies 
have so far been restricted to a few ``promising cases''  
\citep{Fraternali02, Oosterloo07a, Boomsma08}. 
The HALOGAS survey \citep{Heald11,Heald12} aims to detect and
characterize  the extended, low-column density gas in a sample
of 22 nearby galaxies. So far, mass fractions of
the extraplanar HI comparable to the cases of NGC 891 \citep[30\%,][]{Oosterloo07} or NGC 2403 \citep[10\%,][]{Fraternali04} have not been reported
from HALOGAS \citep{Heald11,Zschaechner11,Zschaechner12}.

Single-dish HI surveys of atomic gas in complete, stellar mass-limited
galaxy samples reveal that the majority of disk galaxies lie on a
tight plane that links their atomic gas content with their UV/optical
colours and their stellar surface mass densities \citep[][,hereafter C10]{Catinella10}, in line with the fact that HI-rich galaxies are
systematically bluer and late-type 
\citep[e.g.][]{Roberts94}. Around 10\% of the disk
galaxy population is significantly displaced from this plane in the
sense of having significantly more atomic gas than would be predicted
from their colours and densities. These galaxies have outer disks
that are bluer \citep{Wang11} and younger (higher ratio of
present-to-past averaged star formation). In addition, the ionized gas
in these outer disks is metal-poor \citep{Moran12}. Analysis of the
star formation histories of such galaxies indicate that stars did not
form continuously in these systems -- an elevated rate of star
formation in the past 2 Gyr is required to explain the strong Balmer
absorption seen in the spectra of the outer disk stellar populations
\citep{Huang13}.  All these results provide indirect
evidence that such galaxies may have recently accreted cold gas in
their outer regions. Although these galaxies have unusually high HI
content, their molecular gas masses are normal, suggesting that the
excess atomic gas exists as a largely inert reservoir in the outer
regions of these galaxies \citep{Saintonge11}.

The next step in understanding the nature of these HI-rich galaxies with
young, star-forming outer disks is to investigate how the morphology,
density profiles and kinematics of the gas differ from that of normal
spirals. If the HI gas was accreted recently, one might expect the HI disk
to be significantly more extended than the stellar disk \citep[see][]{Fu10}. We might also see more frequent kinematic signatures of recent
accretion in the form of warps, misaligned or counter-rotating HI
components in the disk, or associated gas clouds.

To this end, we undertook the ``Bluedisk project", a long-term large program
at the Westerbork Synthesis Radio Telescope (WSRT). Uniform, blind HI
surveys covering large areas of the sky such as the Arecibo Legacy Fast
ALFA \citep[ALFALFA,][]{Giovanelli05} survey do not extend as far North as
declinations greater than 38 degrees, where the HI maps produced by
Westerbork have optimal resolution (beam size of $\sim$ 20$\times$15 arcsec$^{2}$).  However, these surveys have taught us
that selection based on optical properties is an efficient means of
targeting HI-rich galaxies \citep{Zhang09,Catinella10,Li12}.  We used the technique outlined in \citet{Li12} to select a
sample of 25 galaxies predicted to be HI-rich, as well as a sample of 25
control galaxies matched in stellar mass, stellar mass surface density,
redshift and inclination. These galaxies were observed at Westerbork over
the period from December 2011 to May 2012.

Our paper is structured as follows: in section 2, we discuss the sample selection
procedure, describe the observational set-up and data reduction strategy,
provide catalogs of HI parameters, and discuss our final scheme for classifying
the galaxies in our sample into those that are ``HI-rich" and ``HI-normal".
In section 3,  we present total intensity HI maps for these two subsets. 
In section 4, we discuss how we measure HI sizes and morphological
parameters. In section 5, we compare these parameters for HI-rich and HI-normal
galaxies. We also examine correlations between HI sizes and morphological
parameters and those measured for the UV and optical light.  
Our findings are summarized and discussed in section 6. 
We assume $H_0= 70$ km s$^{-1}$ Mpc$^{-1}$ throughout the paper.

\section{Data}
\subsection {Sample selection}
Following the work of Zhang et al. (2009), C10 defined a gas-fraction
``plane'' linking HI mass fraction, stellar surface mass density and
NUV-$r$ colour that exhibited a scatter of 0.315 dex in $\log$ M(HI)/M$_*$.
This relation indicates that the HI content of a galaxy scales with its
physical size as well as with its specific  star formation rate. In
subsequent work, Wang et al.  (2011) showed that at fixed NUV-$r$ colour
and stellar surface density, galaxies with larger HI gas fractions have
bluer outer disks. This result motivated us to add a correction term to the
C10 relation  based on the $g-i$ colour gradient of the galaxy.  When $\log
M(HI)/M_* (C10) > -1$, we apply a correction of the form \begin {equation}
\Delta (\log M(HI/M_*) = -0.35\log M_* - 0.51\Delta_{o-i}(g-i) + 3.69
\end {equation} Here $\Delta_{o-i}(g-i)$ is defined as $(g-i)_{out}-(g
-i)_{in}$, where the inner colour is evaluated within a radius R50
enclosing 50\% of the $r$-band light, and the outer colour is evaluated
within R50 and R90, the radius enclosing 90\% of the $r$-band light.  We
note that a slightly retuned version of this relation has been published in
Li et al (2012).

We first selected all galaxies from the DR7 MPA/JHU catalog (http://www.mpa-garching.mpg.de/SDSS/DR7/), which is based
on galaxy spectra from the Data Release 7 of the Sloan Digital Sky Survey \citep{Abazajian09}
 using the following criteria:
$11 > \log M_* > 10$, $0.01 < z < 0.03$, Declination $>$ 30 degrees and
with high $S/N$ NUV detection in the GALEX imaging survey \citep{Martin10}. This yields a
sample of 1900 galaxies, which we use as our parent sample.  Most of these
galaxies have optical diameters of 50 arcsec or greater, so that the HI will
be well resolved by the Westerbork synthesised beam (the minimum half-power
beam width (HPBW) is  ~13'').  From the parent sample,
we selected a sub-sample of 123 galaxies with predicted HI mass fractions
0.6 dex higher in $\log M(HI)/M_*$ than the median relation between HI mass
fraction and stellar mass found in Catinella et al (2010).  This threshold
was chosen because the  estimate of $\log M(HI)/M_*$ given in equation (1)
has a scatter of $\sim 0.25$ dex , so this cut should
generate a reasonably pure sample of true HI-rich galaxies.
25 targets were selected at random for Westerbork observations.

In addition, we selected a sample of 25 ``control galaxies'' that were
closely matched in $M_*$, $\mu_*$ (mass surface density, calculated as 0.5$\times$M$_*/(\pi r(50,z)^2)$, where r(50,z) is the half-light radius in $z$-band), $z$ and inclination, but with predicted
HI fractions between 1 and 1.5 times the median value at the same  value of
$M_*$ and $\mu_*$. The cut was chosen to exclude gas-deficient galaxies,
such as those found in rich groups or clusters. The matching tolerances are
around 0.1 in $\log M_*$, $\log \mu_*$ and $b/a$, and 0.001 in redshift.
The control galaxies  have both redder global colours
and weaker colour gradients in comparison to the targets. Table~\ref{tab:obsample} lists the
optical/UV selection parameters of the 50 galaxies. We also derive star formation rates (SFRs) 
using the spectral energy distribution (SED) fitting technique described 
in \citet{Wang11, Saintonge11}. SFR surface densities 
are calculated as 0.5 SFR$/\pi R_{\rm NUV}(50)^2)$, where $R_{\rm NUV}(50)$
is the NUV half-light radius.

\begin{table*}
\begin{center}
\renewcommand{\arraystretch}{1.}
\begin{tabular}{c c c c c c c c c c c c}
\hline \hline
 \multicolumn{1}{c}{ID} &  \multicolumn{1}{c}{ra} & \multicolumn{1}{c}{dec} & \multicolumn{1}{c}{z} & \multicolumn{1}{c}{$\log M_*/M_{\odot}$} & \multicolumn{1}{c}{NUV--r} & \multicolumn{1}{c}{$\Delta_{o-i}(g-i)$} & \multicolumn{1}{c}{$D_{25}/arcsec$} & \multicolumn{1}{c}{$b/a$} & \multicolumn{1}{c}{$R_{90}/R_{50}$} & \multicolumn{1}{c}{$\log M(HI)/M_*[pred]$} & \multicolumn{1}{c}{flag$_{ana}$}\\
\hline
   1&  123.591766&   39.251354&      0.0277&     10.43&     2.37&     -0.21&     75.96&     0.83&     2.70&     -0.37&	1\\
   2&  127.194809&   40.665886&      0.0245&     10.62&     3.02&     -0.27&     81.30&     0.41&     2.15&     -0.62&	1 \\
   3&  129.277161&   41.456322&      0.0291&     10.42&     2.44&     -0.25&     49.08&     0.80&     2.15&     -0.38&	1 \\
   4&  129.641663&   30.798681&      0.0256&     10.57&     2.36&     -0.29&     88.61&     0.47&     2.19&     -0.40&	1 \\
   5&  132.318298&   36.119797&      0.0252&     10.32&     2.01&     -0.12&     65.27&     0.93&     2.06&     -0.32&	1 \\
   6&  132.344025&   36.710327&      0.0251&     10.84&     2.94&     -0.23&    103.79&     0.62&     2.38&     -0.77&	1 \\
   7&  132.356155&   41.771252&      0.0289&     10.37&     2.57&     -0.23&     69.28&     0.26&     2.44&     -0.40&	0 \\
   8&  137.177567&   44.810658&      0.0267&     10.28&     2.19&     -0.27&     66.18&     0.67&     2.10&     -0.25&	1 \\
   9&  138.742996&   51.361061&      0.0275&     10.76&     2.93&     -0.17&     73.23&     0.46&     2.35&     -0.70&	2 \\
  10&  143.104355&   57.482899&      0.0294&     10.91&     2.43&     -0.21&     60.12&     0.80&     2.00&     -0.62&	2 \\
  11&  152.726593&   45.950371&      0.0240&     10.64&     2.74&     -0.18&     73.74&     0.56&     2.07&     -0.60&	2 \\
  12&  154.042709&   58.427002&      0.0255&     10.63&     2.68&     -0.29&     82.89&     0.36&     2.64&     -0.58&	1 \\
  13&  166.995911&   35.463264&      0.0287&     10.84&     2.74&     -0.28&     83.31&     0.46&     2.74&     -0.78&	0 \\
  14&  176.739746&   50.702133&      0.0238&     10.79&     3.25&     -0.20&     87.33&     0.49&     2.58&     -0.86&	1 \\
  15&  177.247757&   35.016048&      0.0213&     10.82&     2.43&     -0.21&    111.76&     0.63&     2.71&     -0.65&	1 \\
  16&  193.014786&   51.680046&      0.0272&     10.31&     2.01&     -0.22&     54.06&     0.76&     1.94&     -0.21&	1 \\
  17&  196.806625&   58.135014&      0.0275&     10.69&     2.85&     -0.25&     61.00&     0.71&     2.46&     -0.68&	1 \\
  18&  199.015060&   35.043518&      0.0232&     10.34&     2.28&     -0.33&     71.17&     0.57&     2.25&     -0.33&	1 \\
  19&  212.631851&   38.893559&      0.0256&     10.25&     2.15&     -0.27&     57.90&     0.69&     2.16&     -0.27&	1 \\
  20&  219.499756&   40.106197&      0.0261&     10.21&     1.55&     -0.39&     77.39&     0.62&     2.26&     -0.02&	1 \\
  21&  241.892578&   36.484032&      0.0298&     10.41&     2.70&     -0.33&     46.28&     0.63&     2.17&     -0.38&	1 \\
  22&  250.793503&   42.192783&      0.0284&     10.82&     3.07&     -0.29&     86.07&     0.28&     2.76&     -0.79&	1 \\
  23&  251.811615&   40.245079&      0.0295&     10.75&     2.91&     -0.27&     79.17&     0.45&     2.50&     -0.68&	2 \\
  24&  259.156036&   58.411900&      0.0296&     10.71&     2.72&     -0.21&     80.72&     0.53&     2.26&     -0.66&	1 \\
  25&  262.156342&   57.145065&      0.0275&     10.54&     2.50&     -0.36&     58.66&     0.77&     2.52&     -0.52&	2\\
\\
  26&  111.938042&   42.180717&      0.0231&     10.31&     3.81&     -0.17&     52.96&     0.48&     2.56&     -0.78&	1 \\
  27&  120.669388&   34.521431&      0.0288&     10.30&     2.90&     -0.14&     36.99&     0.78&     1.93&     -0.49&	2 \\
  28&  123.309128&   52.458736&      0.0183&     10.54&     3.69&     -0.16&     61.86&     0.84&     2.21&     -0.87&	2 \\
  29&  127.312149&   55.522991&      0.0257&     10.52&     2.90&     -0.25&     74.80&     0.44&     2.17&     -0.56&	0 \\
  30&  138.603149&   40.777924&      0.0280&     10.39&     2.92&     -0.22&     70.32&     0.28&     2.43&     -0.48&	1 \\
  31&  139.190598&   45.812244&      0.0262&     10.18&     2.68&      0.04&     44.73&     0.61&     2.18&     -0.51&	0\\
  32&  139.646255&   32.270008&      0.0269&     10.29&     2.93&     -0.12&     53.68&     0.67&     1.98&     -0.44&	2 \\
  33&  141.539307&   49.310204&      0.0269&     10.74&     3.99&     -0.11&     66.92&     0.53&     2.86&     -1.02&	2 \\
  34&  147.539001&   33.569332&      0.0270&     10.61&     4.40&     -0.13&     50.40&     0.68&     2.81&     -1.14&	0\\
  35&  149.420227&   45.258678&      0.0242&     10.61&     3.32&     -0.14&     76.99&     0.32&     2.66&     -0.77&	1 \\
  36&  149.454529&   51.821190&      0.0249&     10.34&     2.77&     -0.14&     46.80&     0.67&     1.97&     -0.48&	2 \\
  37&  153.797638&   56.672085&      0.0260&     10.39&     2.81&     -0.13&     62.86&     0.86&     1.98&     -0.46&	2 \\
  38&  153.926071&   55.667500&      0.0244&     10.75&     3.52&     -0.18&     57.91&     0.71&     2.12&     -0.86&	2 \\
  39&  162.530365&   36.341831&      0.0239&     10.84&     4.31&     -0.15&     83.13&     0.77&     2.15&     -1.14&	2 \\
  40&  168.563553&   34.154381&      0.0272&     10.36&     2.83&     -0.15&     44.02&     0.78&     1.98&     -0.48&	2 \\
  41&  197.879166&   46.341774&      0.0297&     10.79&     3.45&     -0.20&     54.62&     0.85&     2.05&     -0.87&	0 \\
  42&  198.236252&   47.456657&      0.0281&     10.80&     3.72&     -0.13&     65.32&     0.52&     2.69&     -1.00&	2 \\
  43&  203.374603&   40.529671&      0.0269&     10.26&     2.74&     -0.18&     45.31&     0.67&     2.03&     -0.37&	2 \\
  44&  205.251038&   42.431423&      0.0279&     10.78&     3.34&     -0.16&     61.53&     0.63&     2.20&     -0.85&	2 \\
  45&  222.988846&   51.264881&      0.0259&     10.68&     3.40&     -0.17&     68.83&     0.41&     2.20&     -0.80&	2 \\
  46&  241.528976&   35.981434&      0.0305&     10.68&     3.68&     -0.20&     58.79&     0.44&     2.45&     -0.87&	0 \\
  47&  244.382523&   31.194477&      0.0240&     10.53&     2.91&     -0.17&     65.82&     0.58&     2.28&     -0.62&	1 \\
  48&  246.259842&   40.946644&      0.0287&     10.75&     3.69&     -0.13&     56.45&     0.52&     2.38&     -0.93&	0 \\
  49&  258.650208&   30.733536&      0.0296&     10.42&     2.82&     -0.17&     54.64&     0.54&     2.09&     -0.48&	2 \\
  50&  261.557251&   62.149483&      0.0278&     10.87&     4.27&     -0.11&     82.39&     0.27&     2.57&     -1.10&	1 \\
\hline \hline
\end{tabular}
\caption{The optical$\slash$UV properties of the Bluedisk galaxies, including the galaxy ID used in this project,  Right Ascention, Declination, redshift, stellar mass, NUV--r colour which are corrected for Galactic extinction only, g-i colour gradients, the $g$-band diameter at 25 mag$/$arcsec$^{-2}$, the axis ratio in $r$-band, the concentraion index in $r$-band and the predicted HI mass fraction. The first 25 galaxies were originally selected as HI-rich and the others as control galaxies. The last column, flag$_{ana}$ marks the galaxies according to their classification in Sect. 3: 1 for HI-rich galaxies, 2 for control galaxies and 0 for the galaxies excluded from analysis.}
\label{tab:obsample}
\end{center}
\end{table*}

\subsection {Observations and data reduction}
Our target galaxies were observed with the WSRT 
between December 2011 and June 2012, with an on-source integration
time of 12 h per galaxy. Part of the sample was observed in (non-guaranteed) backup
time, resulting in the loss of one target. The correlator setup was
chosen to cover the HI line while at the same time covering a large portion of the bandpass to obtain sensitive continuum data. One band was reserved for the line observation using a
total bandwidth of 10 MHz and 1024 channels with two parallel
polarisation products (corresponding to a minimum possible channel
width of $2.06\,\rm km\,s^{-1}$), while 7 remaining bands with 20 MHz
each and 128 channels and two parallel polarisation products were
observed in parallel, centered around a frequency of $1.4\,\rm
GHz$. The telescope performance was variable, and some observations
do not have all antennas operational.  Some observations were affected 
by radio interference, including terrestrial
RFI. Most notably, observations performed during the day time
were partly affected by solar interference on the shorter
baselines. This resulted in a quite variable rms noise across our
observations, as listed in Table \ref{tab:HIdata}.  Exploitation of the
continuum observations is the topic of a forthcoming paper. Here, we
describe the reduction of the HI data.

The WSRT HI data were reduced using a pipeline originally developed by
Serra et al. (2012), which is based on the Miriad reduction package
(Sault 1995). The pipeline automates the basic standard data
reduction steps from the HI raw data sets as delivered by the WSRT to
the final data cubes used for this work. 

After reading and converting
the UVFITS data into Miriad format, the system temperature as tracked
by the WSRT is used to perform a relative amplitude calibration, the
data are Hanning smoothed and then flagged to exclude radio interference using a
set of clipping algorithms. The absolute bandpass calibration and a continuum calibration is performed by applying a
calibration on standard calibration sources, which are observed
before and after the target observation.  The resulting
complex bandpass and continuum gain solutions are then copied to
the target data set. 

The continuum phase calibration is
adjusted by means of a self-calibration process using the target data
themselves. The averaged continuum data are inverted (employing
uniform weighting) and cleaned using a clean mask in an iterative
process, in which the threshold level to determine the clean mask and
the clean cutoff level are decreased until convergence is reached (the
mask threshold reaches $5\,\sigma_{\rm rms}$ and the clean cutoff
reaches $1\,\sigma_{\rm rms}$). At the same time, the clean components
are used to adjust the phase calibration of the visibilities by
performing several self-calibration steps.

The final self-calibration solution is then copied to the line data and a
continuum subtraction is performed on the visibilities by subtracting
the visibilities corresponding to the brightest sources in the field
(as determined in the selfcal loop) and then performing a polynomial
continuum subtraction of first or second order (the order being
increased from first to second order if the data inspection indicates
the need to increase the fitting order). After this step, the line
data are averaged, Hanning smoothed, and inverted using several
weighting schemes, to then be iteratively cleaned, again successively
decreasing the clean mask threshold and the clean cutoff
parameter. Other than for the continuum data, the clean regions were
determined using smoothed data cubes, accounting for the fact that the
HI line emission is mostly extended.

We produced data cubes using five different weighting schemes:
a) Robust weighting (Briggs 1995) of 0,
0.4, and 6 without tapering,   
b) Robust weighting  of 0 and 6 and an additional
30-arcsec tapering ( i.e. uv data are multiplied with
a Gaussian tapering function equivalent to a convolution with a
symmetric Gaussian of HPBW of $30^{\prime\prime}$ in the image
domain). All cubes have velocity resolution of 24.8 km/s after Hanning
smoothing. 
The quantitative analysis of the HI properties in this paper (e.g., mass, morphology)
was made using the cubes built with a Robust weighting of 0.4 and no tapering. This was the
most suitable compromise between  sensitivity and
resolution. 
The angular resolution
of the cubes is $\sim 15-20\times 15-20/\sin\left(\delta\right)$ arcsec$^2$,
where $\delta$ is the declination. The noise  ranges between 0.26 and 0.42
mJy$/$beam and the median value is 0.3 mJy$/$beam (90 percent of the cubes
have noise below 0.34 mJy$/$beam).  If the resulting data cubes are used
without further smoothing, the 5$\sigma$
column density threshold within one velocity resolution element is $8-14
\times10^{19}\times \sin\left(\delta\right)$ cm$^{-2}$.

We look for HI emission in the cubes using the source finder
developed by Serra et al. (2012). The finder is based on a
smooth-and-clip algorithm, with additional size-filtering to reject
noise peaks. We refer to Serra et al. (2012) for a more complete
description of the finder. In this work, we use a set of Gaussian
convolution kernels with FWHMs of 0, 48, 96, 144 and 192 arcsec 
in the spatial domain
combined with a set of convolution kernels with FWHM of $24.7,\, 49.5,\, 98.9,\, 197.9,\, 395.7\, \rm
km\,s^{-1}$ in the velocity domain, using a clip level of $4
\sigma_{\rm rms}$. The output
of the finder is a binary mask where pixels containing emission are
set to 1 and pixels with no emission are set to 0. 

\subsection{Moment-0 images and error estimation}
In this section, we describe how we extract two-dimensional  maps of
the overall gas distribution (moment-0 maps) for each of our targeted galaxies.  
We work with the masked HI cubes, i.e. cubes for which  
pixels where no HI emission is detected are set to  0. 
The value of a given pixel P1 in the two-dimensional HI map is calculated as the 
the sum over velocity channels that contain detected HI flux.  

\begin{figure}
\includegraphics[width=7cm]{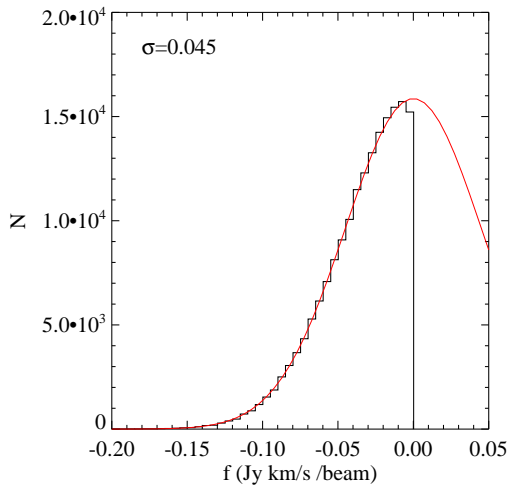}
\caption{An example of the distribution of the negative pixels used to estimate the error
on the flux at a given sky position. The red line shows the best-fit Gaussian
function and the width  $\sigma$ is denoted in the top-left corner.} 
\label{fig:noisefit}
\end{figure}

To calculate the error on the calculated flux for pixel P1, we define a set of "sky pixels"   
where the mask values are 0 over the same range of channels that contribute
to the flux calculated for pixel P1. We sum over the channels to produce a distribution
function of  "sky flux values" from these  empty regions.
The negative part of this distribution is very well
fit  by a Gaussian function (Figure~\ref{fig:noisefit}) and can be used to estimate the error
on the flux in P1. From the width of          
the Gaussian, we calculate $\sigma$, the error on
pixel P1. We repeat the above process for all non-zero pixels in the total
HI image and the end product of this process is the error image.

We make contour maps from the total HI image with a smoothing width of 3
pixels. We truncate the image at the  
contour where the median signal-to-noise ratio of the pixels along the contour reaches 2. This
corresponds to a median threshold of 0.46 $\times 10^{20}$ atoms cm$^{-2}$;
all of the images reach below column densities of 0.7 $\times 10^{20}$ atoms cm$^{-2}$ except galaxy 16 which
reaches a column  of 0.87 $\times 10^{20}$  atoms cm$^{-2}$ (Table~\ref{tab:HIdata}). 

\begin{figure}
\includegraphics[width=7cm]{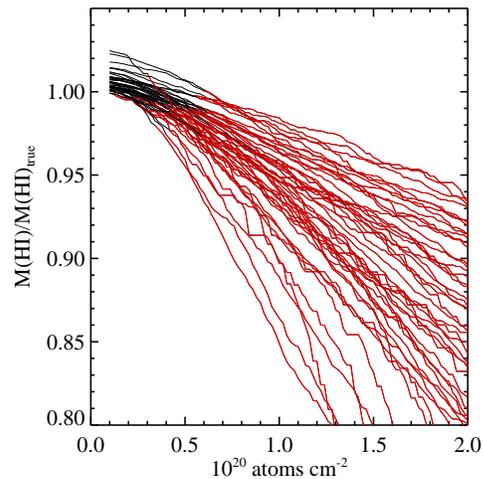}
\caption{The fraction of ``true'' HI mass of the Bluedisk galaxies detected above a
given column density limit as a function of the column density limit. The red parts of the lines are where the column
density limits are above our adopted $S/N>2$ threshold (see Section 2.3).} 
\label{fig:flux_thresh}
\end{figure}

In Figure~\ref{fig:flux_thresh}, we show how the HI mass of the
Bluedisk galaxies varies as a function of the column density limit if all  HI
emission below the limit is excluded. The ``true'' total HI  mass is defined
from summing flux from all pixels set to 1 in the mask created by the source finder.
We can see that at least 97 
percent of the total HI mass is included within our adopted $S/N > 2$ contour level threshold. 
We note that the morphological analysis in the following sections is applied to  
pixels in the 2D map that lie above this threshold. 
Table~\ref{tab:HIdata} lists the total HI mass as well as the $S/N>2$ column density
threshhold for all the galaxies in our sample.

\begin{table*}
\begin{center}
\renewcommand{\arraystretch}{1.}
\begin{tabular}{ccccccccccc}
\hline \hline
 \multicolumn{1}{c}{ID} & \multicolumn{1}{c}{date} & \multicolumn{1}{c}{noise} & \multicolumn{1}{c}{beam} & \multicolumn{1}{c}{$\log$ M(HI)} & \multicolumn{1}{c}{$\log$ M(HI)/M$_*$}   & \multicolumn{1}{c}{threshold (mom-0)} & \multicolumn{1}{c}{R1} & \multicolumn{1}{c}{R50}  & \multicolumn{1}{c}{R90} & \multicolumn{1}{c}{$\mu_{HI,25}$}\\
 \multicolumn{1}{c}{ } &   \multicolumn{1}{c}{ }   & \multicolumn{1}{c}{mJy beam$^{-1}$} & \multicolumn{1}{c}{arcsec$^2$} & \multicolumn{1}{c}{M$_{\odot}$} & \multicolumn{1}{c}{ } &  \multicolumn{1}{c}{10$^{20}$ atoms cm$^{-2}$} & \multicolumn{1}{c}{arcsec} &  \multicolumn{1}{c}{arcsec} & \multicolumn{1}{c}{arcsec} & \multicolumn{1}{c}{M$_{\odot}$ kpc$^{-2}$}  \\
\hline
   1 & 2012 FEB 24  &  0.28 &  26.9$\times$17.0 & 10.09	&-0.34  &0.44 &64 &32 &57  &6.59\\
   2 & 2011 DEC 23  &  0.28 &  25.9$\times$17.2 &  9.94	&-0.68  &0.53 &52 &23 &47  &6.70\\
   3 & 2012 JAN 09  &  0.29 &  25.8$\times$17.2 &  9.89	&-0.53  &0.49 &45 &21 &40  &6.76\\
   4 & 2012 JAN 04  &  0.32 &  31.4$\times$16.4 & 10.26	&-0.31  &0.64 &67 &28 &54  &6.85\\
   5 & 2011 DEC 20  &  0.31 &  27.2$\times$17.5 & 10.20	&-0.12  &0.39 &80 &40 &75  &6.80\\
   6 & 2012 JAN 12  &  0.28 &  28.1$\times$17.0 & 10.31	&-0.53  &0.40 &87 &44 &76  &6.65\\
   7 & 2012 JAN 13  &  0.29 &  25.5$\times$17.3 & 10.00	&-0.37  &0.53 &41 &26 &86  &6.63\\
   8 & 2012 JAN 15  &  0.28 &  23.9$\times$17.8 & 10.20	&-0.08  &0.47 &69 &36 &64  &6.76\\
   9 & 2012 JAN 20  &  0.32 &  22.5$\times$15.1 &  9.91	&-0.85  &0.60 &45 &20 &38  &6.70\\
  10 & 2012 JAN 22  &  0.29 &  20.5$\times$17.7 & 10.02	&-0.89  &0.56 &45 &28 &56  &6.78\\
  11 & 2012 MAY 10  &  0.29 &  21.9$\times$16.3 &  9.71	&-0.93  &0.51 &40 &21 &36  &6.67\\
  12 & 2012 MAR 04  &  0.31 &  19.0$\times$16.3 & 10.12	&-0.51  &0.62 &67 &33 &66  &6.72\\
  13 & 2012 JAN 16  &  0.29 &  28.7$\times$17.0 & 10.23	&-0.61  &0.45 &64 &34 &80  &6.69\\
  14 & 2012 FEB 11  &  0.26 &  22.0$\times$17.5 & 10.14	&-0.65  &0.48 &69 &38 &83  &6.70\\
  15 & 2012 MAR 10  &  0.29 &  29.2$\times$16.9 & 10.39	&-0.43  &0.41 &119&60 &110 &6.73\\
  16 & 2012 MAR 26  &  0.27 &  21.7$\times$17.6 & 10.00	&-0.31  &0.87 &44 &22 &45  &6.96\\
  17 & 2012 APR 18  &  0.35 &  21.3$\times$18.5 & 10.34	&-0.35  &0.65 &96 &57 &98  &6.67\\
  18 & 2012 APR 19  &  0.32 &  25.3$\times$20.0 & 10.07	&-0.27  &0.51 &57 &27 &51  &6.95\\
  19 & 2012 APR 10  &  0.41 &  25.5$\times$16.0 & 10.17	&-0.08  &0.60 &65 &36 &70  &6.81\\
  20 & 2012 JUN 15  &  0.30 &  25.2$\times$16.1 & 10.21	& 0.00  &0.63 &58 &28 &52  &6.97\\
  21 & 2012 FEB 20  &  0.32 &  26.7$\times$16.5 &  9.88	&-0.53  &0.42 &49 &31 &54  &6.50\\
  22 & 2012 APR 29  &  0.32 &  26.6$\times$15.9 & 10.00	&-0.82  &0.60 &52 &24 &46  &6.62\\
  23 & 2012 MAY 13  &  0.36 &  23.5$\times$15.0 &  9.94	&-0.81  &0.56 &48 &23 &39  &6.61\\
  24 & 2012 JAN 21  &  0.30 &  20.0$\times$17.6 & 10.21	&-0.50  &0.52 &70 &36 &61  &6.56\\
  25 & 2012 APR 22  &  0.28 &  20.0$\times$17.0 &  9.76	&-0.78  &0.46 &40 &21 &40  &6.69\\
  26 & 2011 DEC 21  &  0.32 &  25.3$\times$17.3 &  9.50	&-0.81  &0.49 &33 &15 &28  &6.72\\
  27 & 2011 DEC 11  &  0.32 &  28.5$\times$16.0 &  9.27	&-1.03  &0.31 &24 &14 &25  &6.46\\
  28 & 2011 DEC 13  &  0.29 &  22.2$\times$17.0 &  9.12	&-1.42  &0.26 &35 &23 &36  &6.36\\
  29 & 2012 JAN 27  &  0.27 &  21.9$\times$18.3 & 10.37	&-0.15  &0.55 &75 &45 &103 &6.83\\
  30 & 2011 DEC 07  &  0.32 &  26.5$\times$16.5 & 10.07	&-0.32  &0.56 &55 &25 &53  &6.75\\
  31 & 2012 JAN 27  &  0.27 &  23.5$\times$17.7 & 10.06	&-0.12  &0.34 &46 &33 &163 &6.96\\
  32 & 2011 DEC 09  &  0.33 &  29.4$\times$16.3 &  9.56	&-0.73  &0.35 &40 &23 &40  &6.41\\
  33 & 2012 JAN 31  &  0.27 &  22.6$\times$17.9 &  9.20	&-1.54  &0.09 &29 &17 &29  &6.17\\
  34 & 2012 FEB 22  &  0.31 &  28.9$\times$15.9 &  0.00	&--  & --  &-- &-- &--  &--\\
  35 & 2012 FEB 07  &  0.26 &  24.0$\times$17.6 &  9.93	&-0.68  &0.44 &61 &33 &73  &6.55\\
  36 & 2011 DEC 25  &  0.28 &  21.2$\times$17.5 &  9.39	&-0.95  &0.44 &29 &16 &28  &6.65\\
  37 & 2012 JAN 07  &  0.28 &  20.5$\times$17.6 &  9.70	&-0.69  &0.55 &37 &19 &32  &6.68\\
  38 & 2012 JAN 10  &  0.27 &  20.3$\times$17.9 &  8.75	&-2.00  &0.12 &18 &15 &26  &5.90\\
  39 & 2012 MAR 01  &  0.31 &  27.0$\times$16.1 &  9.14	&-1.70  &0.29 &31 &28 &44  &5.99\\
  40 & 2012 JAN 06  &  0.29 &  29.4$\times$17.0 &  9.65	&-0.71  &0.39 &38 &24 &46  &6.55\\
  41 & not observed &  --   &         --        &  --   &--     & --  &-- &-- &--  &--\\
  42 & 2011 DEC 05  &  0.31 &  22.6$\times$17.8 &  9.57	&-1.23  &0.28 &38 &23 &43  &6.36\\
  43 & 2011 DEC 17  &  0.29 &  25.8$\times$17.1 &  9.59	&-0.67  &0.31 &38 &22 &39  &6.58\\
  44 & 2011 DEC 08  &  0.31 &  24.0$\times$16.5 &  9.28	&-1.50  &0.21 &27 &16 &25  &6.30\\
  45 & 2011 DEC 26  &  0.29 &  21.3$\times$17.2 &  9.61	&-1.07  &0.32 &36 &19 &34  &6.54\\
  46 & 2012 MAR 11  &  0.31 &  28.1$\times$16.7 & 10.11	&-0.57  &0.56 &43 &29 &74  &6.74\\
  47 & 2012 MAR 03  &  0.32 &  29.7$\times$15.3 & 10.01	&-0.52  &0.39 &66 &41 &75  &6.60\\
  48 & 2012 MAR 04  &  0.37 &  25.3$\times$16.1 &  8.25	&-2.50  &0.44 &13 &15 &25  &5.85\\
  49 & 2011 DEC 10  &  0.31 &  31.9$\times$16.8 &  9.74	&-0.68  &0.40 &42 &23 &41  &6.48\\
  50 & 2012 MAR 01  &  0.28 &  19.6$\times$16.9 &  9.80	&-1.07  &0.60 &45 &20 &37  &6.50\\
\hline \hline
\end{tabular}
\caption{HI derived data products: the galaxy ID, observational date, r.m.s noise per channel (24.9 km s$^{-2}$), beam size, HI mass, total HI image $S/N>2$  threshold, R1, HI half-light radius, HI 90-percent-light radius and averaged surface HI density within R25.}
\label{tab:HIdata}
\end{center}
\end{table*}

\subsection{Predicted and  observed HI mass fractions}
Our  sample consists of 25 ``HI-rich'' galaxies and 25 ``control'' galaxies 
selected using the  HI mass fraction predicted using UV/optical photometry. Some of the disks 
turn out to be very extended and to overlap ( both spatially and in velocity) 
with one or  more neighbouring galaxies identified in the
SDSS and GALEX images (for example galaxy 7). If  $r_{neighbour} < r_{target}+3$ mag,  
the galaxy is flagged as a ``multi-source'' system; these include
galaxies  7, 13, 29, 31 and 46  (see Figure A3 is the Appendix). 

\begin{figure*}
\includegraphics[width=14cm]{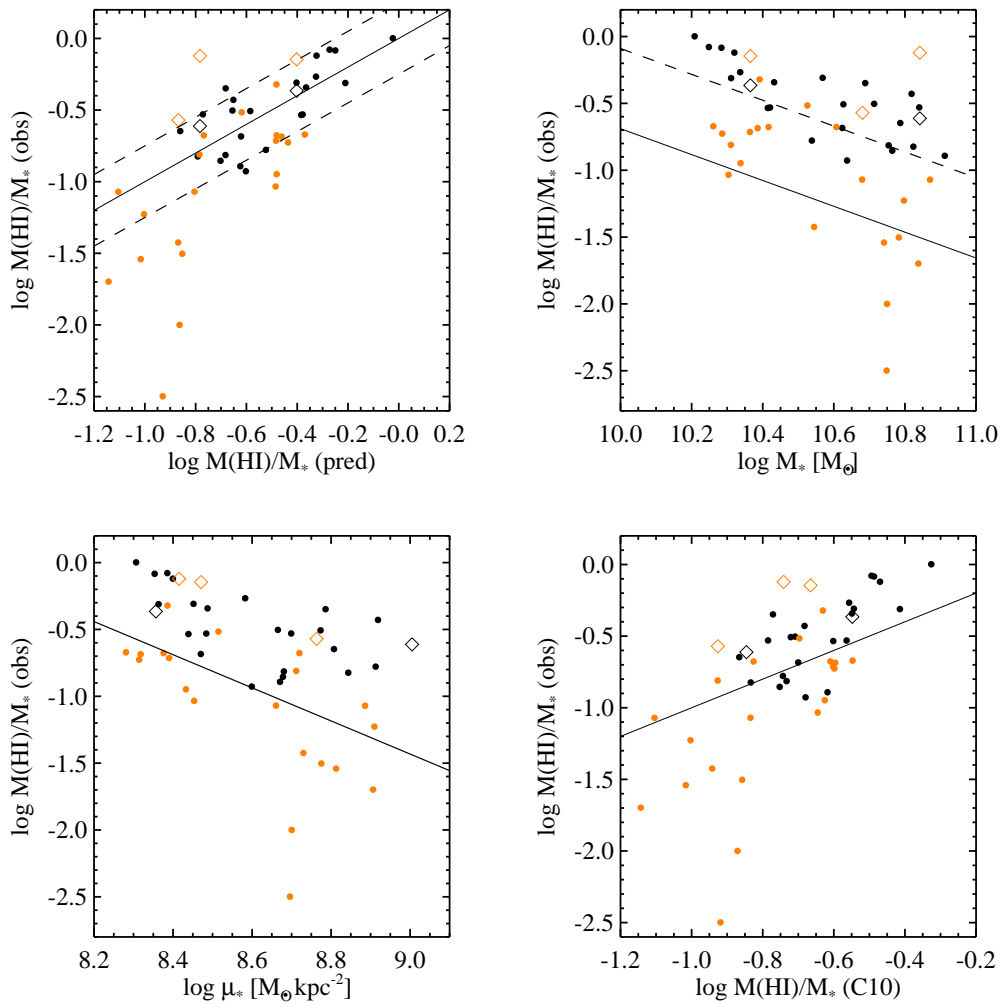}
\caption{Relation between the observed HI mass fraction and predicted HI mass fraction, stellar mass, mass surface density and C10 HI mass fraction. The black symbols are the ``HI-rich'' galaxies, the orange symbols are the ``control'' galaxies and the diamonds mark the ``multi-source'' systems. In the top-left panel, the solid line is the y$=$x line and the dashed lines are vertically 0.25 dex away from the y$=$x line. In the top-right panel, the solid line shows the median relation between HI mass fraction and stellar mass found by C10, and the dashed line is vertically 0.6 dex above the solid line.  In the bottom-left panel, the solid line shows the median relation between HI mass fraction and mass surface density found by C10. In the bottom-right panel, the solid line is the y$=$x line.} 
 \label{fig:fhi_pred}
\end{figure*}

In the top-left panel of Figure~\ref{fig:fhi_pred}, we compare the HI mass fraction predicted
by our UV/optical photometry-based estimator with the actual HI mass fraction measured using
our data. The galaxies predicted to be gas-rich are plotted as blue dots and the control
galaxies are plotted as red dots. The multi-source systems are enclosed by a 
green diamond. As can be seen, the estimator works very well
for the  ``HI-rich'' galaxies -- there is a scatter of only 0.15 dex between  
the predicted and observed values of $\log$ M(HI/M$_*$.  The estimator works 
considerably less well for the ``control'' sample. The observed HI mass fraction is generally         
lower than the predicted one. In fact {\em all} galaxies with predicted  HI mass fractions 
$\log$ M(HI)/M$_* < -0.85$  lie systematically below the predicted value, with the difference reaching more than
a factor 10 in two extreme cases. In the top right panel, the observed HI mass fractions are plotted
as a function of the stellar mass of the galaxy. We see that all the strongly outlying points
(those with $\log$ M(HI)/M$_* < -1.5$ in the top left panel) are galaxies with high stellar masses.  

Although the scatter in HI mass fraction in the control sample is larger than envisaged, 
the control sample is systematically gas-poor compared to the HI-rich sample at fixed
stellar mass and at fixed stellar surface mass density. The solid lines in the top right and
bottom left panels show the median relation between HI mass fraction and
stellar mass and stellar surface density from
Catinella et al (2010). Consistent with their definition, the HI-rich galaxies lie $\sim 0.6$
dex above these relations on average, while the control galaxies lie much closer to the median.
Finally, the bottom-right panel compared the observed HI mass fraction to that predicted
by the ``plane'' relating HI mass fraction with the stellar surface mass density $\mu_*$
and NUV-r colour discussed in Catinella et al (2010). As can seen, most of the ``HI-rich'' galaxies
lie above the plane, demonstrating that our criterion of requiring a galaxy to have an
unusually blue outer disk has selected unusually HI-rich galaxies by this metric as well.  
The majority of galaxies in the control sample lie below the plane. 

Finally, we note that two of the ``multi-source'' systems fall in the ``HI-rich'' category and for these, the HI fractions
are consistent with predictions. However,  3 control galaxies are also multi-source systems
and  all of these  have observed HI mass fractions that are larger than the predictions by 0.3-0.5 dex.

\section{Sample definitions and visual inspection of the HI intensity maps}
\label{sec:gallery}

Since this paper deals with HI sizes and morphologies, we will exclude the multi-source systems
from the analysis. In addition, 
galaxy 48 has a very complicated offset HI cloud that is kinematically connected only on one side of the galaxy
(see gallery in the Appendix), and is also excluded from the current study.
No observations exist for  galaxy 41;  galaxy 34 is not detected in HI (we have confirmed
this non-detection using the Arecibo telesope). In total, we are left with a sample of 42 galaxies, which 
we separate into two parts according to whether they lie above or below
the HI plane defined in C10. Hereafter, the galaxies which have $\Delta f(HI)>0$ are referred to as 
HI-rich galaxies and the rest of the sample are referred as  control galaxies.
This definition is  different to that adopted when we designed the observational sample,
but it now properly reflects the actual measured HI content of each system. 
In summary, our sample consists of 23 HI-rich galaxies and 19 control galaxies.

An atlas containing HI contour maps overlayed with the optical images is presented in the Appendix. 
The outermost contour corresponds to the $S/N=2$ threshold column density (see Section 2.3).
The HI-rich galaxy sample is shown in Figure~\ref{fig:G_HIrich},  
while the control sample is shown in Figure~\ref{fig:G_ctrl}. 
Comparison of the two figures shows that the HI-rich galaxies tend to have 
HI disks that are very extended compared to their optical disks. Most of the HI-rich 
galaxies have low column density outer contours that are irregular. 
In some cases, the HI appears to be very clumpy (galaxies 15, 17, 20, 47), but there 
are also HI-rich galaxies with smooth and symmetric HI disks (galaxies 1, 6, 22, 26). 
In contrast, the control galaxies have much smaller HI disks; the HI disks of some 
of the control galaxies (galaxies 33, 38, 39 and 44) even end within the optical disk. 
The smallest HI disks also tend to be mis-aligned with respect to the optical disks. 
Galaxies 38, 39 and 42 have highly asymmetric HI distributions. 
The outer contours of the HI disks of most of the control galaxies are smooth 
compared to the HI-rich galaxies, but there are also clearly-disturbed systems 
(galaxies 9, 10, 39 and 42). 

In the next section, we will attempt to put these ``visual impressions" on more 
quantitative footing.

\section{Quantitative analysis of the HI images}
All the HI parameters studied in the following sections 
are measured using the high resolution HI total intensity maps. 
The typical beam has a major axis of 22 arcsec and a minor axis of 16 arcsec ($\sim$ 12.0 and 8.5 kpc). 

\subsection{Size measurements}
We quantify the size of the HI disks
using three different measures:  R50(HI), R90(HI) and R1. 
R50(HI) is the radius enclosing half of the total HI flux , R90(HI) 
is the radius enclosing 90 percent of the HI flux. R1 is the radius 
where the face-on corrected angular averaged HI column density reaches 
1 M$_{\odot}$ pc$^{-2}$ (corresponding to 1.25$\times10^{20}$ atoms cm$^{-2}$).  Column densities are measured along elliptical 
rings, with the position angle and ellipticity determined from the $r-$band
image, and are corrected to be face-on by scaling the column density 
in each pixel by cos$\theta\sim$b$/$a, where b$/$a is the axis ratio
of the galaxy measured in the $r$-band, and $\theta$ is the 
corresponding inclination angle. We note that there could be differences between 
the inclination of the optical and the HI disks, but visual inspection shows that the differences are usually small. In future work, we will fit tilted-disk models
to the velocity fields in order to derive the inclination of the HI disk. 

\begin{figure*}
\includegraphics[width=18cm]{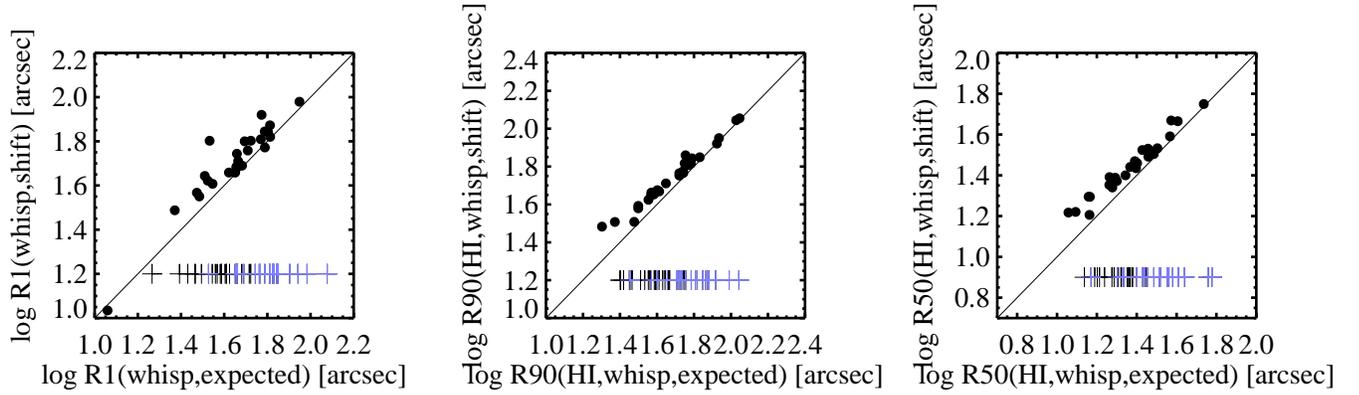}
\caption{WHISP galaxies are shifted and convolved with the WSRT beam to have similar 
appearance to the Bluedisk galaxies. R(shift) are the sizes measured from 
the shifted and convolved images, and R(expected) are the intrinsic sizes expected 
at the redshifts of the Bluedisk galaxies. The crosses show the sizes of the Bluedisk galaxies.} 
 \label{fig:size1}
\end{figure*}

One might worry that HI-sizes of Bluedisk galaxies will be over-estimated with respect to their optical
sizes, because of resolution effects. In order to quantify this for our Bluedisk sample, 
we have transformed the images of nearby, large angular size galaxies by placing them at the
same redshift as the Bluedisk galaxies and convolving them with the WSRT beam.
We selected 51 galaxies from the WHISP survey  \citep[Westerbork observations of neutral 
Hydrogen in Irregular and SPiral galaxies,][]{vanderHulst01,Swaters02}  with optical images available 
from the SDSS and with stellar masses above 10$^{9.8}$ M$_{\odot}$. The WHISP galaxies 
have a median distance of z$\sim$0.008 and a median apparent size of 2.3 arcmin. 
We make use of the full resolution total intensity maps, which were obtained with
a typical beam of  16$\times$10 arcsec$^2$. Since the HI disk sizes are much larger 
than the beam size, the sizes measured from the originial WHISP maps can be viewed as 
intrinsic sizes. Then the WHISP galaxies are shifted to the median redshift of 
the Bluedisk galaxies (0.026) by rebinning the pixels, and convolved with a Gaussian kernel so that they end up with a PSF of 22$\times$16 arcsec$^2$.
We measure the apparent sizes R50, R90 and R1 from the shifted WHISP galaxies (size(shift)), 
and compare them with the apparent sizes that are expected at the redshift of 0.026 
from the intrinsic values measured from the original WHISP images (size(expected)).  
We can see from Figure~\ref{fig:size1} that  size(shift) correlates tightly 
with size(expected), and that for most galaxies, size(shift) is larger than size(expected)
by $<$0.1 dex. At the smaller size end where R90$<$10$^{1.4}$ arcsec 
or R50$<$10$^{1.2}$ arcsec, the over-estimation can be as large as 0.2 dex, 
but almost all the Bluedisk galaxies have an apparent size larger than that.
We thus conclude that we can robustly measure the sizes R50, R90 and R1 from 
the Bluedisk total intensity maps.

\subsection{Morphology measurements}
\subsubsection{CAS parameters}
Gini, M20 and Asymmetry ($A$) are traditional morphological parameters
based on the  concentration-asymmetry-smoothness (CAS)
system \citep{Lotz04,Conselice03}, which have been widely applied in
optical studies of galaxies. 
The Gini parameter measures the smoothness of the light,
M20 measures the central concentration of the
light and $A$ measures the 180 degree rotational
difference around the center. Individually, or in
combination, these parameters have been demonstrated to be
effective in detecting signatures of galactic mergers or
interactions. 

In this paper, we measure Gini, M20 and $A$ using the HI total intensity maps. 
Most of the calculation steps are similar to        
\citet{Lotz04}, with the following adaptations: 
(1) we use a lower limit of 0.7$\times$10$^{20}$ atoms
cm$^{-2}$ to select the pixels used in the calculation.
As described in the previous section,
0.7$\times$10$^{20}$ atoms cm$^{-2}$ is a safe lower limit for all the
images with a Robust weighting of 0.4 and no tappering; (2) we do not apply a background correction to $A$
(while the optical analysis does) because the background noise is a
channel dependent term. 
We note that the optical criteria used to identify interactions/mergers
will not be directly applicable here. Our intention is to carry out a {\em relative}
comparison between HI-rich Bluedisk galaxies and the control sample.

\subsubsection{Morphological parameters that are sensitive to lower column-density gas}
\begin{figure*}
\includegraphics[width=14cm]{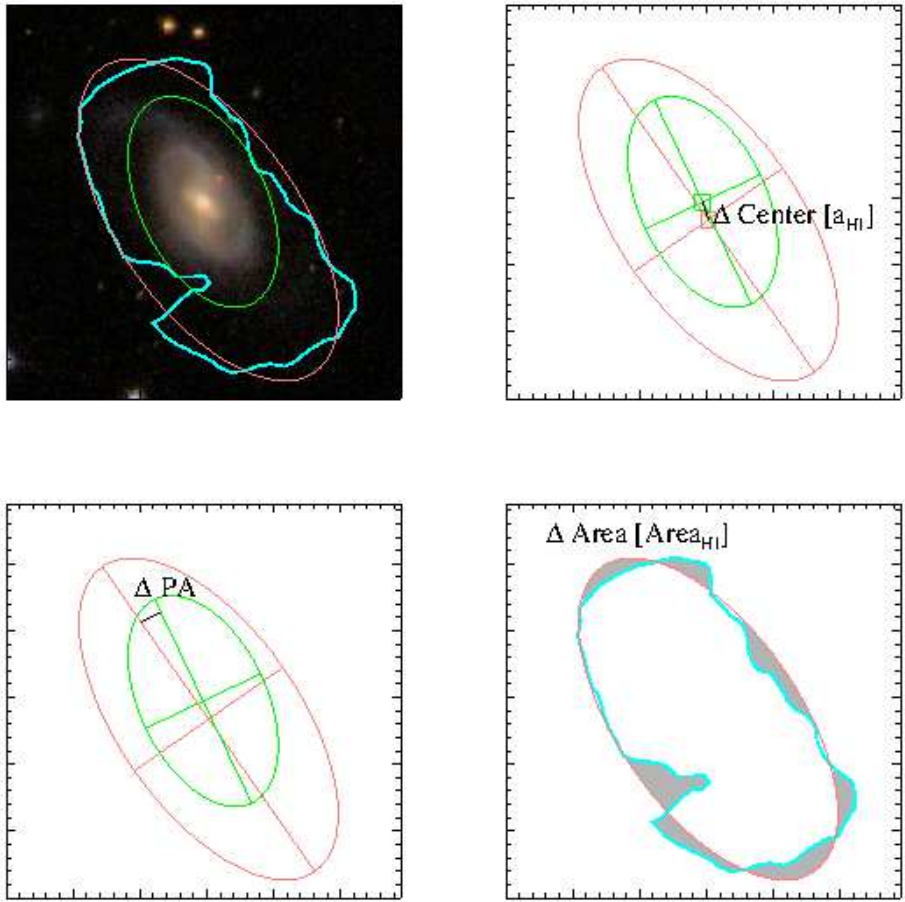}
\caption{An illustration of how $\Delta Center$, $\Delta PA$ and $\Delta Area$ 
are measured for one galaxy. The left-top panel displays the optical image 
overlayed with the 0.7$\times$10$^{20}$ atoms cm$^{-2}$ HI contour (cyan line). 
The pink ellipse is the best-fit ellipse for the region enclosed by the cyan contour 
and the green ellipse indicates the ellipse-fit to the $r$-band image from
which we measure the  center, position angle and ellipticity of the optical disk. 
The top-right panel illustrates how $\Delta$Centeris measured as the distance 
between the centers of the pink and the green ellipses, normalized by the 
semi-major axis of the pink ellipse. The bottom-left panels shows how 
$\Delta PA$  is measured as the difference between the major axis orientations
of the pink and green ellipses. The bottom-right panels show how
$\Delta$Area is measured as the area of the grey regions, normalized by the 
area enclosed by the cyan line. } 
 \label{fig:morph_def}
\end{figure*}

Gini, M20 and $A$ are mainly sensitive to the
regions of the galaxies containing high column-density gas, and thus may not be sensitive
diagnostics of recent accretion events. We quantify the irregularity of the
low column-density outer disks by defining 3 new parameters: $\Delta$Center, 
$\Delta$PA and $\Delta$Area, which are all measured from the
0.7$\times$10$^{20}$  surface density contour of the HI maps. 
For each galaxy, we fit an ellipse to the 0.7$\times$10$^{20}$ contour 
by using the uniformly weighted second order moment of the pixel positions inside 
the contour (see the left-top panel of Figure~\ref{fig:morph_def}). 

Optical centers, position angles and ellipticities are measured from the SDSS $r$-band 
images using SExtractor \footnote[3]{http://www.astromatic.net/software/sextractor}, 
using the standard method of measuring the flux weighted second order moments 
of the spatial distribution of the flux \citep{Bertin96}. 
Our new morphological parameters are defined as follows:

\begin{enumerate}
\item $\Delta$Center is calculated as the distance between the center 
of the HI ellipse and the center of the $r$-band ellipse, normalized by the semi-major axis 
of the HI ellipse (see the right-top panel of Figure~\ref{fig:morph_def}). 
\item $\Delta$PA is calculated as the difference between the position angle of the 
HI ellipse and the position angle of the $r$-band ellipse 
(see the left-bottom panel of Figure~\ref{fig:morph_def}). 
The position angle of face-on disks cannot be accurately estimated from images alone, 
so galaxies which have the minor-to-major axis ratio larger than 0.85 either in the optical 
or in the HI are excluded. This affects  2 galaxies from the HI rich sample 
and 2 galaxies from the control sample. 
\item $\Delta$Area is calculated as the difference in  area enclosed
by the  0.7$\times$10$^{20}$ HI contour and the best-fit ellipse, 
normalized by the total area inside the contour (see the right-bottom panel of Figure~\ref{fig:morph_def}). 
\end {enumerate}

$\Delta$PA and $\Delta$Center both measure the mis-alignment of the HI disk
with respect to the stellar disk. The mis-alignment could be due to clumpiness in the HI distribution. When the HI disk is more extended than the optical
disk, the mis-alignment is very likely to be caused by a warp \citep{Jozsa07}. 
Specifically, a significant $\Delta$Centre would be indicative of an asymmetric warp, and
a large $\Delta$PA would be indicative of a symmetric, S-shaped warp. 
There have been studies showing that large HI warps  are more frequently found in
galaxies in rich environments \citep{Garcia02}.  

Finally, $\Delta$Area measures the distortion of the outermost 
part of a disk from a symmetric elliptical shape, i.e. the clumpiness of the outer HI disk. 

\subsection{Error estimation for size and morphological measurements}
We estimate the errors on the HI size$\slash$morphological parameters  by
means of Monte-Carlo simulations. We construct a set of  ``perturbed'' cubes  
and then calculate the variation 
of the size$\slash$morphological parameters.  The r.m.s value of
the noise  for each cube is listed in Table~\ref{tab:HIdata}, 
but we must take into account the fact that the pixels in the data cube
are not independent. We simulate the noise by assuming that it has a  Gaussian distribution 
with $\sigma_{gauss}$ approximately equal to the r.m.s of the data cube.  We construct 200
noise cubes and then convolve them  
with the WSRT beam in the spatial domain and a Gaussian 
with FWHM of 2.2 pixels in the velocity domain \footnote {In practice, we tune the value of $\sigma_{gauss}$ 
until the r.m.s of the convovled noise cubes matches that of the actual data cube.}. 
We create 200  ``perturbed'' data cubes by adding each noise cube to the real data cube. So the perturbed data cube has a noise that is $\sim\sqrt{2}$ times the noise in the actual data cube. 
We then project the 200 ``perturbed'' data cubes to form
200  HI images, using the same masks as  for the original data cube.
\footnote {We have tested that re-running the mask finding routine on each of the perturbed
data cubes makes essentially no difference to final result, because we are not working
in the limit of very noisy data.}
We measure sizes and morphological parameters for  
these 200 perturbed HI images. We find the mean of the 200 measurements does not vary much (smaller than the errors shown below) from the parameters measured from true data cubes, suggesting no significant systematical effect caused by noise. The standard deviation around the mean 
is adopted as the error on each parameter. The perturbed data cube has a larger noise than the actual data cube, so the estimated error can be viewed as a lower limit of the true error. This is confirmed by enlarging the noise in the noise cube, and we find the errors of parameters increase nearly linearly with the noise of the perturbed cube.

The Bluedisk galaxies have typical 1 $\sigma$ error of 0.24 arcsec in R50,  
0.55 arcsec in R90, 0.86 arcsec in R1, 0.042 in M20, 0.007 in Gini, 0.017 in A, 
0.013 in $\Delta$Area, 1.4 deg in $\Delta$PA, and 0.006 in $\Delta$Center.

\section{Results}

\subsection{The HI mass-size relation}
It is well known that there is a  tight correlation between D1, the diameter of the
HI disk, and the total mass in atomic gas (M(HI)) in the galaxy. The relation changes 
very little from normal spiral galaxies \citep{Broeils97},  
cluster spiral galaxies \citep{Verheijen01}, late-type dwarf galaxies \citep{Swaters02} 
to early-type disk galaxies \citep{Noordermeer05}. 

\begin{figure}
\includegraphics[width=6cm]{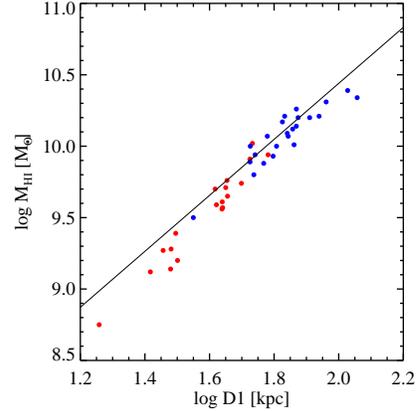}
\caption{The  M(HI)- D1 relation. The solid line is from Broeils et al. 1997. 
The blue dots are the HI rich galaxies and the red dots are the control galaxies. The typical 1 $\sigma$ error for D1 is 0.86 arcsec (Sect. 4.3).} 
 \label{fig:D_HIM}
\end{figure}

In the left panel of Figure~\ref{fig:D_HIM}, the solid line shows the best-fit relation 
between D1 and M(HI)  from \citet{Broeils97}, and our Bluedisk galaxies 
are plotted on top. The HI-rich galaxies in our sample lie on the same D1-M(HI) relation as 
normal spiral galaxies, though offset to higher HI mass and larger HI size compared
to the control sample. The control galaxies lie exactly on the  \citet{Broeils97} 
relation when M(HI)$>$10$^{9.3}$M$_{\odot}$, but deviate towards higher D1 values at fixed
M(HI)  by 0.05-0.1 dex when M(HI)$<$10$^{9.3}$M$_{\odot}$.  These galaxies all have 
R1$<$35 arcsec, and according to Figure~\ref{fig:size1} their sizes are over-estimated 
by $\sim$0.1 dex. We note that  
5 out of the 6 galaxies with the lowest values of M(HI) lie far below the 
C10 HI plane  ($\Delta$f(HI)$<-0.48$). The only exception, galaxy 27,
has an HI distribution that  is strongly off-center compared to the optical disk 
(see  Section~\ref{sec:gallery}). 

\begin{figure*}
\includegraphics[width=16cm]{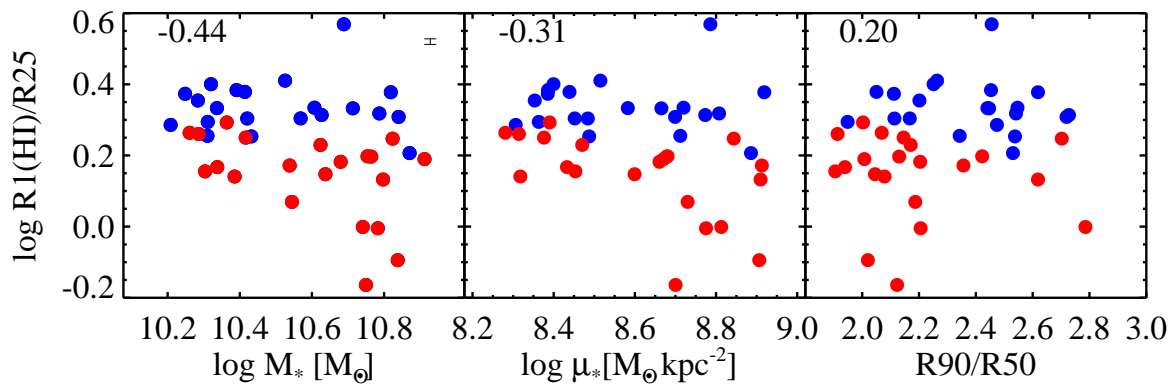}
\caption{The relation between the size ratio R1$/$R25 and stellar mass (M$_*$), 
stellar mass surface density ($\mu_*$) and concentration (R90$/$R50). 
The blue dots are the HI rich galaxies and the red dots are the control galaxies. 
The correlation coefficients for the whole sample are indicated at the top of
each panel. The typical 1 $\sigma$ error bar for R1$/$R25 is plotted at the corner of the first panel.}  
 \label{fig:size_prop}
\end{figure*}

In Figure~\ref{fig:size_prop}, we plot the ratio of HI and optical sizes (R1$/$R25) 
as a function of the stellar mass of the galaxy, stellar surface mass density
and concentration index of the $r$-band light.  We can see that R1$/$R25 does
not correlate with stellar concentration, which is related to the bulge-to-disk ratio of the galaxy.
R1$/$R25 is only weakly correlated with stellar mass and mass surface density. 
At  fixed stellar mass, surface density and  concentration, the HI rich 
galaxies have larger R1$/$R25 than the control galaxies. The difference
is about 0.2 dex in log(R1$/$R25) on average. 

\subsection{Comparison of disk structure in HI, UV and optical}
In this section, we compare structural parameters derived for  the HI distribution, 
including size, concentration, surface density and asymmetry index with those
derived from the UV and optical light. As we progress to longer wavelengths, we 
investigate the structure of progressively older stellar populations.
The goal of this exercise is to investigate the degree to which the structure of the  HI
is correlated with the structure of the old stars.

\begin{figure*}
\includegraphics[width=16cm]{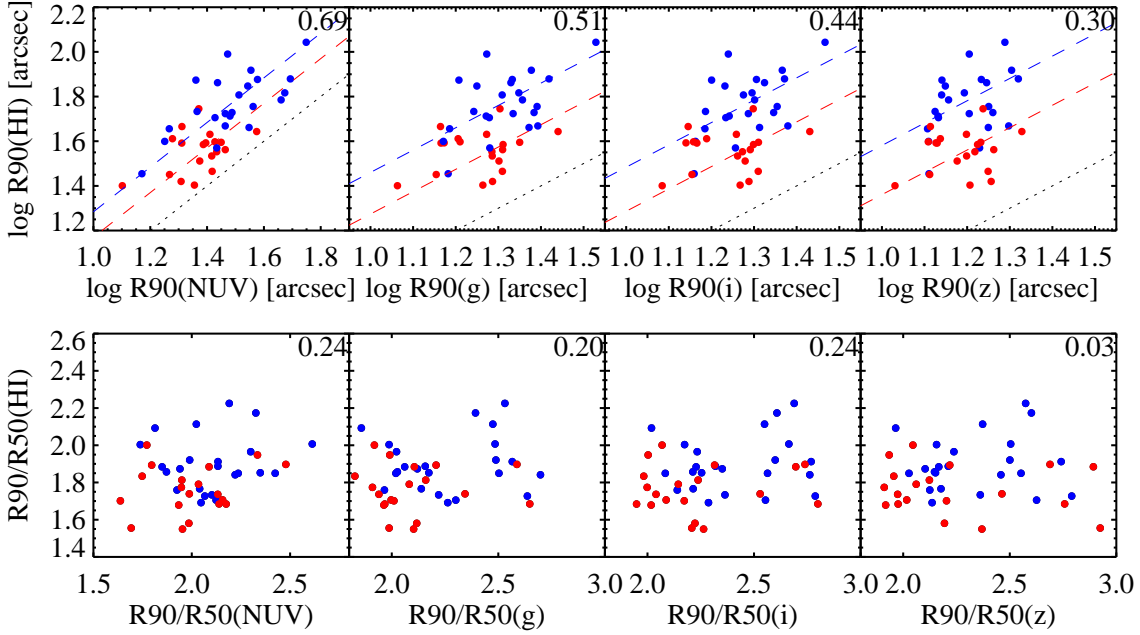}
\caption{Correlations between HI, NUV, g, i and z-band sizes. The blue dots are the HI-rich 
galaxies and the red dots are the control sample. The dotted line is the x$=$y line and 
the dashed lines indicate the median ratio of HI-to-optical/NUV size. The numbers on the
top left corners of each panel are the correlation coefficients for the whole sample. The typical 1 $\sigma$ error bars for R90(HI) and R90$/$R50(HI) are 0.005 dex and 0.003.} 
 \label{fig:Diameter}
\end{figure*}

We compare the size  R90(HI)  with  R90 measured from the NUV, $g$, $i$ and $z$ band 
images  in the top row of Figure~\ref{fig:Diameter}.  In the top left panel
we see that R90(HI) correlates well with R90(NUV), with a correlation coefficient of
$\sim 0.7$. The HI-rich galaxies are offset from the control galaxies both in HI
size and in UV-size. In the next three panels, we see that  
the correlation between R90(HI) and R90(optical) becomes progressively worse
towards longer wavelengths and HI-rich galaxies are no longer significantly offset along
the x-axis. 
The second row of the figure repeats the same progression for the concentration index R90$/$R50.
The correlation between HI concentration and UV/optical concentration is significantly weaker
than that for size. The progression from the UV to the z-band
in terms of the strength of the correlations is not seen.  

\begin{figure*}
\includegraphics[width=16cm]{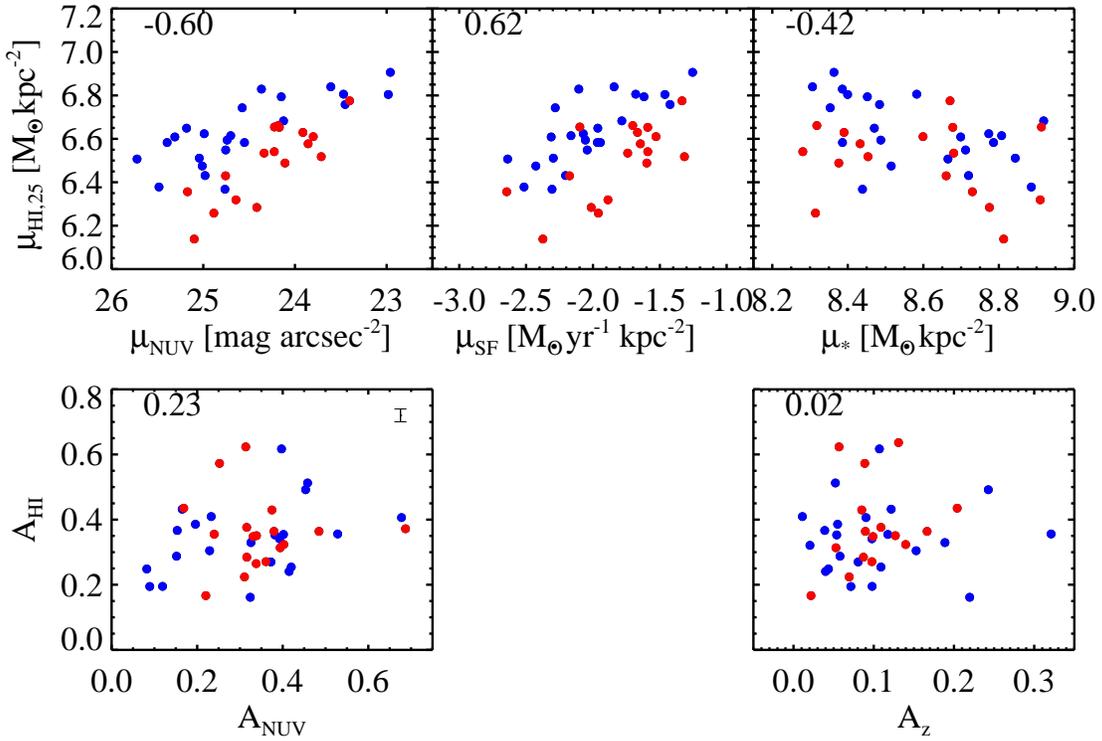}
\caption{Comparison of the surface density and asymmetry between HI, NUV (left panel), star formation (middle panel) and optical z band$\slash$stellar mass(right panel). The HI surface density here is measured as the averaged surface density within the optical R25. The blue dots are the HI rich galaxies and the black dots are the control galaxies. The numbers on the left-top corner is the correlation coefficient for the whole Bluedisk sample. The typical 1 $\sigma$ error bar for A(HI) is plotted at the corner of the left-bottom panel.} 
 \label{fig:MuAsym}
\end{figure*}

The top row of Figure~\ref{fig:MuAsym} shows how the
average HI surface density within R25 are correlated with
average UV surface density, star formation 
rate surface density and stellar surface mass density.  
The right-hand panel shows that there is an overall anti-correlation 
between $\mu_{HI}$ and stellar mass surface density.
There is a suggestion of a  ``break'' in the relation between
$\mu_{HI}$ and $\mu_{*}$ at $\log \mu_* \sim 8.6$, but the sample size is too small
to confirm this for sure. We note that $\log \mu_* \sim 8.6$ corresponds to
the stellar surface mass density where Catinella et al (2010), Saintonge et al (2011)
and Kauffmann et al (2012) identified a sharp transition to a population of
``quenched'' galaxies. Below this surface mass density threshold, the HI and stellar surface densities 
do not appear to correlate with each other. Above this threshold, there is an apparent anti-correlation.
We caution, however, that the stellar surface density includes the contribution of both the bulge
and the disk in this regime.

$\mu_{HI}$ correlates with both $\mu_{NUV}$ and $\mu_{SF}$. Interestingly, the HI-rich galaxies have similar star formation rate surface density as the control sample. 

In the bottom panels, we look at the correlations between the HI asymmetry index $A$ and
the the asymmetry index measured from the GALEX UV images and the SDSS $z$-band images.  
The correlation index hints at a potential correlation between A$_{HI}$ and A$_{NUV}$, though this is difficult to substantiate in our data. A$_{HI}$ is not correlated with A$_{z}$.

In summary, therefore,  HI-rich galaxies lie on the same scaling relation between HI disk size
and HI mass found for normal spirals. However, they are significantly displaced with respect to
the control sample in relations that link optical and HI quantities. At fixed stellar mass,
optical concentration index, stellar surface density and star formation rate surface density, HI-rich
galaxies have significantly more extended HI disks than the control galaxies.
The size of the HI disk only correlate  with that of 
of the optical disk for galaxies with stellar surface mass 
densities less than $\sim 3 \times 10^8 M_{\odot}$ kpc$^{-2}$.  
The radial extent of the UV light  is found to be the best
proxy for the radial extent of the HI disk.  Other structural properties derived from the UV images,
such as concentration and asymmtery, 
correlate only weakly with those derived from the HI maps.    

\subsection{Analysis of morphological parameters}
\begin{figure*}
\includegraphics[width=16cm]{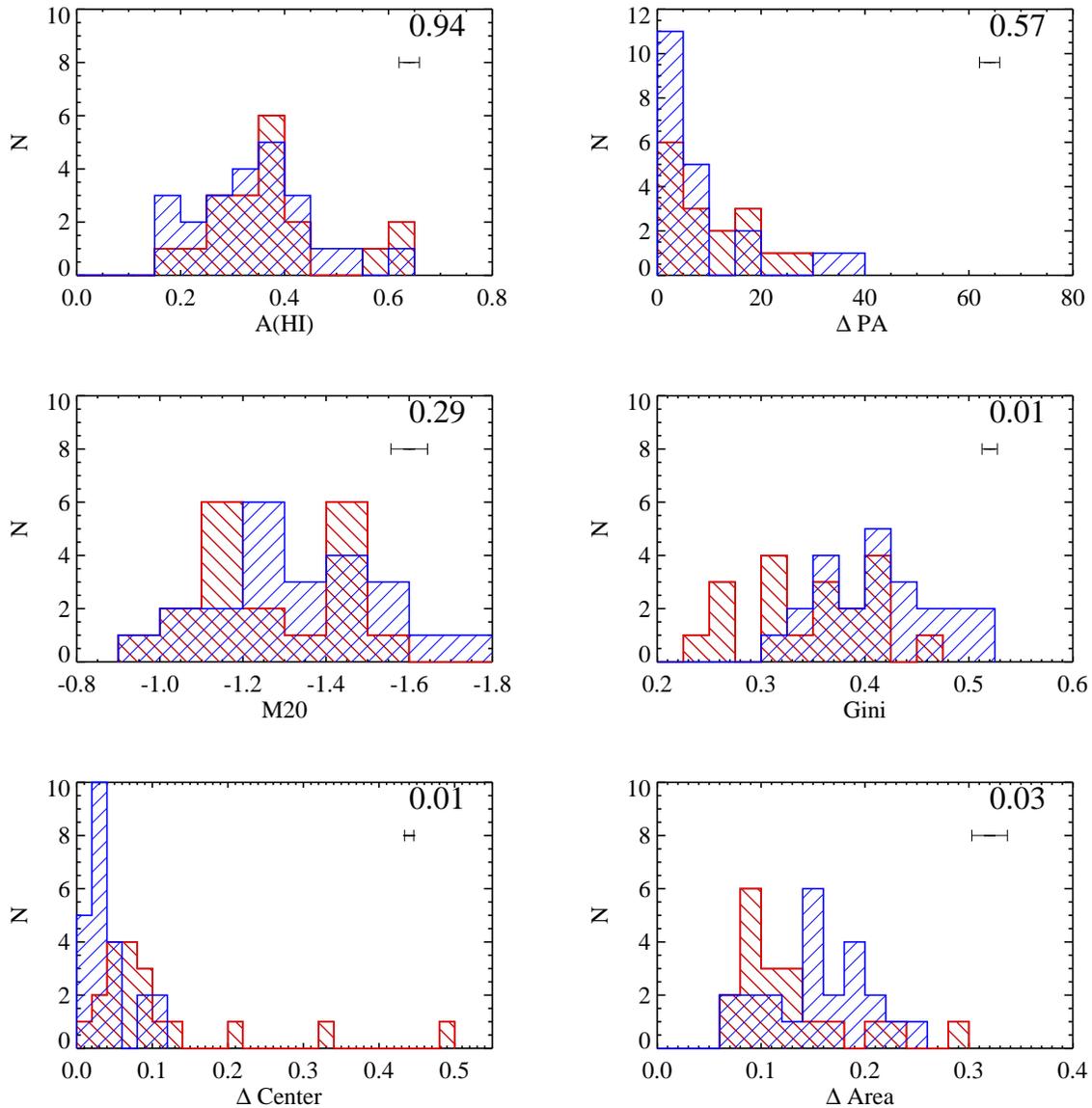}
\caption{Histograms showing the distributions of 6 different morphological parameters.  HI-rich
galaxies are shown in blue and control galaxies in red. KS test probabilities that the red and
blue histograms are drawn from the same underlying distribution are indicated in each panel. The typical 1 $\sigma$ error bars for the morphological parameters  are also plotted in the top-right corners.} 
 \label{fig:morph_hist}
\end{figure*}

Figure~\ref{fig:morph_hist} presents distributions of the six morphological parameters discussed
in Section 4. The blue histograms show results for the HI-rich galaxies, while the red 
histograms show results for the control sample. The KS-test probability that the 
blue and red histograms are drawn from the same underlying distribution is
indicated in the top right corner of each panel.

We see that HI-rich galaxies do not differ from the control sample in their distribution of
HI asymmetry or concentration indices. There is no evidence that the position angle of the HI
disk is more offset with respect to the optical disk for the HI-rich sample.

The morphological parameters that differs most significantly between the HI-rich and control samples 
is the Gini coefficient. The median value of Gini is offset to significantly higher
values in the HI-rich sample, indicating that the HI distribution is more clumpy. 
There is also a significant difference between $\Delta$Center for the two samples.
The control sample exhibits a tail of galaxies where the center of the HI distribution and
the center of the optical light distribution are significantly offset from each
other. This tail is missing in the HI-rich sample, and corresponds to the small HI disks (see Sect. 3). If we only compare the distributions when $\Delta$Center$<$0.15, the KS test probability for similarity between the two samples is still 0.03. Finally, there is weak  evidence 
from differences in the $\Delta$Area histograms, that HI-rich galaxies may have more irregular
outer contours {\em on average} than the control galaxies.   

\begin{figure*}
\includegraphics[width=16cm]{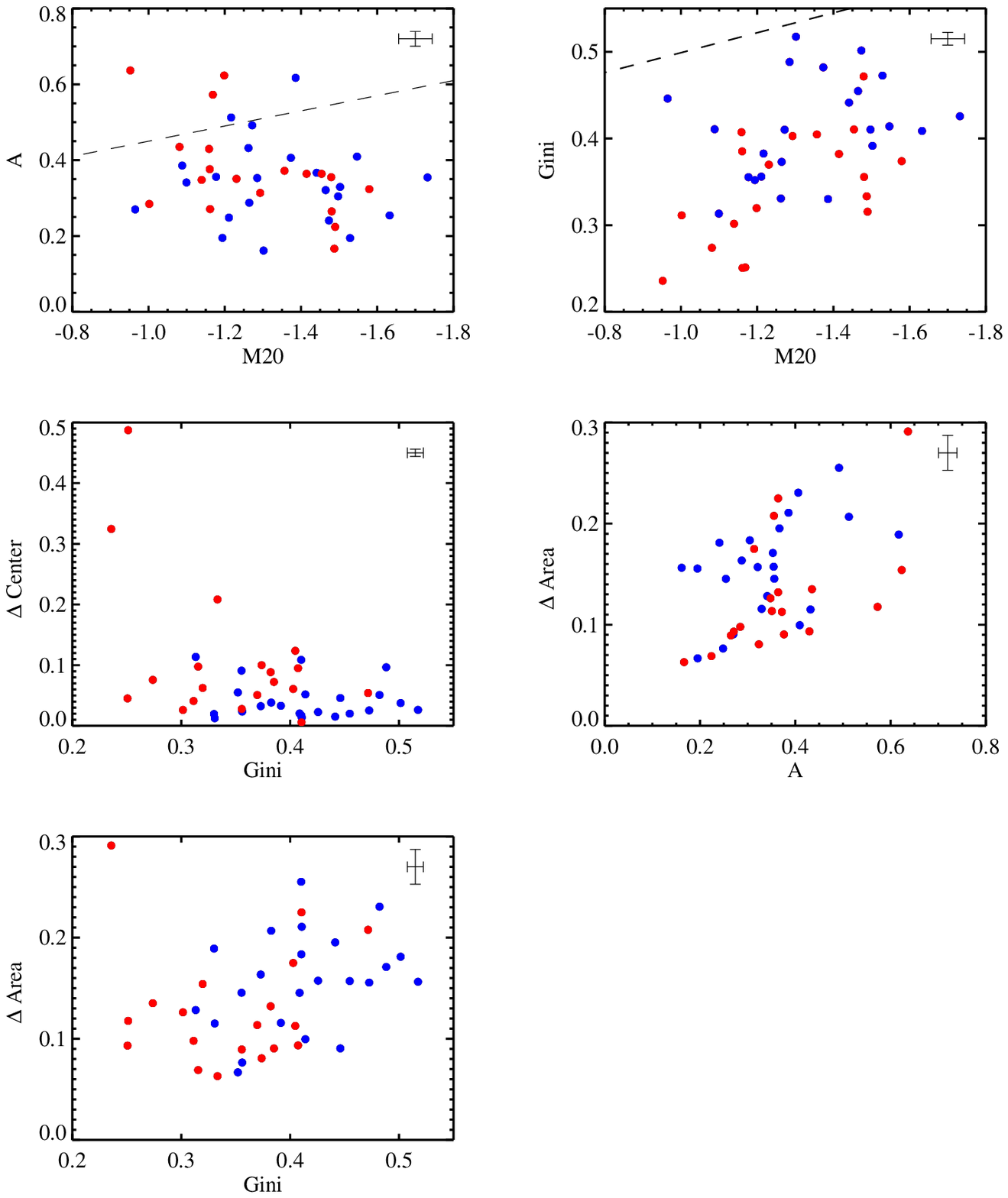}
\caption{ Scatter plots showing HI-rich (blue) and control (red) galaxies
in 2-dimensional  morphological parameter spaces. The dashed line is taken from
\citet{Holwerda11c}; most of the prominently interacting  galaxies in the WHISP
sample lie above this line. The typical 1 $\sigma$ error bars for the morphological parameters are plotted in the top-right corners.  } 
 \label{fig:morph_all}
\end{figure*}

In Figure~\ref{fig:morph_all}, we show a variety of scatter-plots of one morphological
quantity as a function of another one. The panel that shows the most striking
separation between HI-rich galaxies and the control sample is the middle left
one where $\Delta$(Center) is plotted as a function of Gini coefficient.
The segregation of the HI-rich population to a very narrow region of parameter
space enclosing high Gini and low $\Delta$ Center values is very striking. The HI-rich galaxies have a median $\Delta$Center of 0.032 with a scatter of 0.030, while the control galaxies have a median $\Delta$Center of 0.072, with a scatter of 0.11 The HI-rich galaxies are also more frequently scattered to larger values of $\Delta$Area at a fixed asymmetry than the control galaxies.
We conclude that gas-rich galaxies have extended and clumpy disks, 
which are positioned almost exactly
on the center-of-stellar mass of the system. 

In contrast, the top right panel shows that HI-rich and control galaxies do not segregate
at all the plane of Gini versus M20. This is the  diagram traditionally used 
by optical astronomers to identify galaxies that are interacting with a companion \citep{Lotz04}. 
To assess whether  the HI-rich galaxies have large value of Gini because of clumpy light distribution or high central concentration), we measure M20 and Gini for both HI-rich and and control samples within 3$\times$R25, 2$\times$R25 and R25. The KS test probabilities for the two samples to have the same M20 from these three measurements are 0.29, 0.92 and 0.42, and the KS test probabilities for Gini are 0.02, 0.13 and 0.1. Within 2$\times$R25, the HI-rich and control galaxies have the same central concentration (M20), but very different Gini. The light between R25 and 2$\times$R25 shows the most striking difference between the HI-rich and control galaxies.

In a series of papers, Holwerda et al (2001a,b,c,d) applied the CAS
morphological
parameter system to  HI maps of  nearby galaxies. In
particular, he investigated whether these parameters could distinguish
galaxies that were visually identified as  actively interacting in
HI images. The combination of asymmetry and M20 was found to
be an efficient diagnostic of mergers. One advantage of applying
the analysis to HI rather than optical images, is that a major
merger event is predicted from simulations to be visible  for a  longer
timescale \citet{Holwerda11c}. The criterion that Holwerda et al
adopted to identify interacting HI disks in the  Asymmetry versus
M20 plane is  plotted in the top left panel of Figure~\ref{fig:morph_all}
as a dashed line; galaxies above the line are identified as
interacting.  2 HI-rich galaxies and 3 control galaxies meet that
criteria (galaxies 15, 17, 38, 39 and 42), which is comparable to the interaction fraction of 13\%
in the WHISP sample  calculated by \citet{Holwerda11d}.

\section{Summary and discussion}
This paper presents the sample selection, observational set-up,  data
reduction strategy, and a first comparison of the sizes and structural properties
of the atomic gas disks for a sample of  25 unusually HI-rich galaxies and a control
sample matched in stellar mass, stellar mass surface density, inclination and redshift. 

Our main results may be summarized as follows:
\begin {itemize} 
\item  HI-rich galaxies lie on a direct extrapolation of the HI mass versus HI size relation
exhibited by  normal spirals, extending it HI masses of 
$\sim 2 \times 10^{10} M_{\odot}$ and radii of $\sim 100$ kpc. The scatter about this relation for the HI-rich and control
samples is about the same.
\item  HI-rich galaxies have  HI-to-optical size ratios that are displaced to
significantly larger values at fixed stellar mass, concentration index, stellar surface mass density
and star formation rate surface mass density.
\item The sizes and structural parameters of  HI disks correlate very weakly with those of
the optical disk, particularly for galaxies with high stellar
surface mass densities. The tightest correlations are found at ultra-violet wavelengths. 
between HI sizes/surface mass densities an UV sizes/surface brightnesses. This is true
both for the HI-rich and the control samples.
\item Of the 6 morphological parameters investigated in this paper, the Gini coefficient proved to be the
best discriminator of the HI-rich population.   
\item In HI-rich galaxies, the center of the HI disk tends to correspond more closely with the center
 of the optical disk than in the control sample.
\item There is no evidence that the atomic gas disks of HI-rich galaxies are more asymmetric 
than in the control sample, though the outermost regions tend to be more irregular. 
\end {itemize}

One of the key goals of the Bluedisk project was to study the morphological and dynamical evidence of
recent gas accretion for galaxies with excess HI gas.  
Our results argue against a picture where gas-rich galaxies have experienced recent major interactions.  
The most striking difference between the HI-rich galaxies and the control sample are their
clumpy HI disks that are generally very precisely centered on the central peak of 
stellar mass distribution and that are more extended with respect to the optical
light compared to the control galaxies. However, these extended HI disks lie on a direct extrapolation of
the  HI size versus mass relation for ``normal" spirals.   

The similarity of the HI disks suggests that the excess gas must come in
with a broad range of angular momentum and in a relatively
well-ordered way. It is possible that the ``order" results from the gas
being  initially in  equilibrium with the dark matter halo.
However, we also have to consider whether  gas-rich satellites accreted in
a misaligned configuration will be tidally disrupted
and  settle relatively quickly on the plane, without causing detectable
perturbations to the existing disk. These questions need to be investigated
with hydrodynamical simulations in a cosmological context.

These results suggest that galactic disks grow in a roughly ``self-similar'' manner at late times,
such that their outer HI profiles always have roughly the same shape. The weak connection between
HI properties and the properties  of the older stellar population can arise if stellar components  
grow through stellar accretion in addition to gas accretion, as might
be expected to be the case in high stellar surface density, bulge-dominated 
galaxies.  

A more detailed investigation of
the HI profile shapes of the galaxies in our sample will the subject of an upcoming paper
(Wang et al 2013, in preparation), as will be a detailed comparison with existing semi-analytic
models of galaxy formation that predict the gas and stellar surface density profiles of
ensembles of galaxies in a $\Lambda$CDM Universe (Fu et al 2013, in preparation).   
In these models, the growth of low redshift galactic disks is predicted to be fuelled by accretion
of gas in the halo that was previously heated and  ejected by supernova-driven winds (Weinmann et al 2010). 

We also plan to search for/rule out kinematic evidence of gas accretion in the forms of warps, 
misaligned or counter rotating HI components in the disk and associated gas clouds and
to study the gas distributin in and around neighbouring galaxies around the HI-rich
and control samples. The  multi-source HI systems will be included in this analysis.
The reader is referred to one extraordinary galaxy (galaxy 48) with extremely wide-ranging 
HI clouds connected to one side of the galaxy; such systems may provide snapshots 
of the most prominent accretion events. The hope is that the results
of this analysis can feed into preparations for future large-scale HI surveys 
with wide-field instruments like Apertif (Verheijen et al 2008).

Finally, we would like to note that this analysis excluded 7 out of 50
(i.e. 15\%) of the targeted galaxies, because they were interacting
with companions, undetected, unobserved or otherwise disturbed. This is a substantial
fraction of the original sample and we note that some of our conclusions
about
the ``quiesence" of the gas-rich population may change once we come up
with ways to parameterize the full population in a fair way.
This will be the subject of future investigation.

\section*{Acknowledgements}
We thank L. Shao, B. Catinella, R. Reyes, E. Elson for useful discussions. 

Milan den Heijer was supported for this
research through a stipend from the International Max Planck Research
School (IMPRS) for Astronomy and Astrophysics at the Universities of
Bonn and Cologne

GALEX (Galaxy Evolution Explorer) is a NASA Small Explorer, launched in April 2003, developed in cooperation with the Centre National d'Etudes Spatiales of France and the Korean Ministry of Science and Technology. 

Funding for the SDSS and SDSS-II has been provided by the Alfred P. Sloan Foundation, the Participating Institutions, the National Science Foundation, the U.S. Department of Energy, the National Aeronautics and Space Administration, the Japanese Monbukagakusho, the Max Planck Society, and the Higher Education Funding Council for England. The SDSS Web Site is http://www.sdss.org/.

\bibliographystyle{mn2e}

\appendix
\section{Atlas of the Bluedisk galaxies}
\begin{figure*}
\includegraphics[width=7cm]{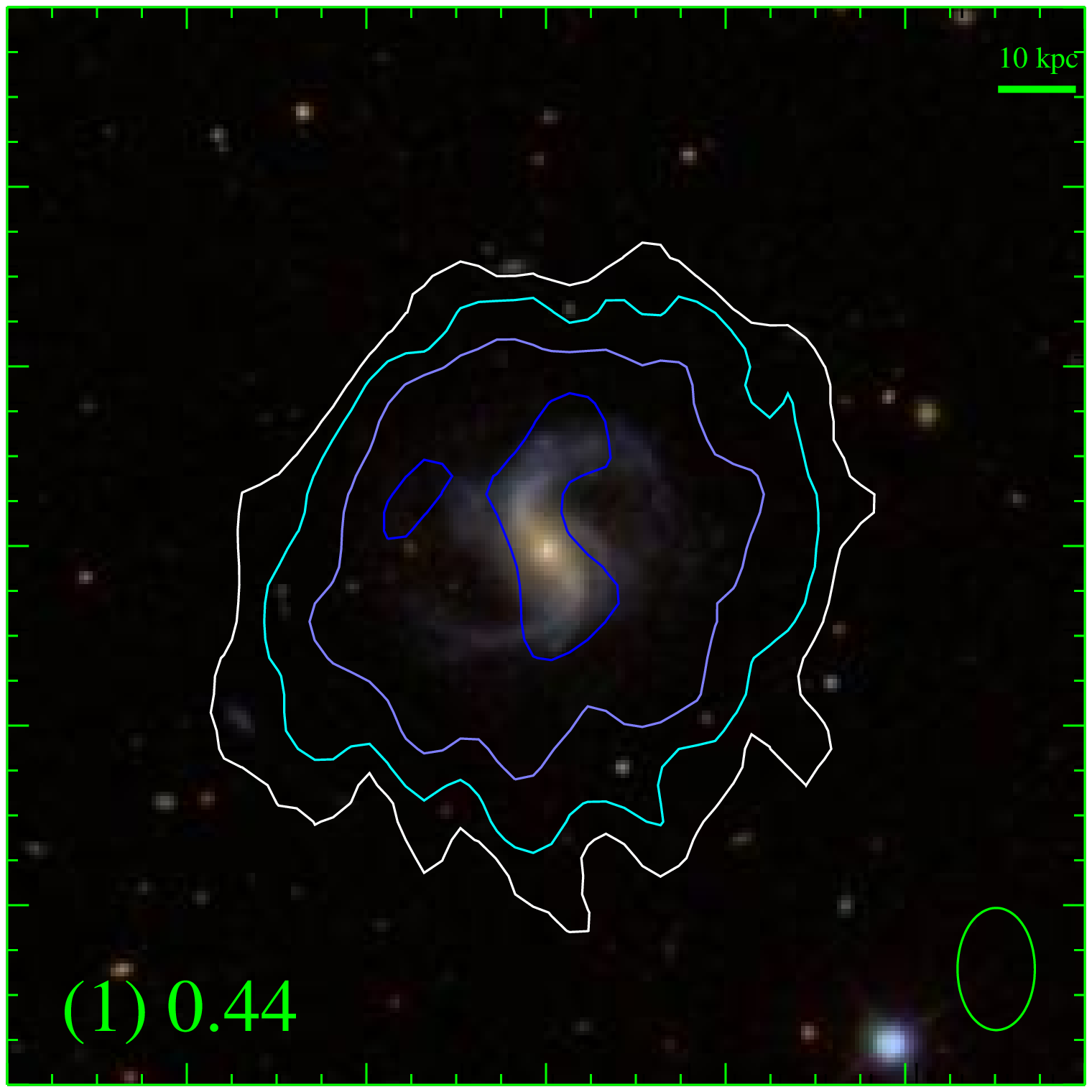}
\includegraphics[width=7cm]{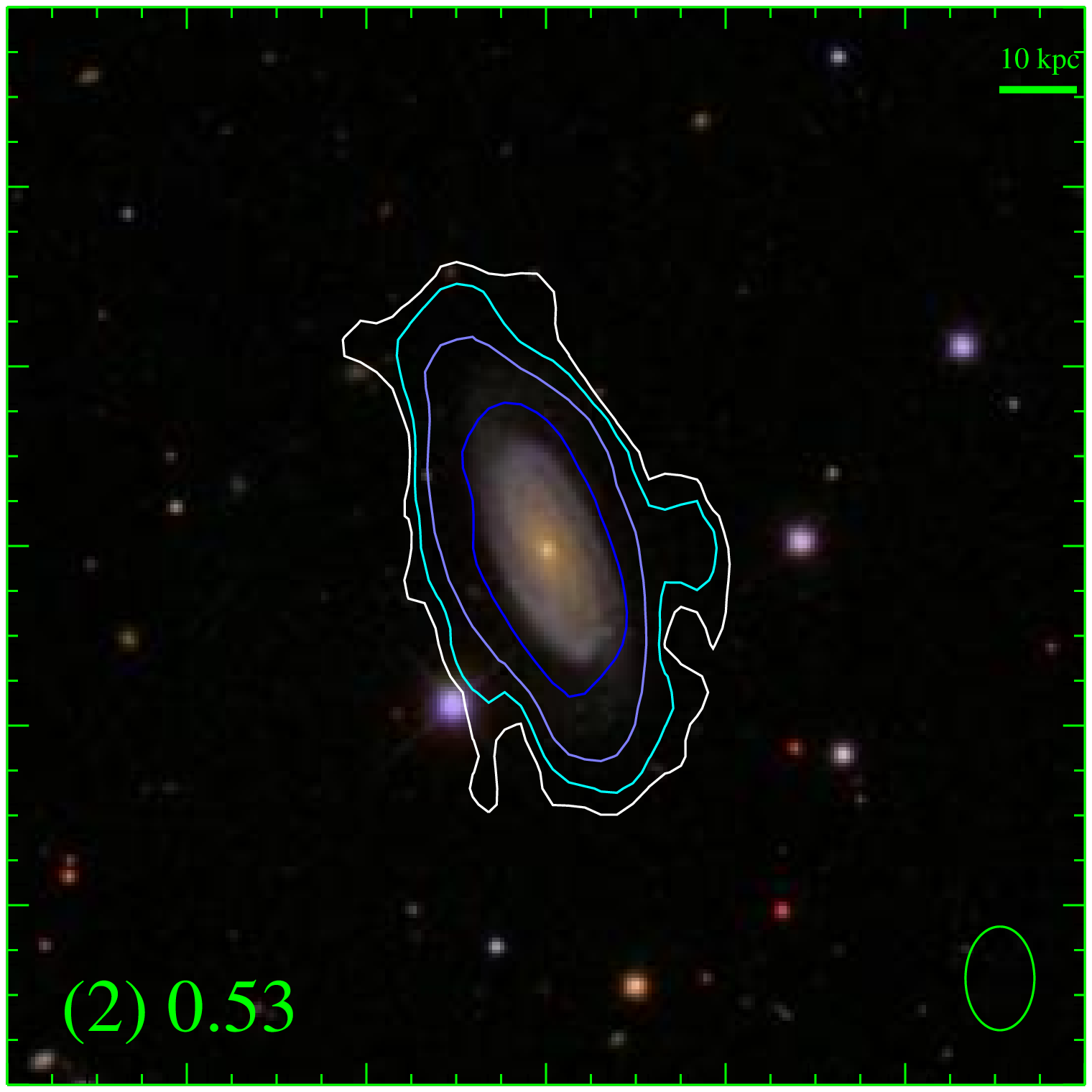}

\includegraphics[width=7cm]{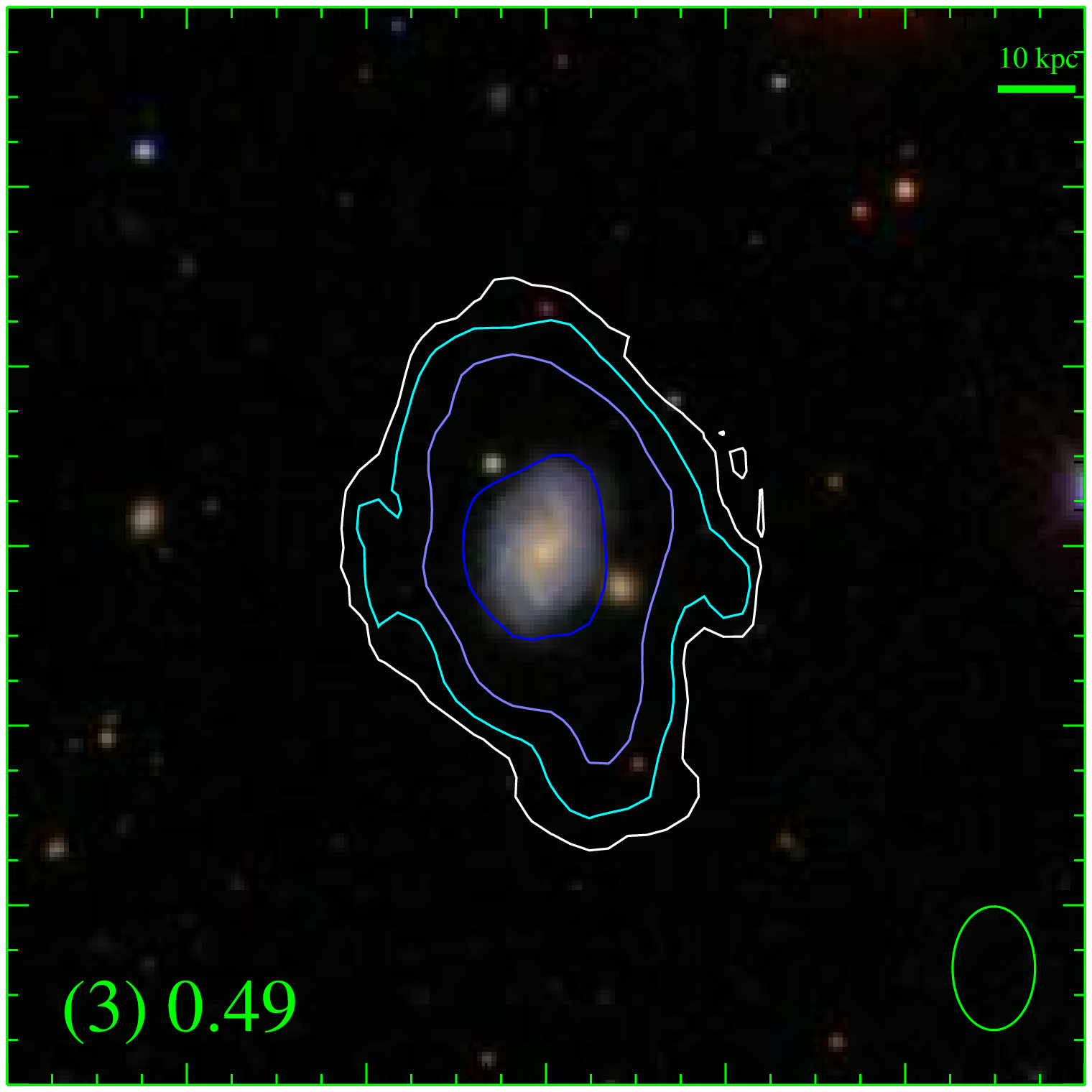}
\includegraphics[width=7cm]{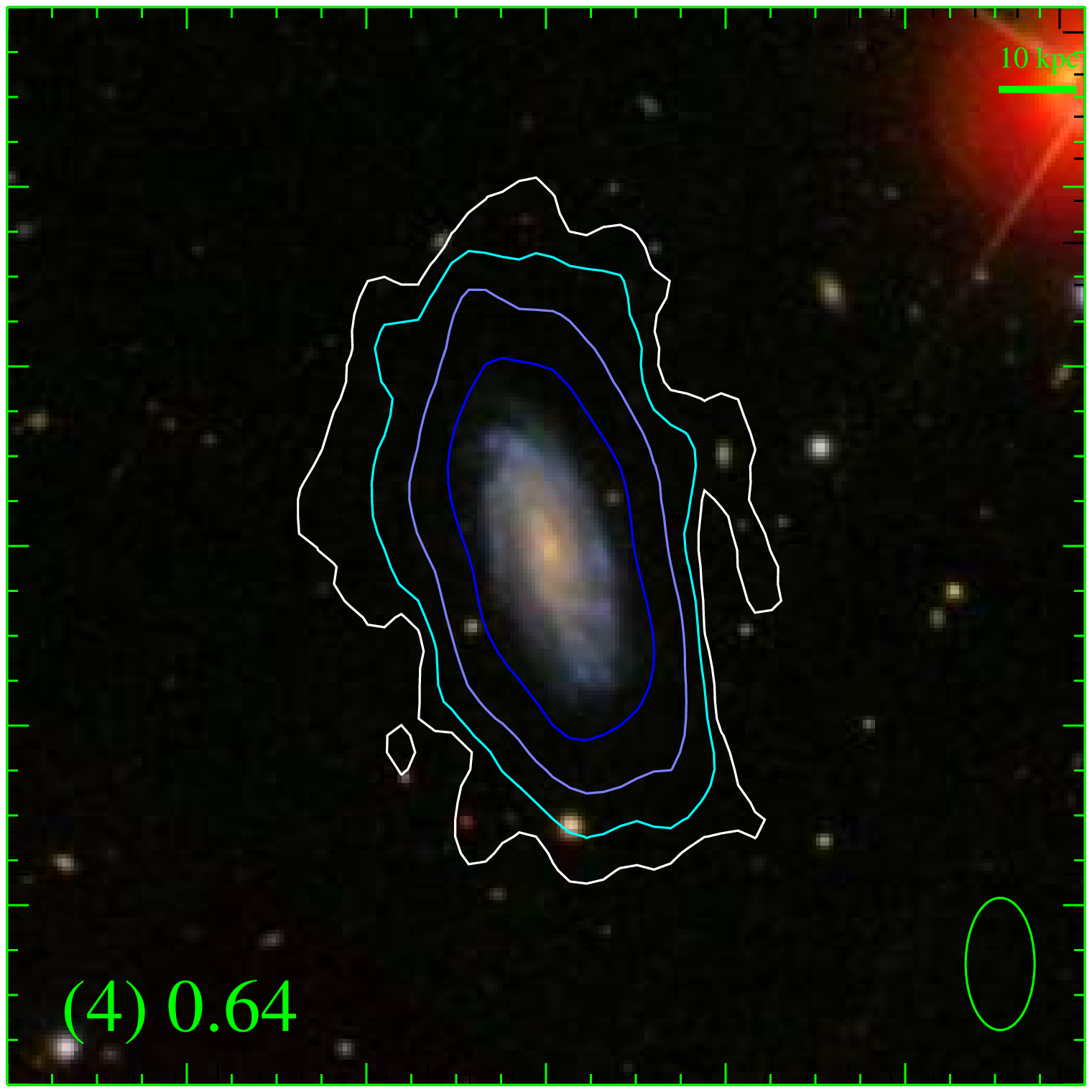}

\includegraphics[width=7cm]{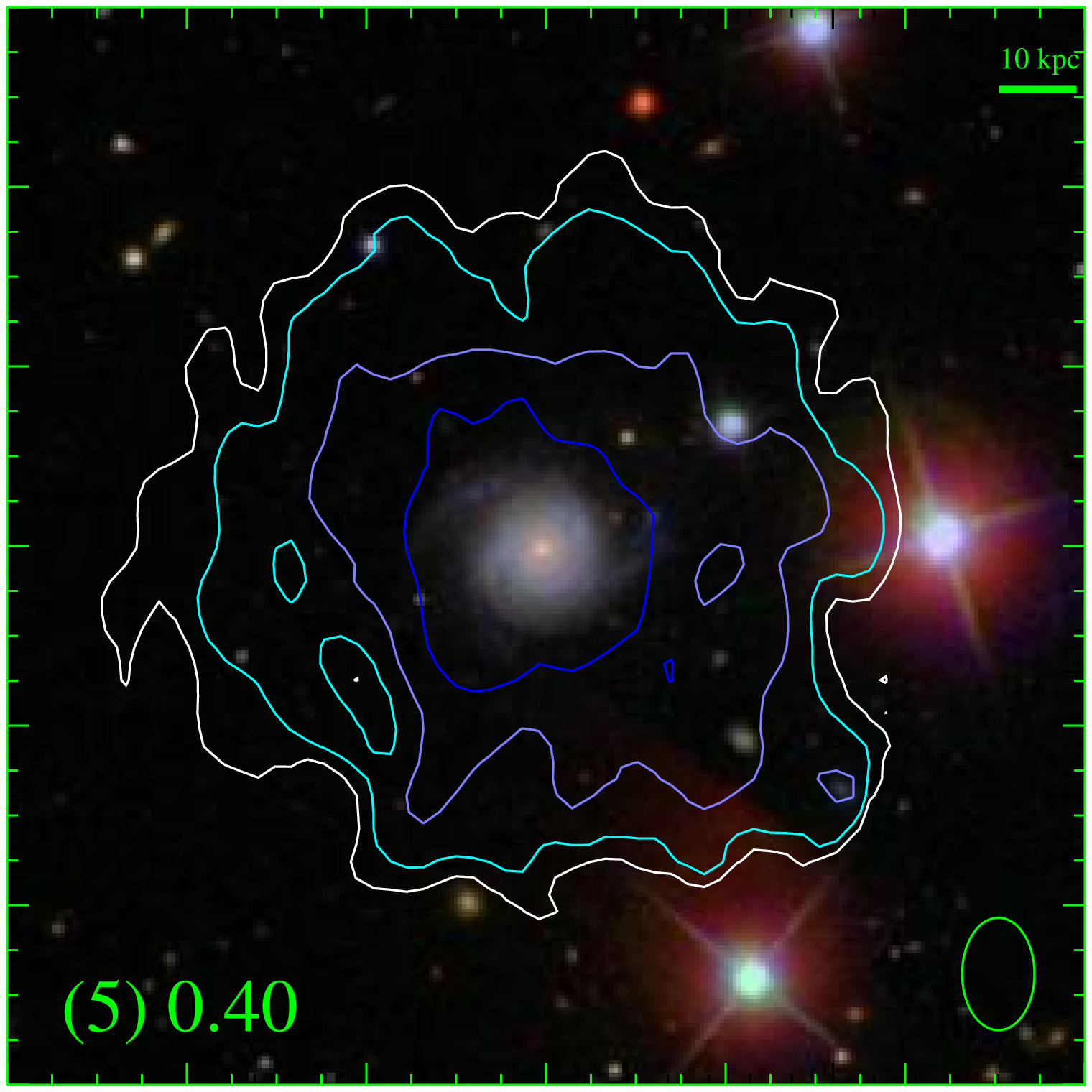}
\includegraphics[width=7cm]{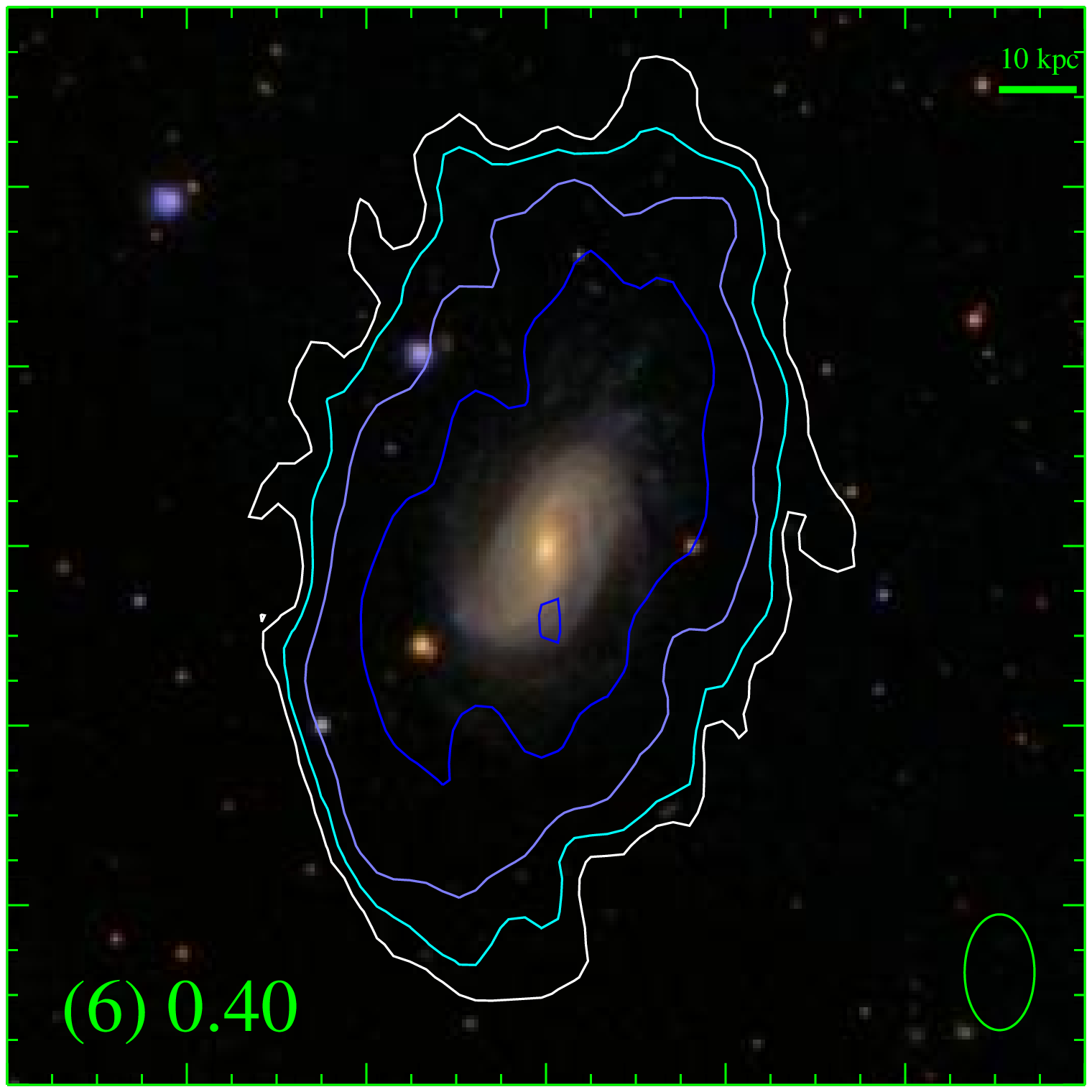}
\caption{HI column density contours on the optical images for the HI-rich galaxies. All the maps have a size of 140 kpc, and the length of 10 kpc is displayed at the top-right corner.  The galaxy ID is denoted in brackets at bottom-left corner of each map. The outmost contour has a column density equivalent to the estimated detection threshold of the total HI image(see Section 2.2) and is denoted in unit of 10$^{20}$ atoms cm$^{-1}$ in the bottom-left corner of each map. The contour levels increase with a step of 2.5 times. The shape of the beam is plotted at the bottom-right corner of each map. All the maps are displayed as north up and east left. To be continued} 
 \label{fig:G_HIrich}
\end{figure*}

To get a visual impression of the difference between the HI-rich and control samples, we display the HI surface density maps overlayed on the SDSS optical images (obtained from the SDSS DR7 CAS  visual tools \footnote[2]{http://skyserver.sdss.org/public/en/tools/chart/list.asp}) for the two samples separately in Figure~\ref{fig:G_HIrich} and \ref{fig:G_ctrl}. All the maps have the same size of 140 kpc. The outmost contour has a level of the estimated detection threshold of the total HI image (see Section 2.3).  The galaxies that are excluded from analysis in this work are also displayed in Figure~\ref{fig:G_multi}.

\addtocounter{figure}{-1}
\begin{figure*}
\includegraphics[width=7cm]{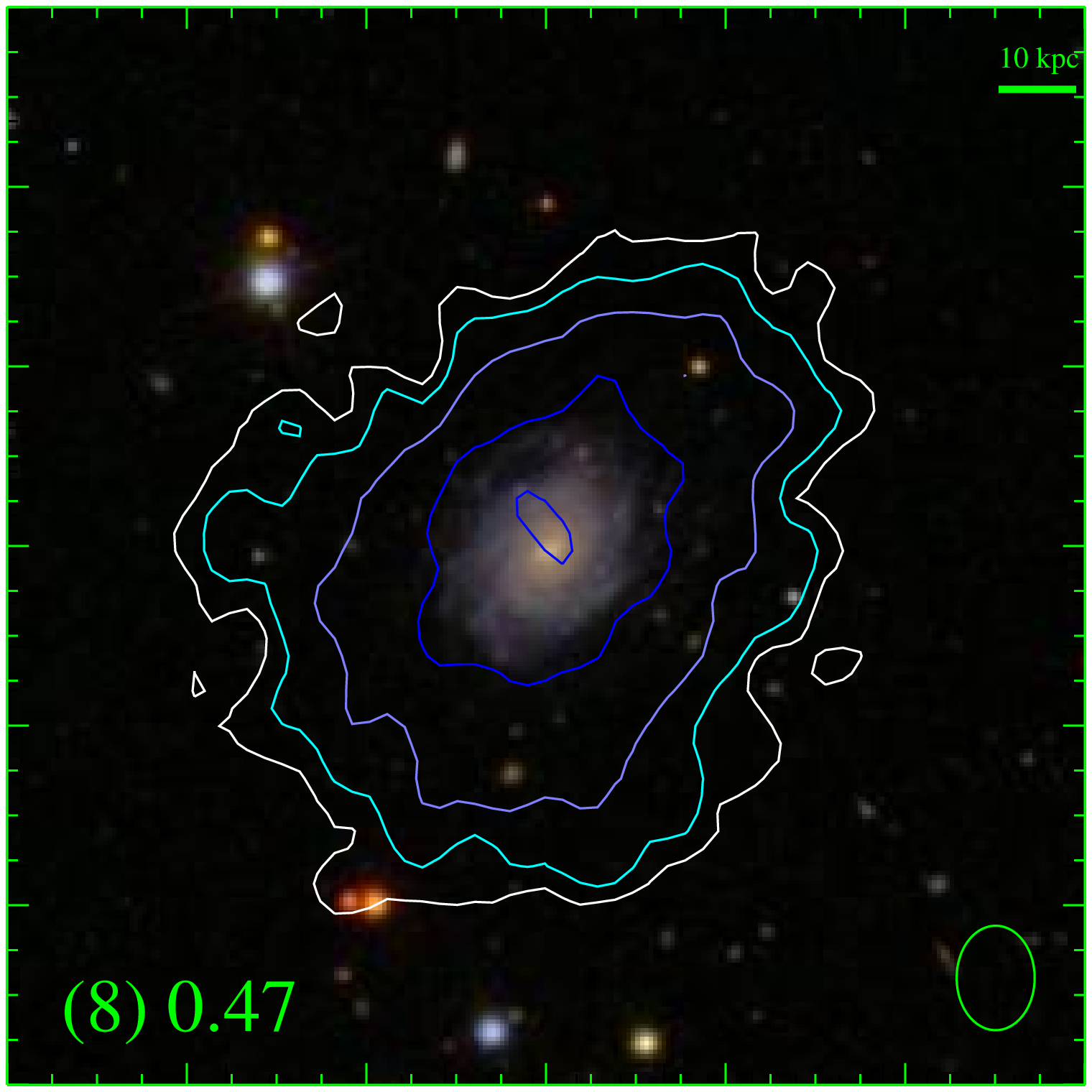}
\includegraphics[width=7cm]{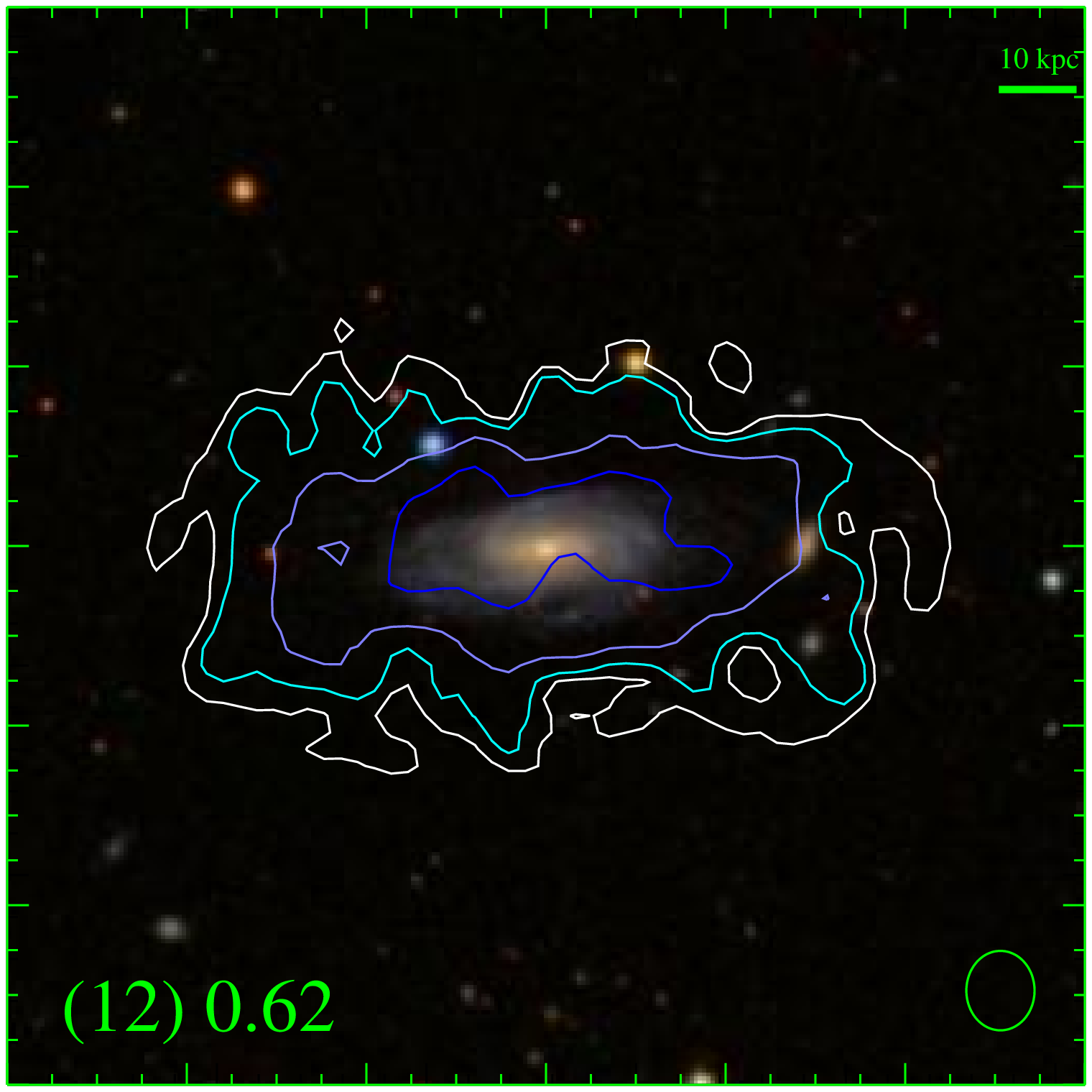}

\includegraphics[width=7cm]{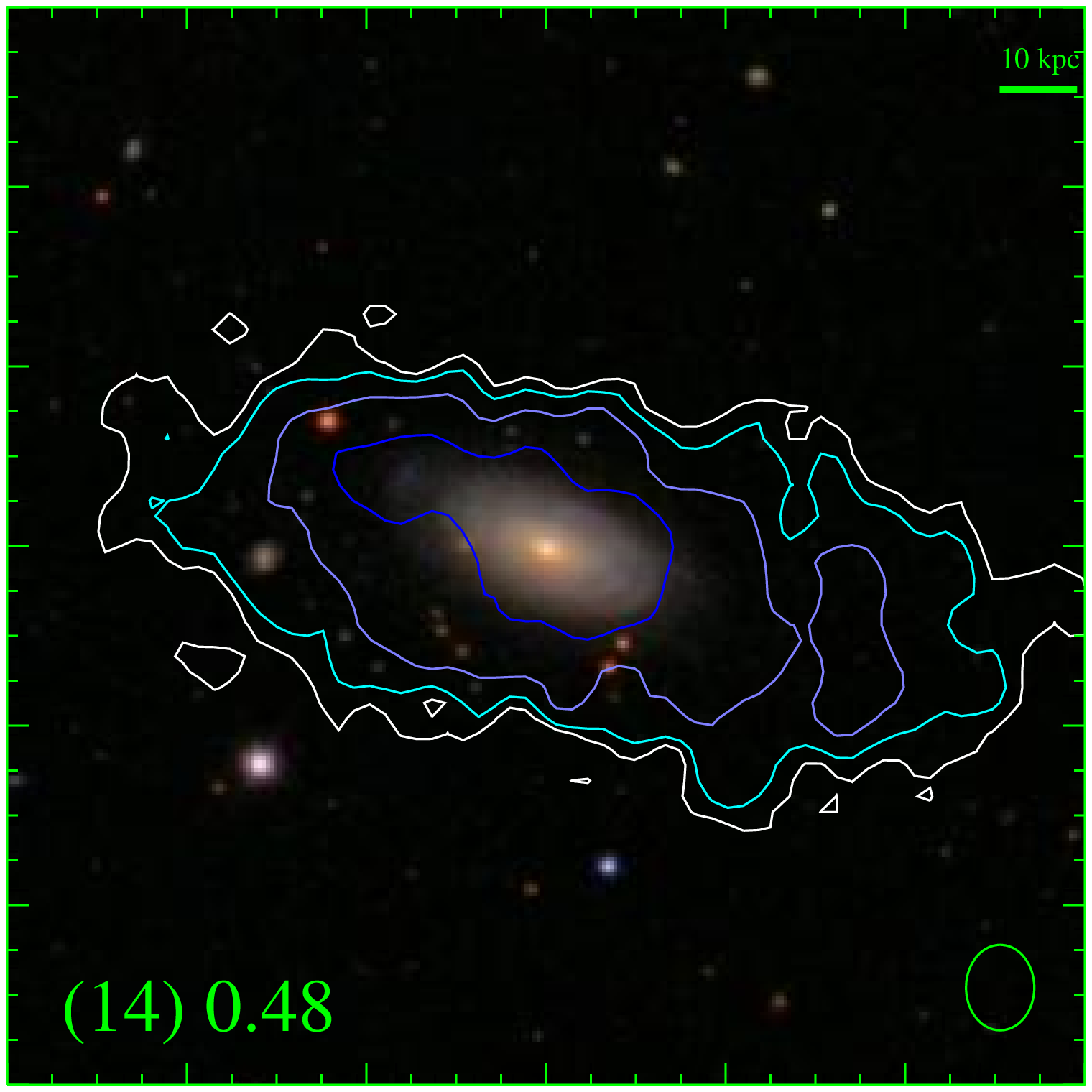}
\includegraphics[width=7cm]{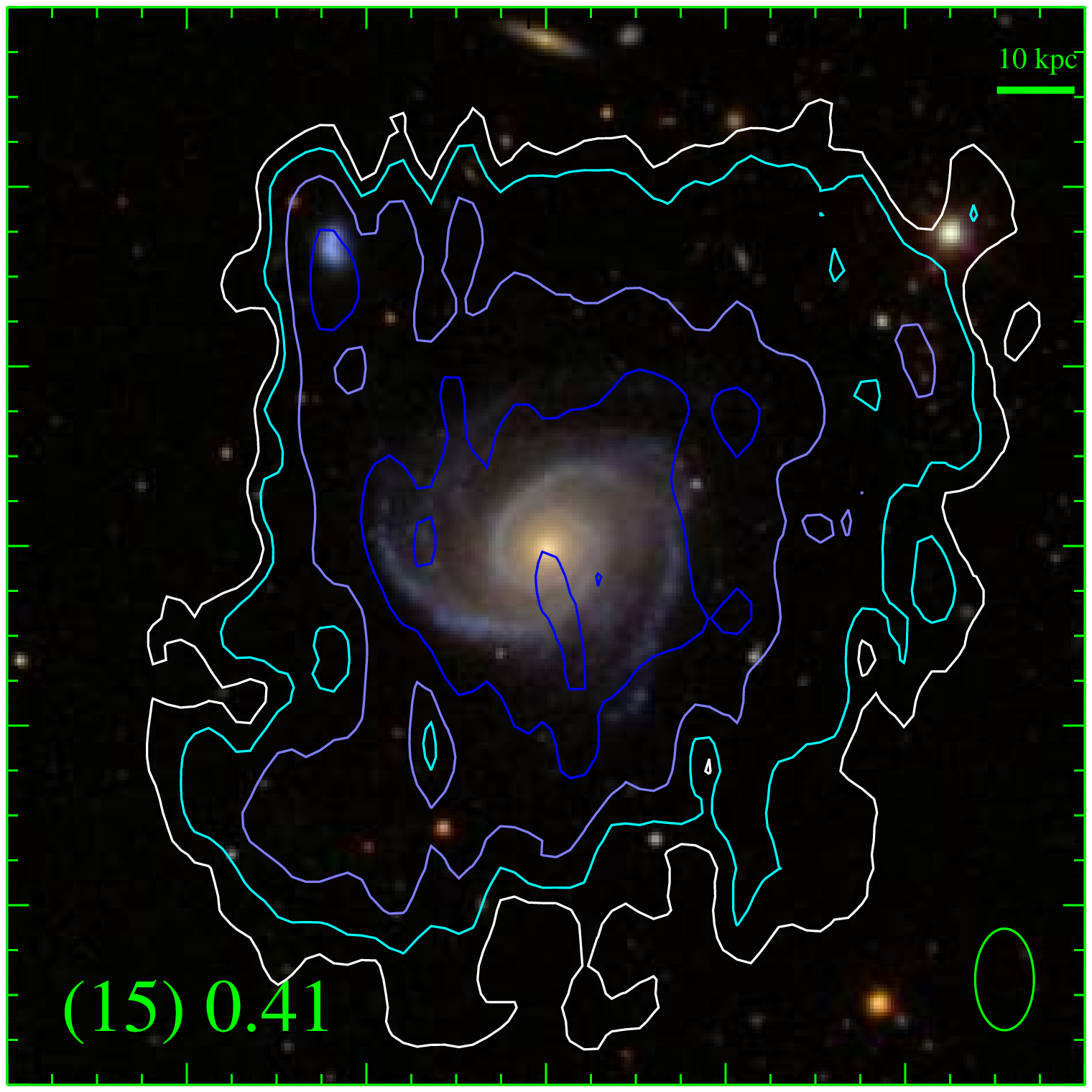}

\includegraphics[width=7cm]{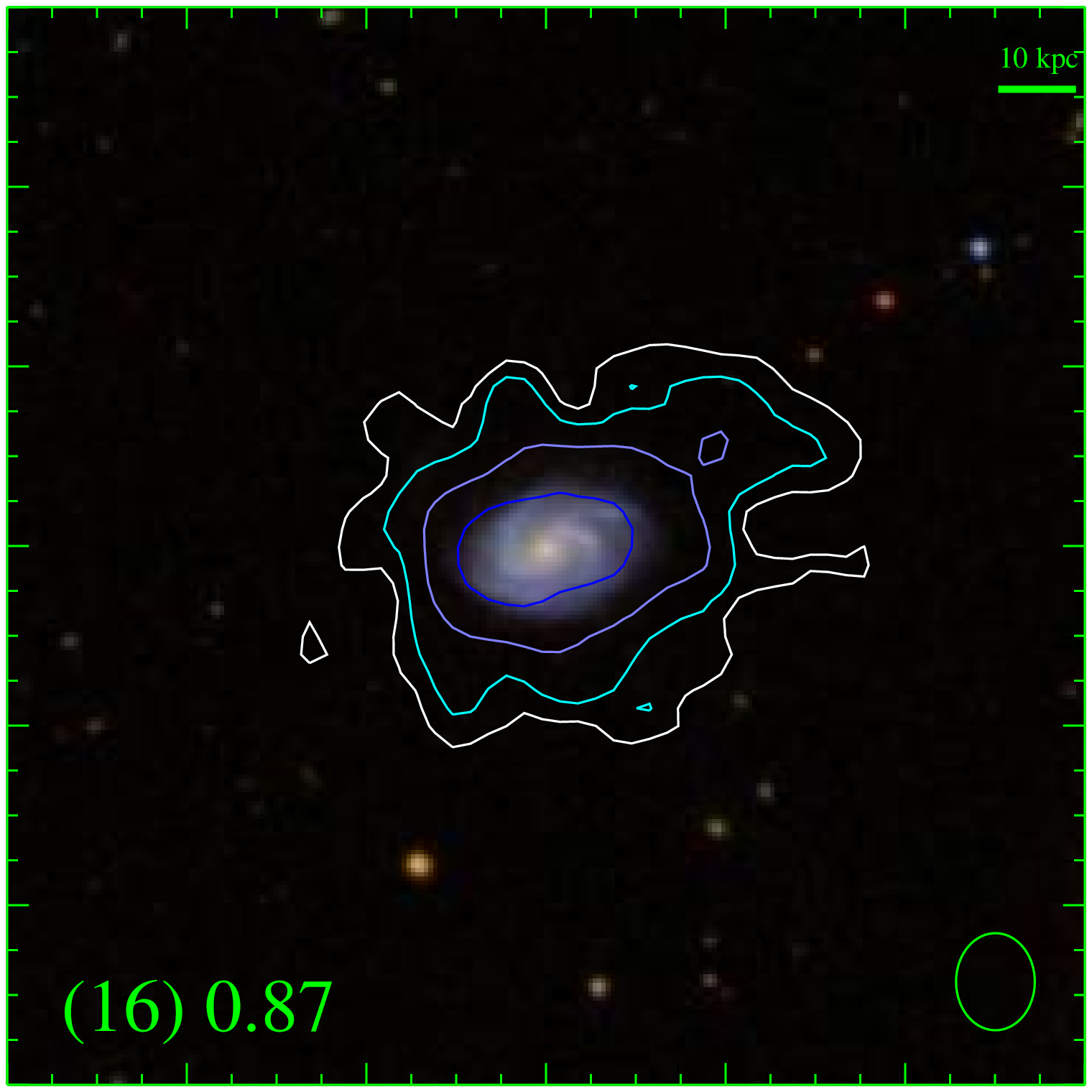}
\includegraphics[width=7cm]{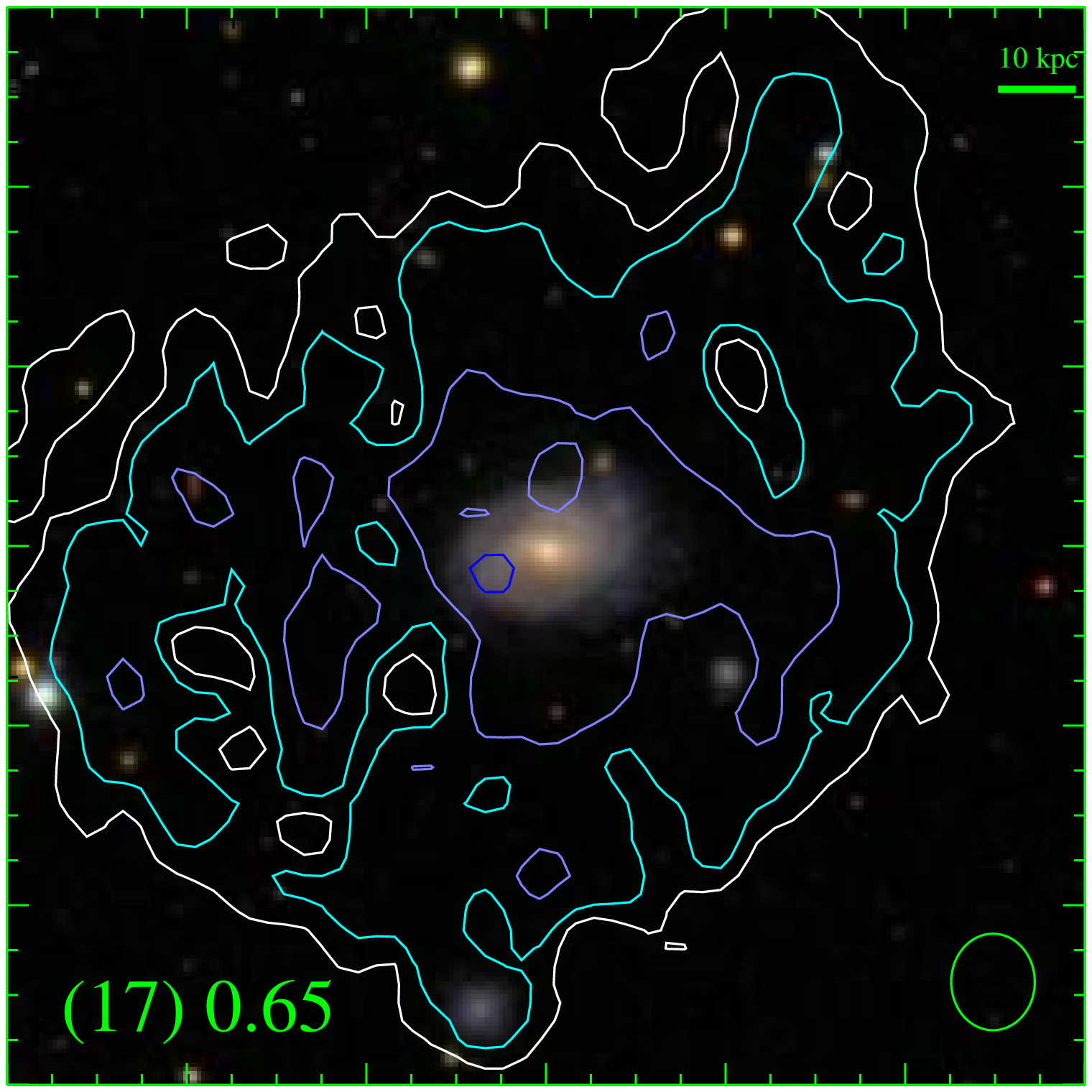}
\caption{To be continued} 
\end{figure*}

\addtocounter{figure}{-1}
\begin{figure*}
\includegraphics[width=7cm]{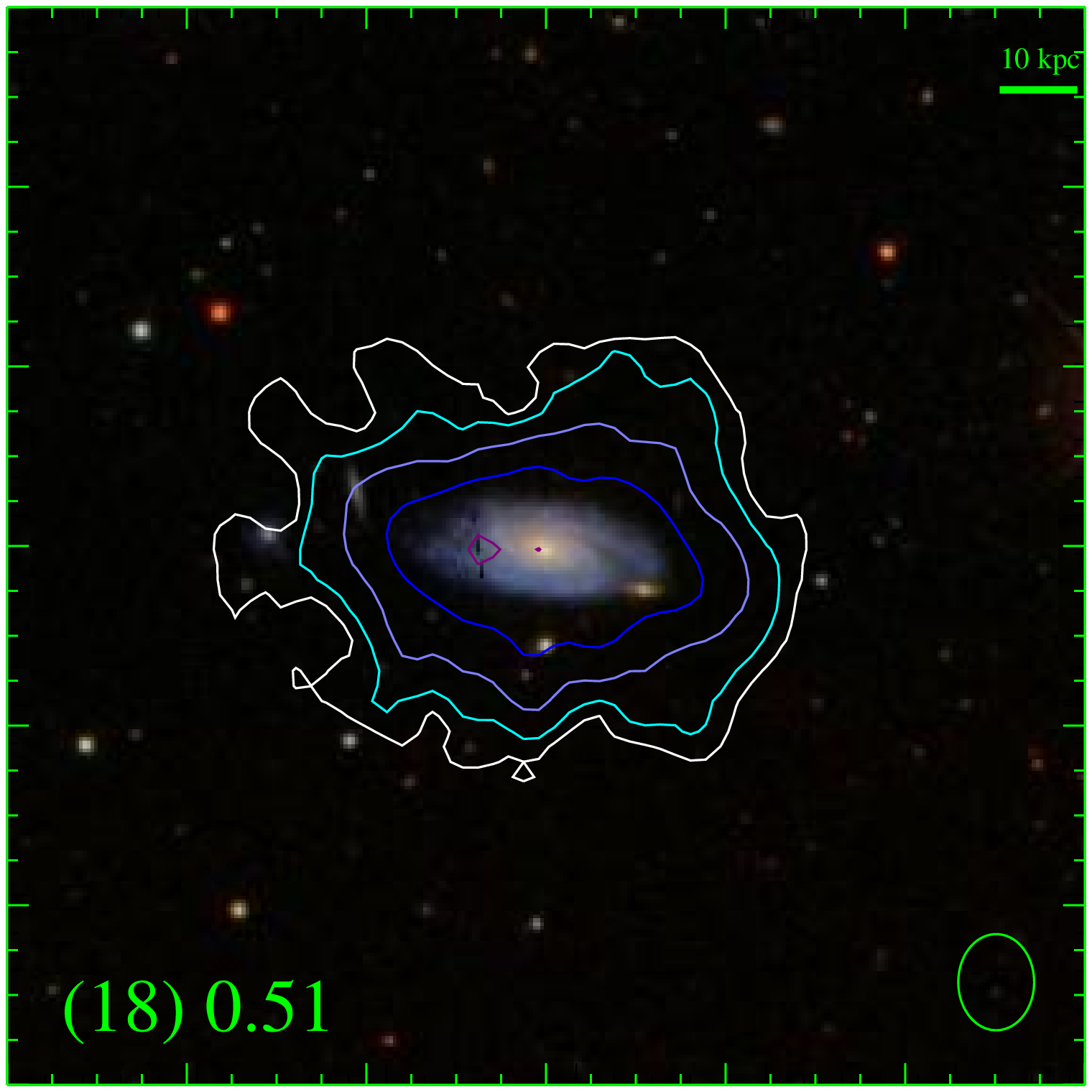}
\includegraphics[width=7cm]{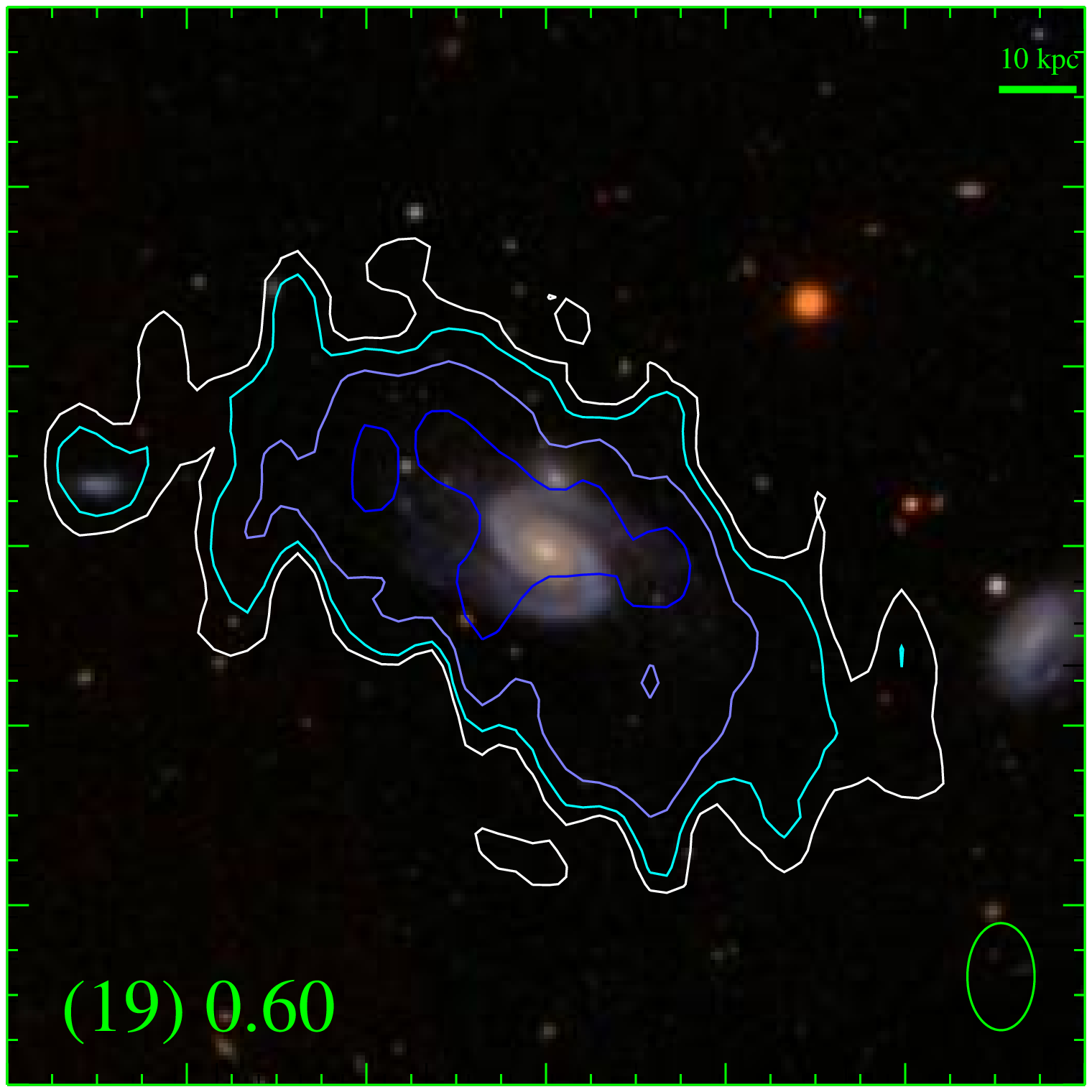}

\includegraphics[width=7cm]{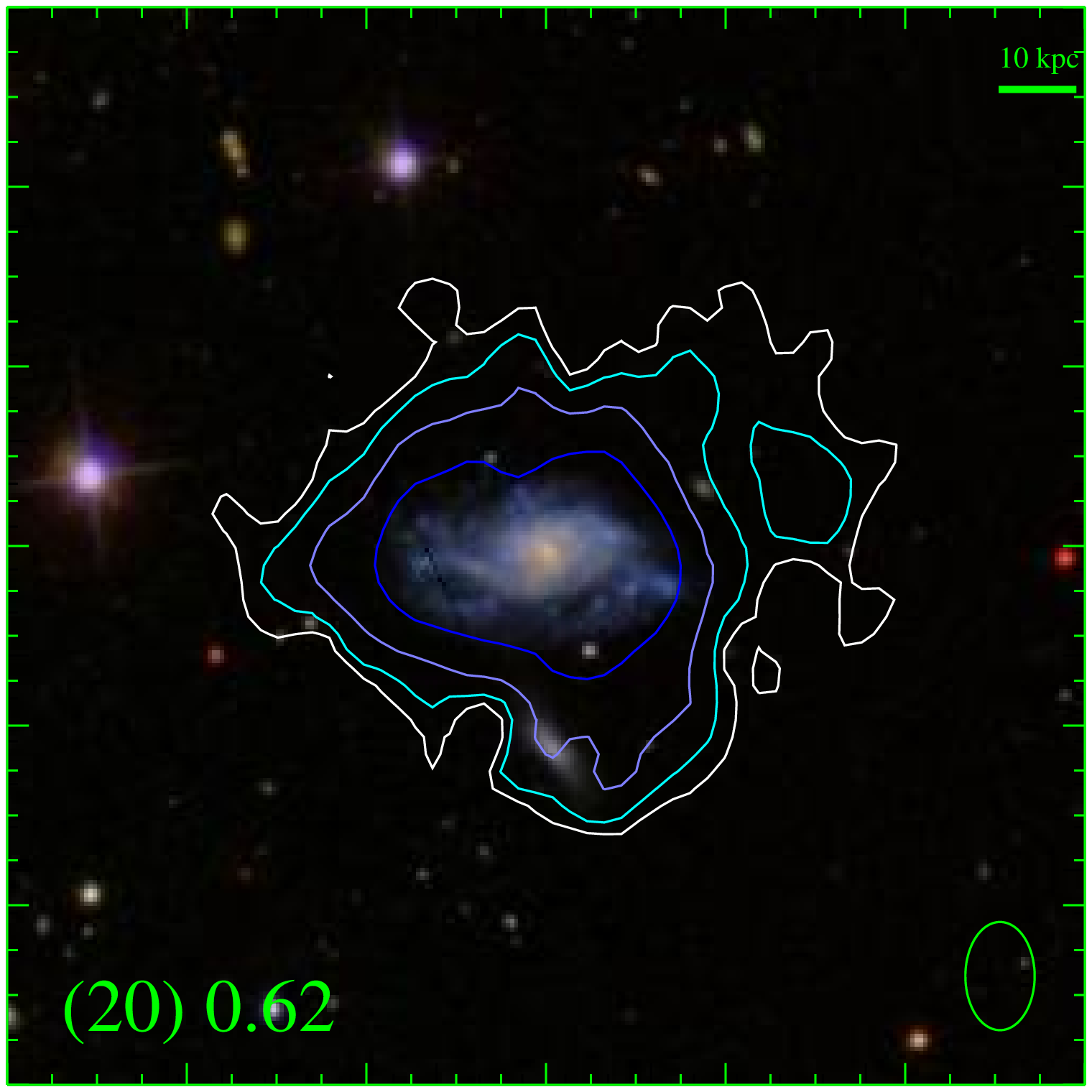}
\includegraphics[width=7cm]{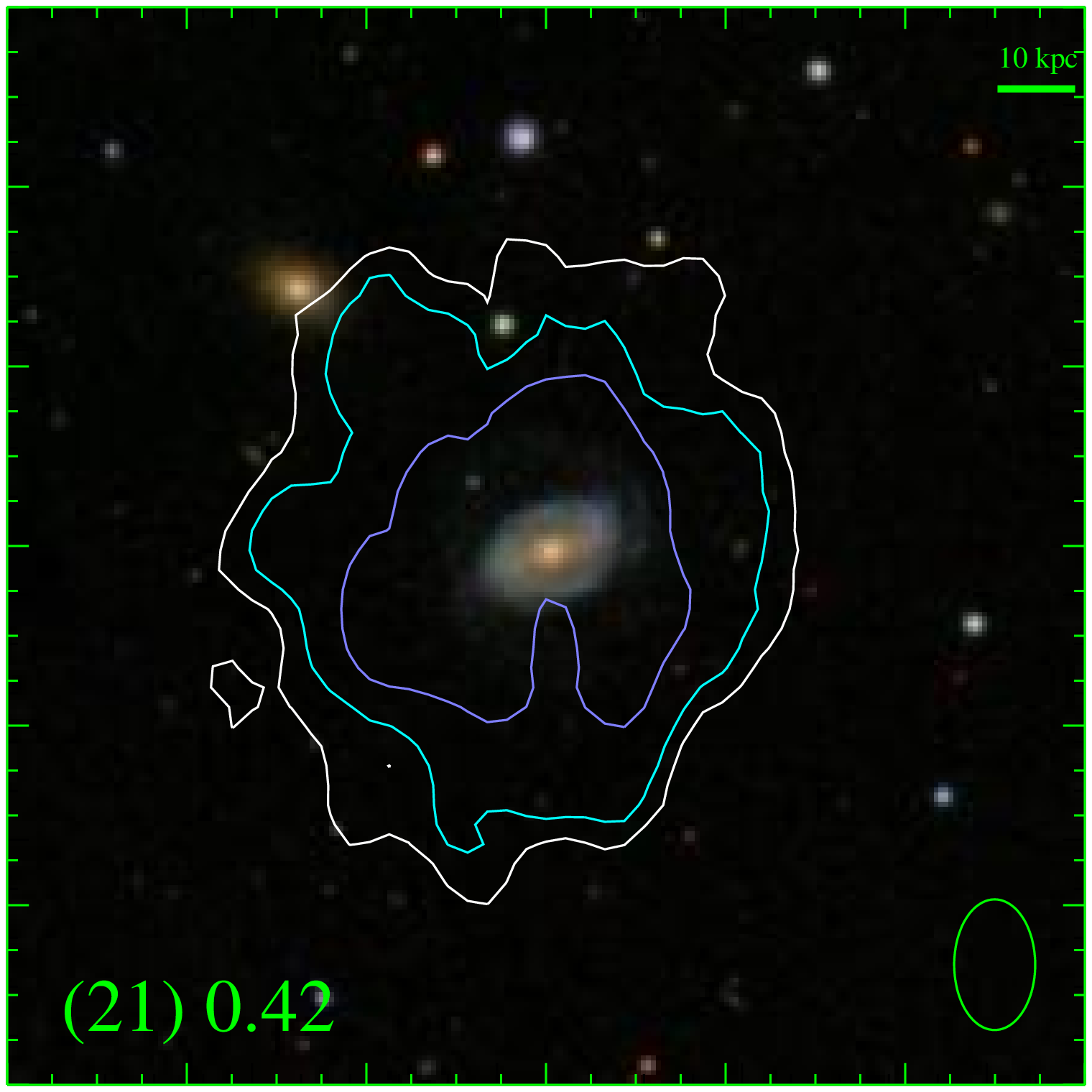}

\includegraphics[width=7cm]{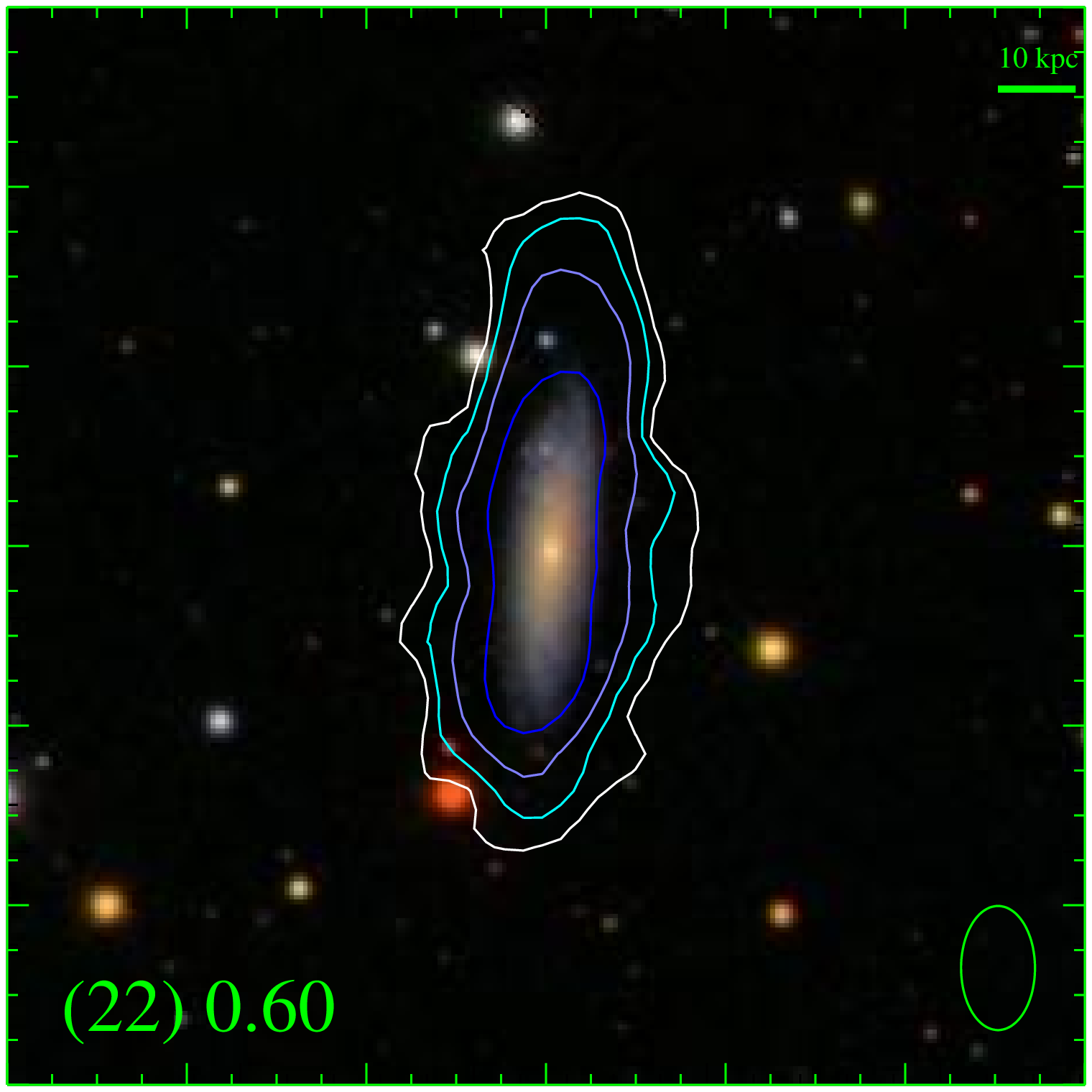}
\includegraphics[width=7cm]{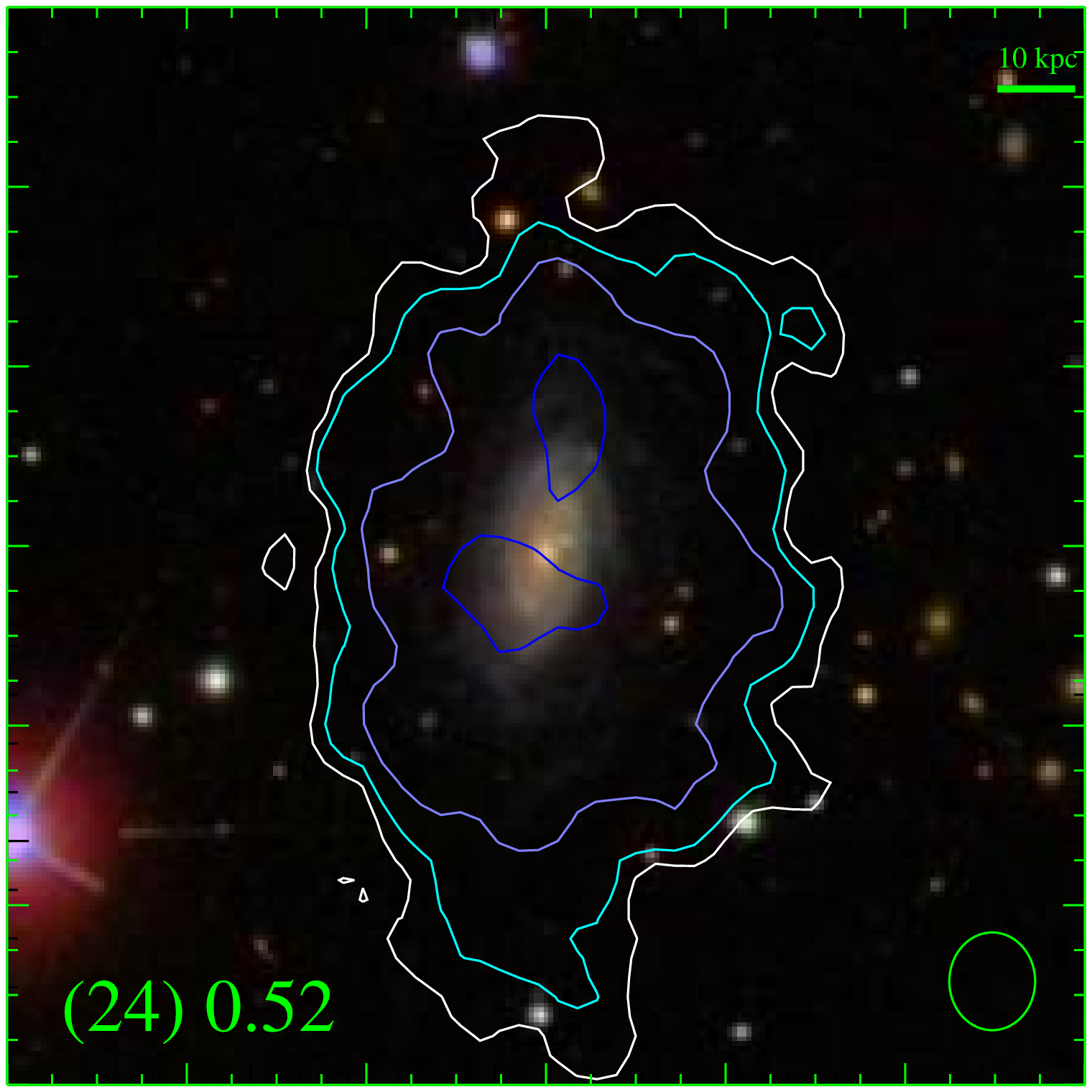}
\caption{To be continued} 
\end{figure*}

\addtocounter{figure}{-1}
\begin{figure*}
\includegraphics[width=7cm]{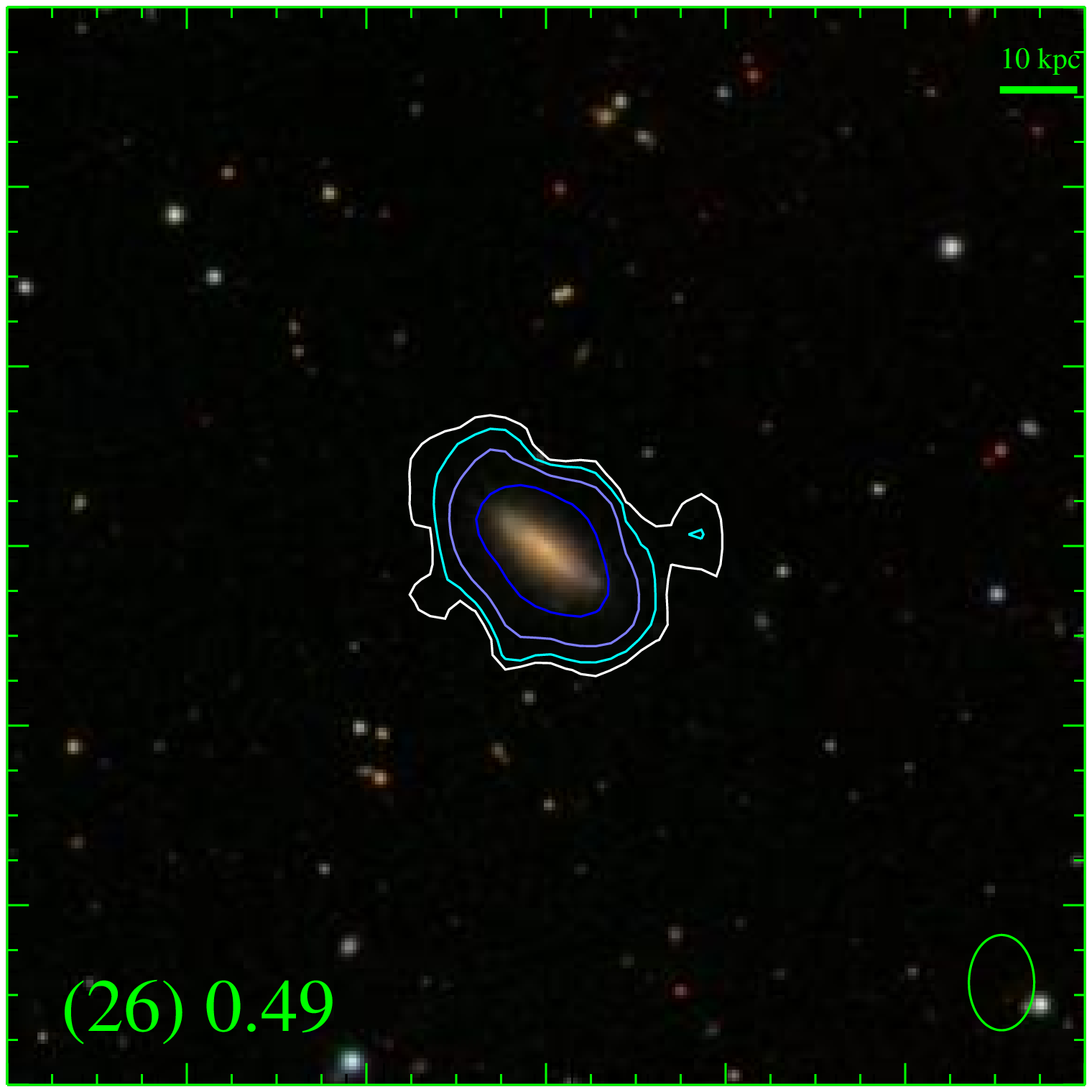}
\includegraphics[width=7cm]{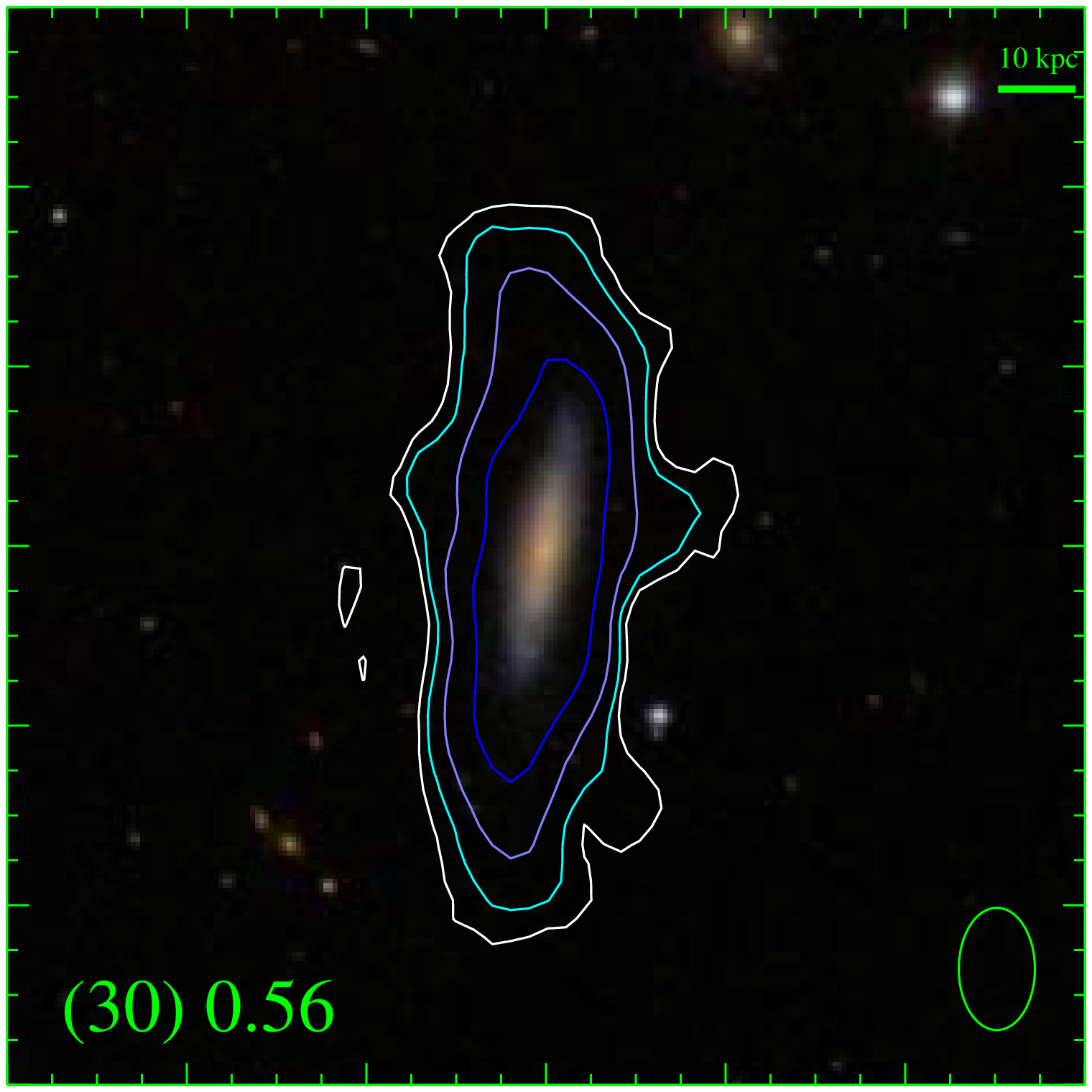}

\includegraphics[width=7cm]{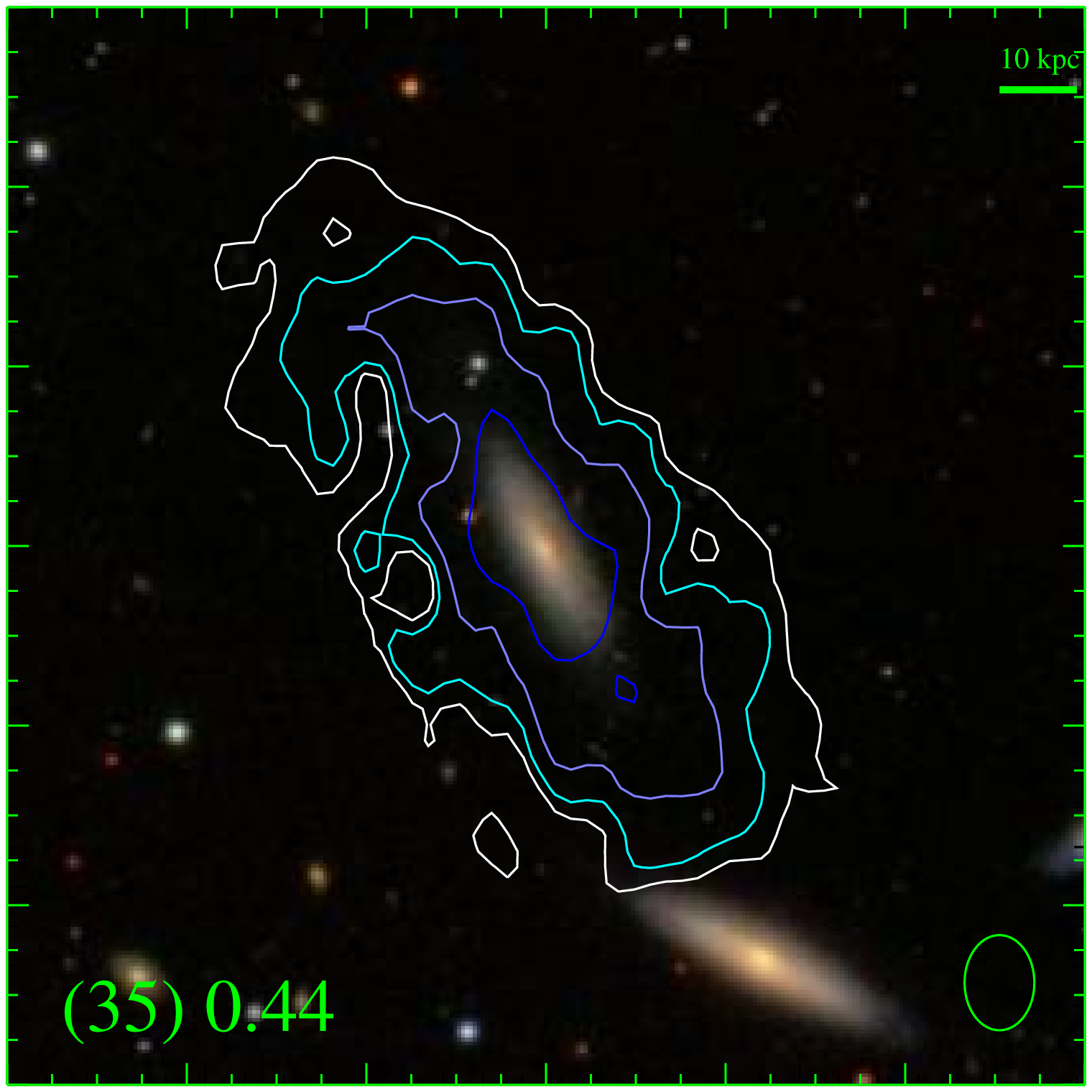}
\includegraphics[width=7cm]{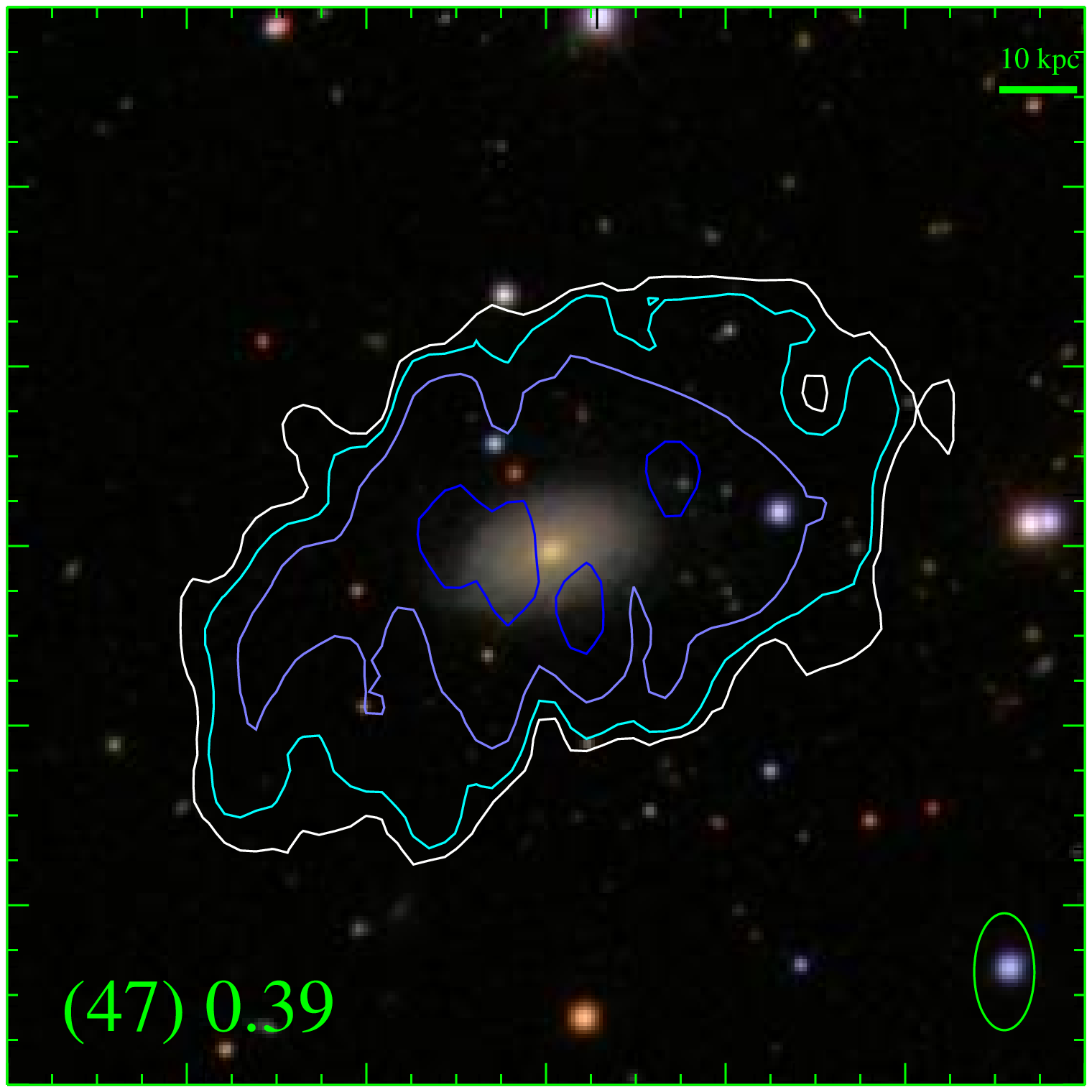}

\includegraphics[width=7cm]{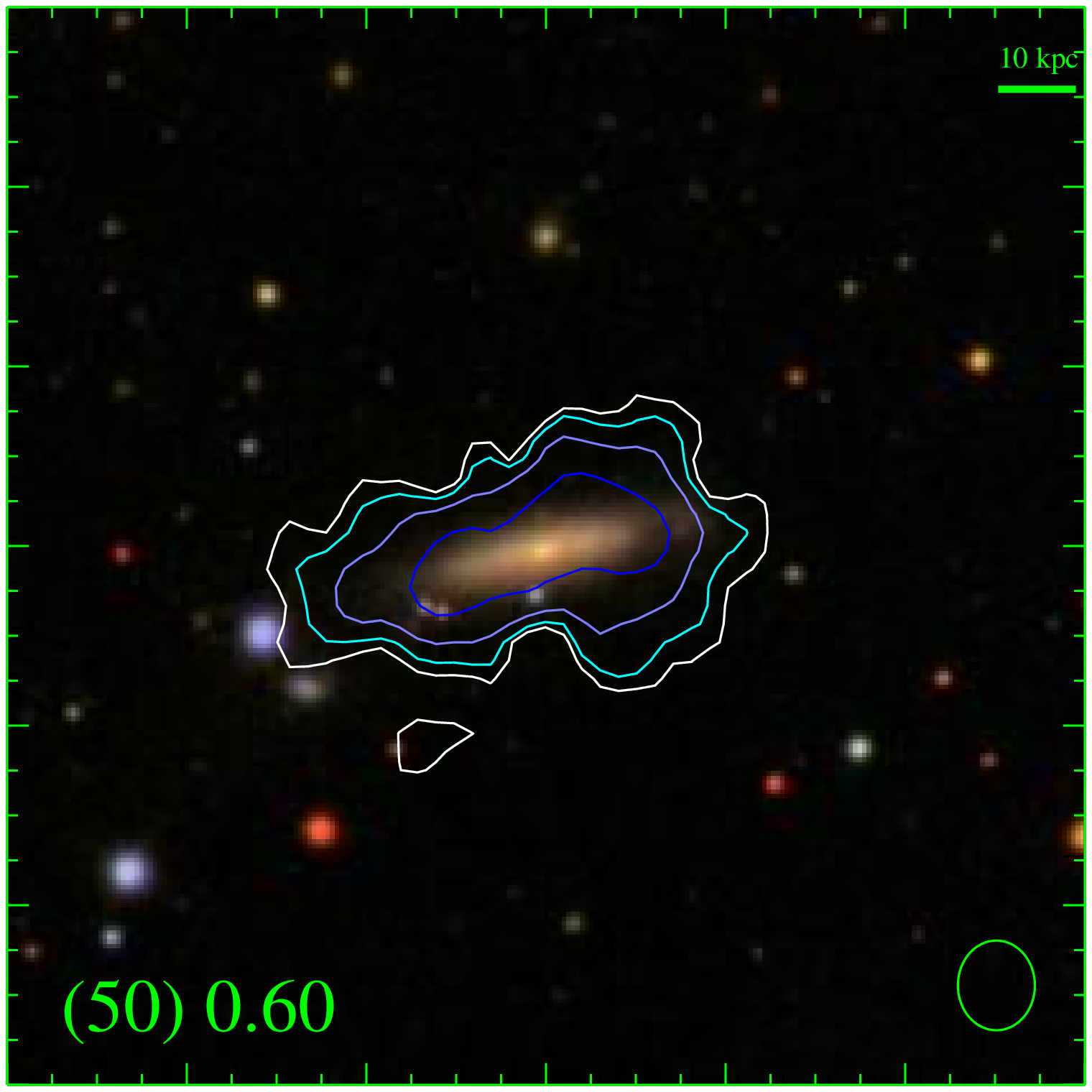}
\hspace{7cm}
\caption{-} 
\end{figure*}

\begin{figure*}
\includegraphics[width=7cm]{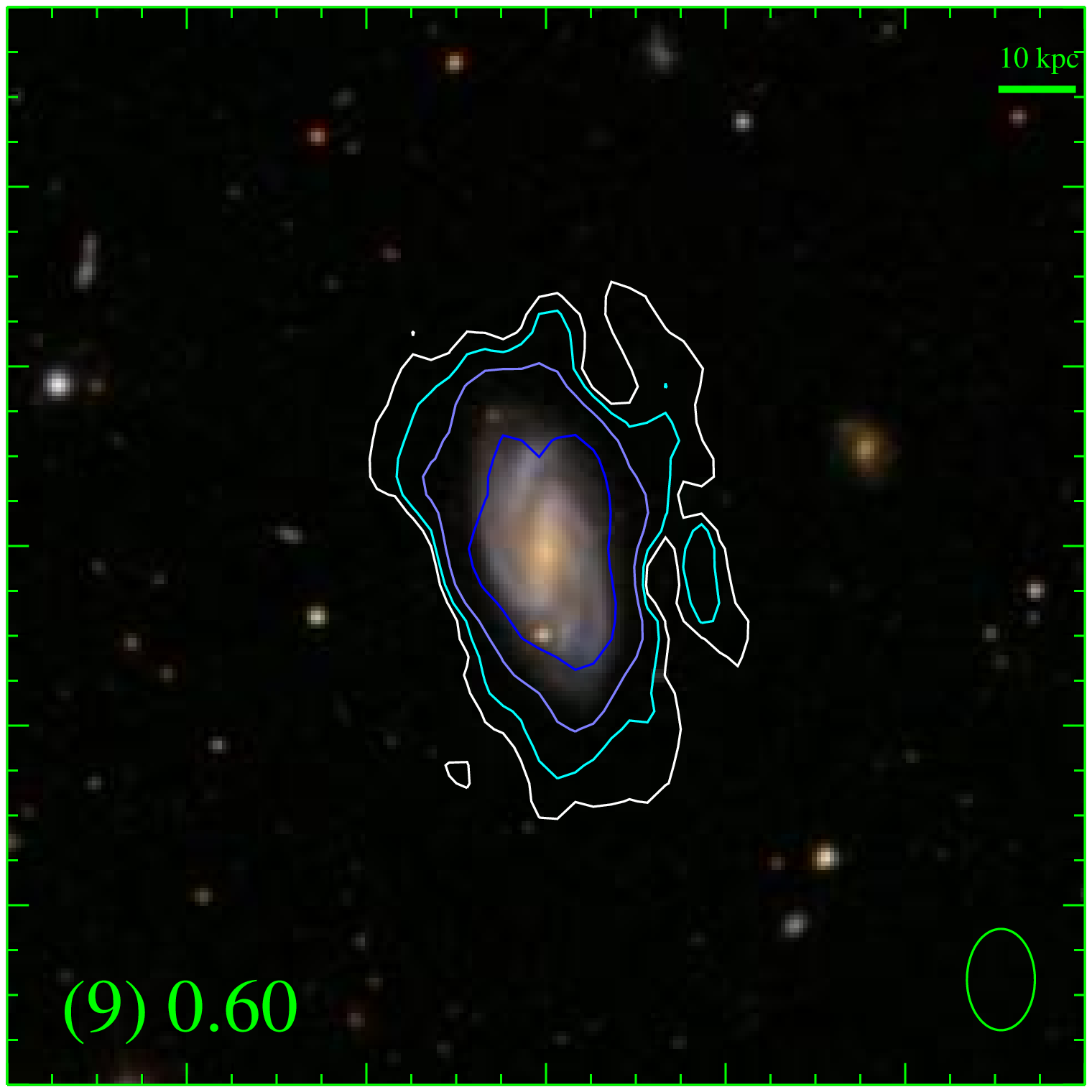}
\includegraphics[width=7cm]{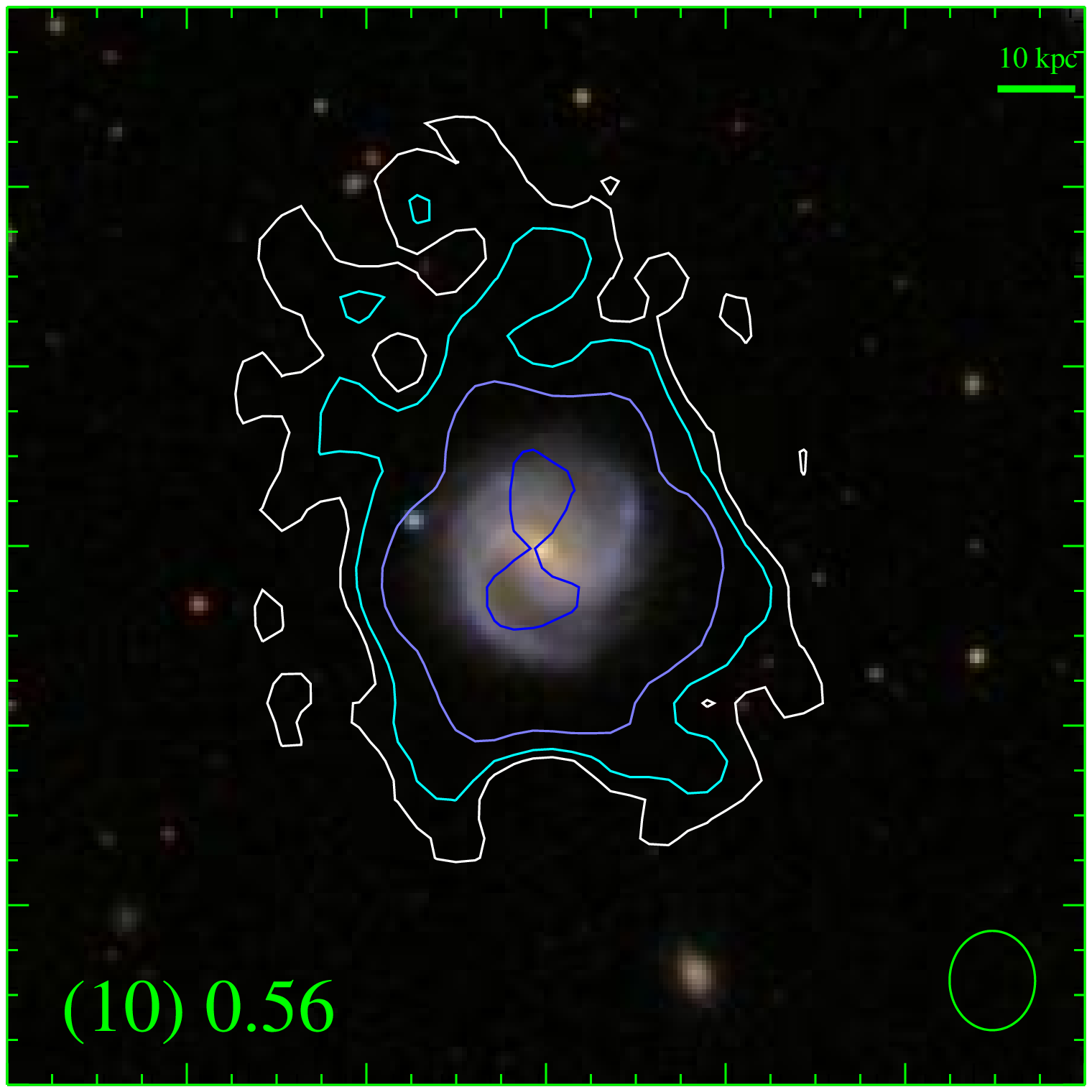}

\includegraphics[width=7cm]{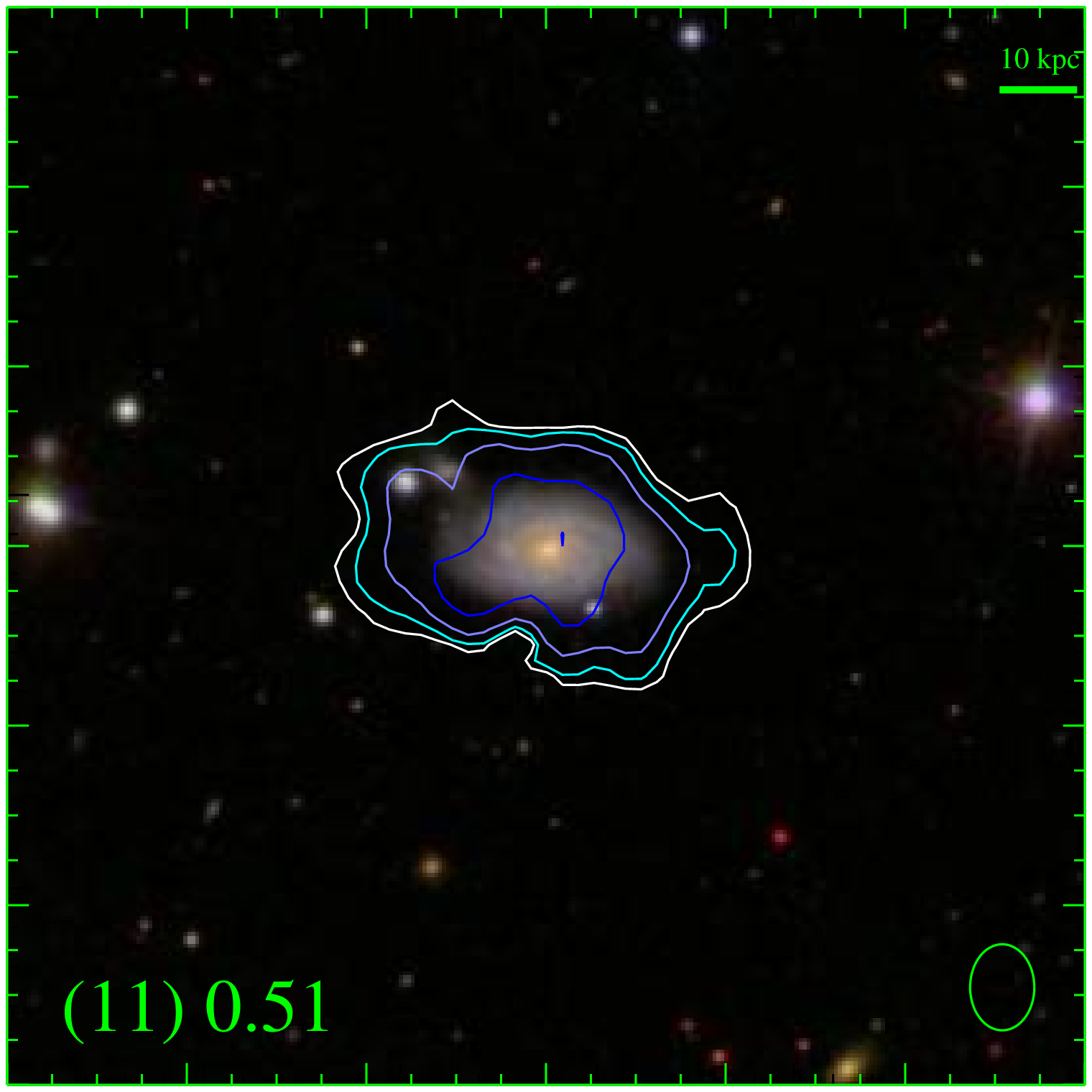}
\includegraphics[width=7cm]{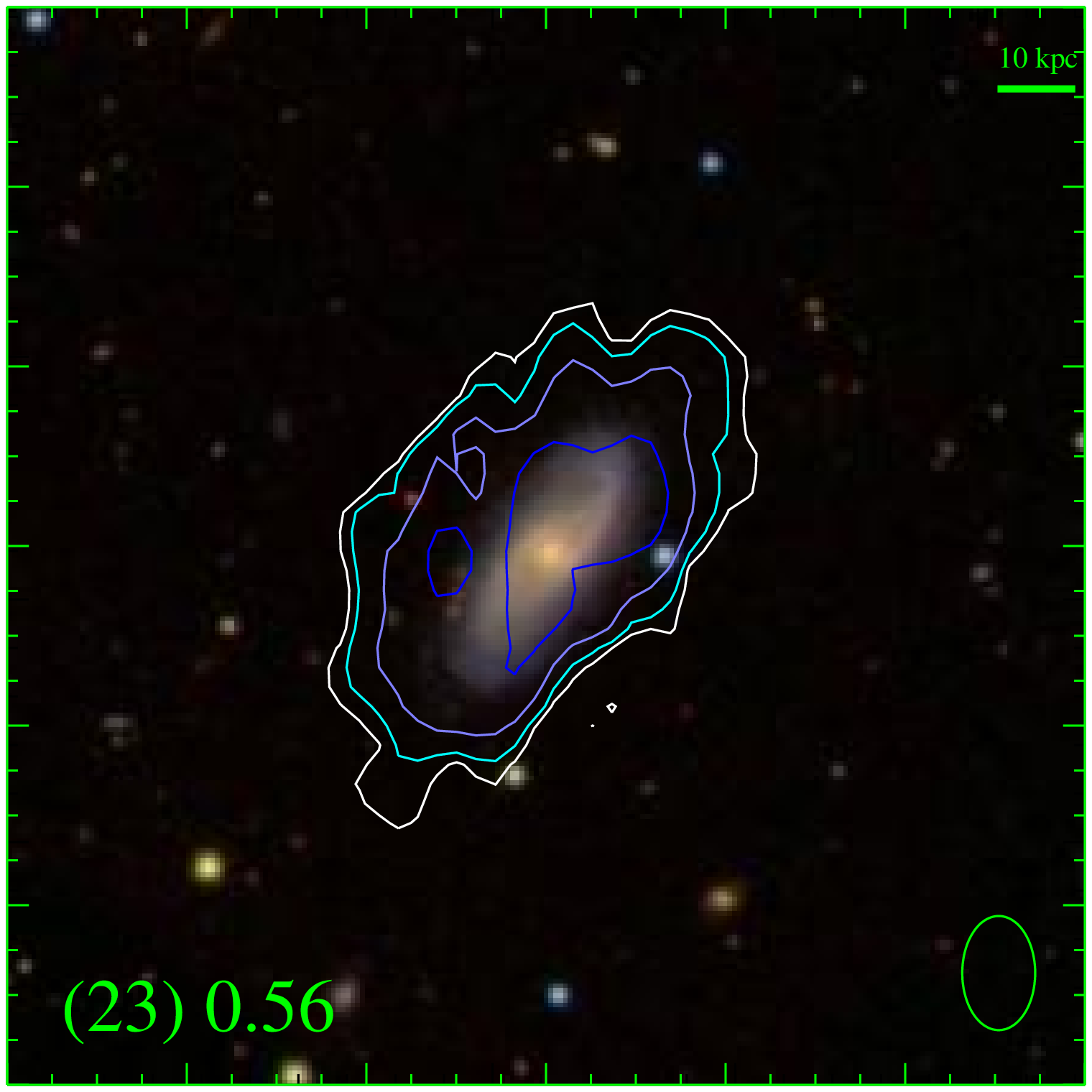}

\includegraphics[width=7cm]{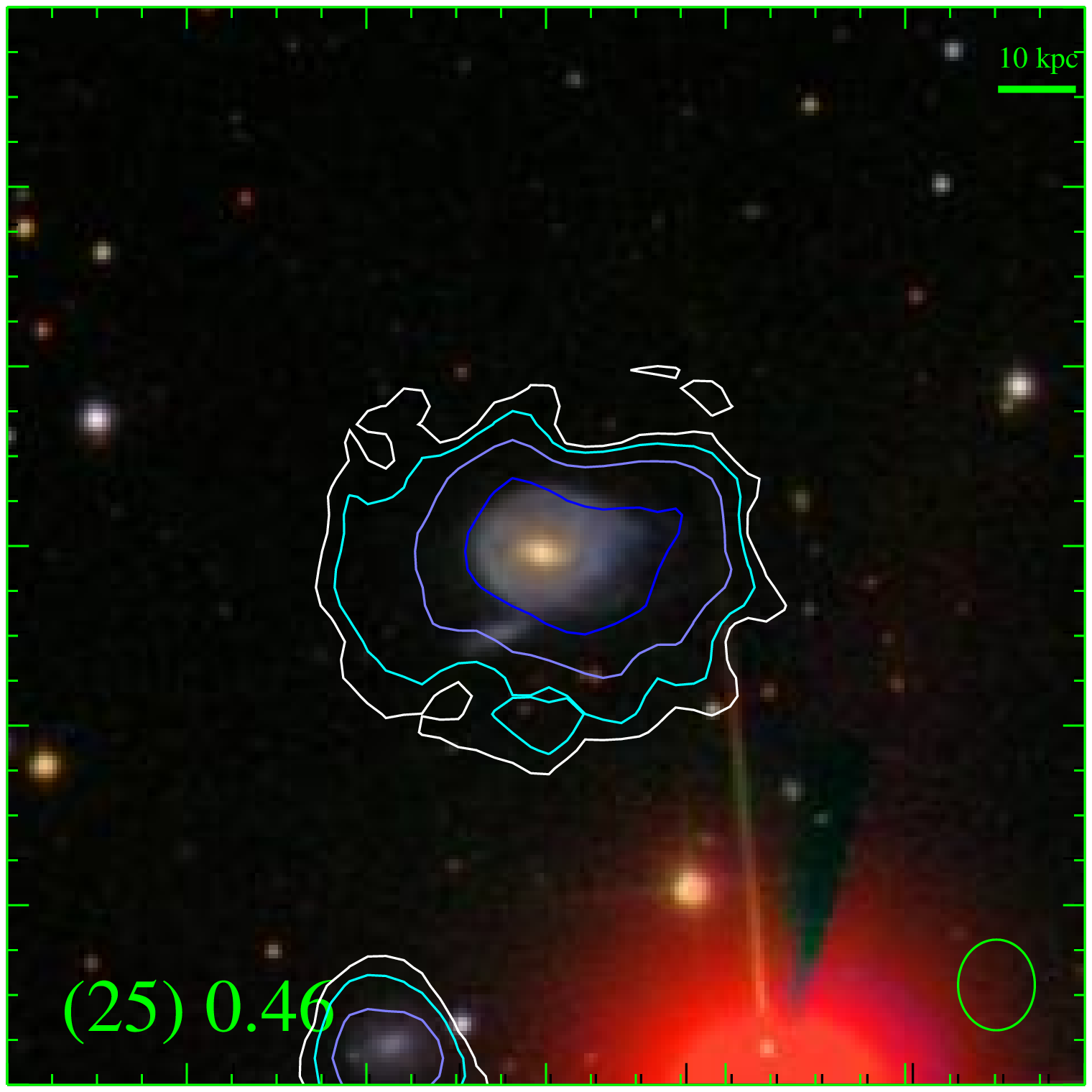}
\includegraphics[width=7cm]{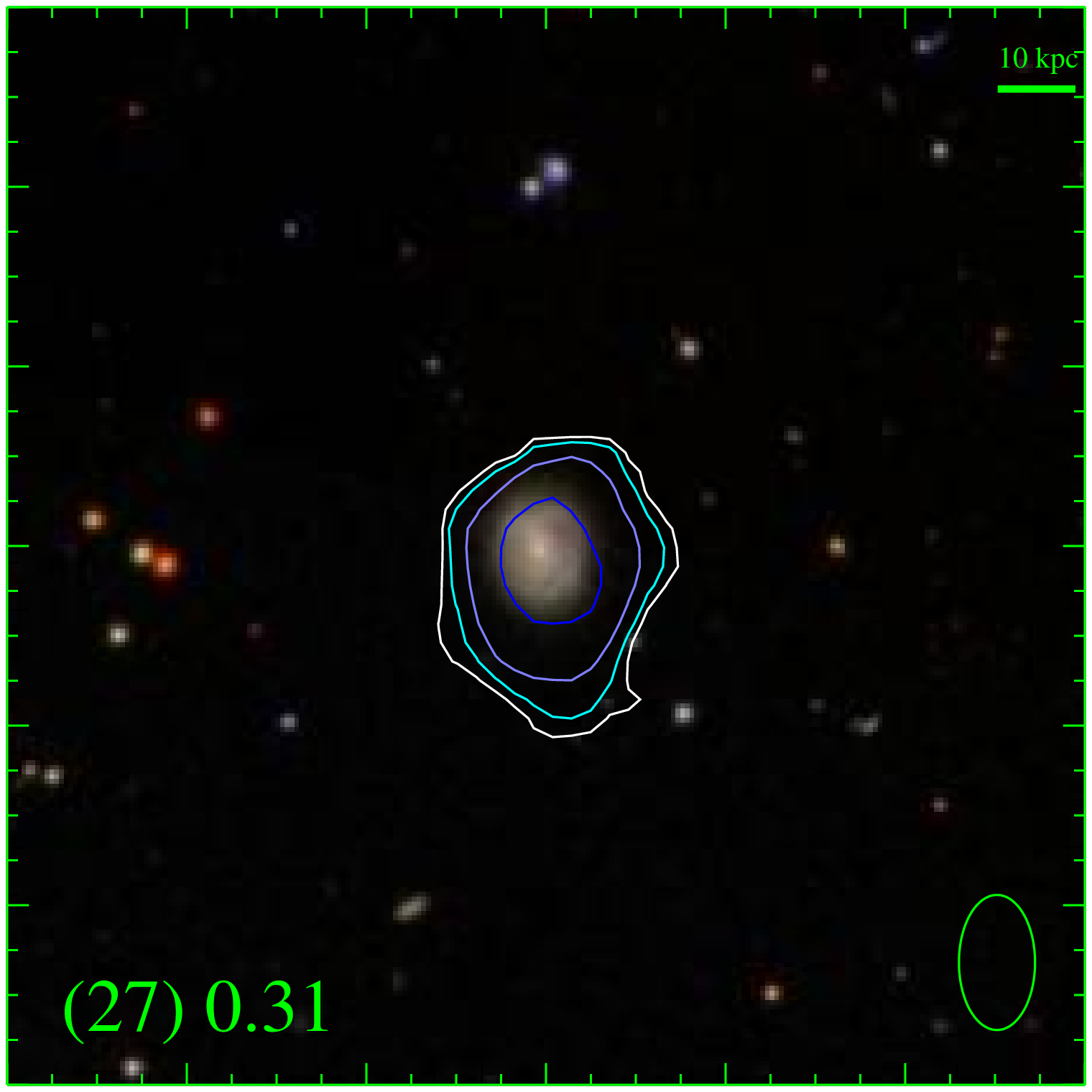}
\caption{HI column density contours on the optical images for the control galaxies. See caption of Figure~\ref{fig:G_HIrich}. To be continued} 
 \label{fig:G_ctrl}
\end{figure*}

\addtocounter{figure}{-1}
\begin{figure*}
\includegraphics[width=7cm]{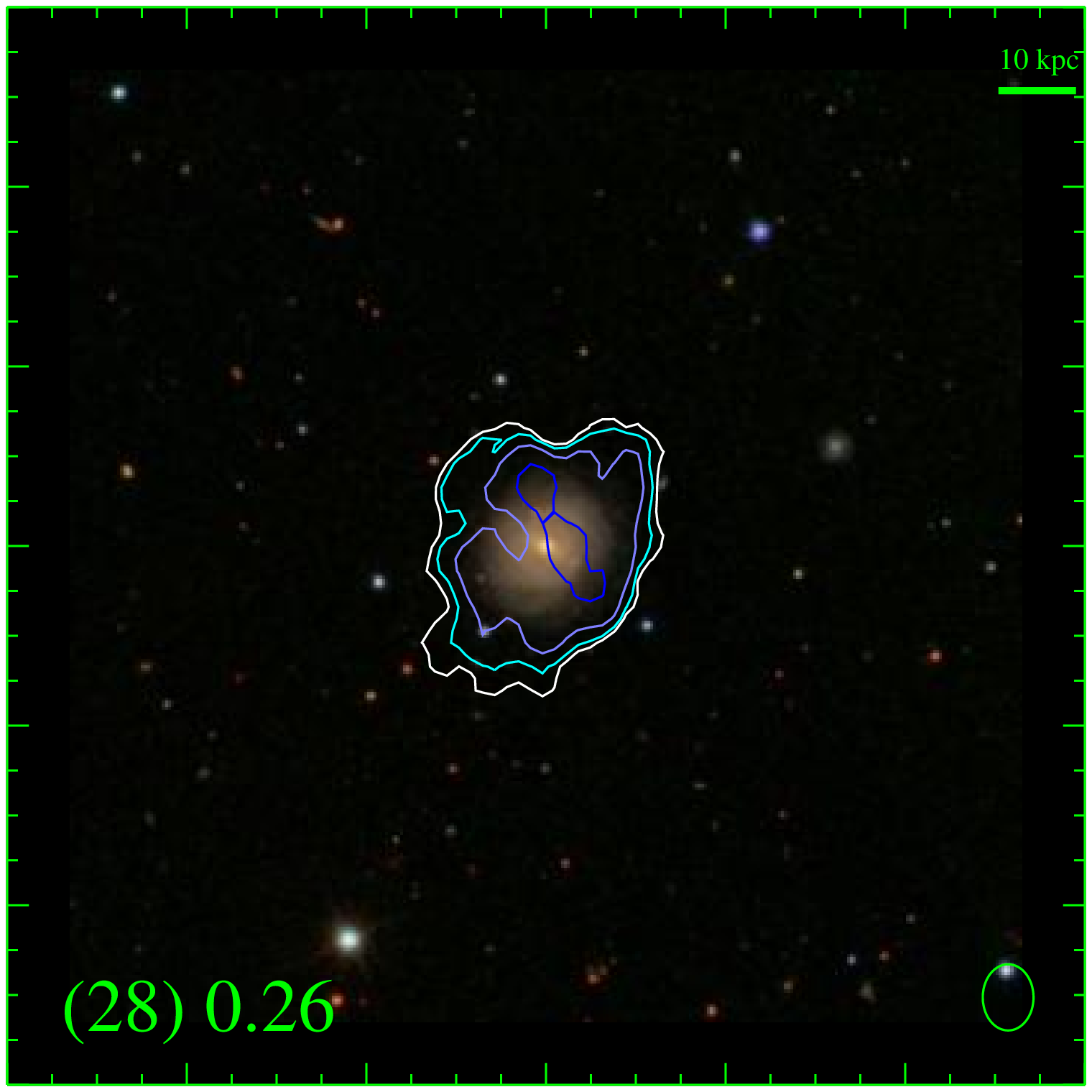}
\includegraphics[width=7cm]{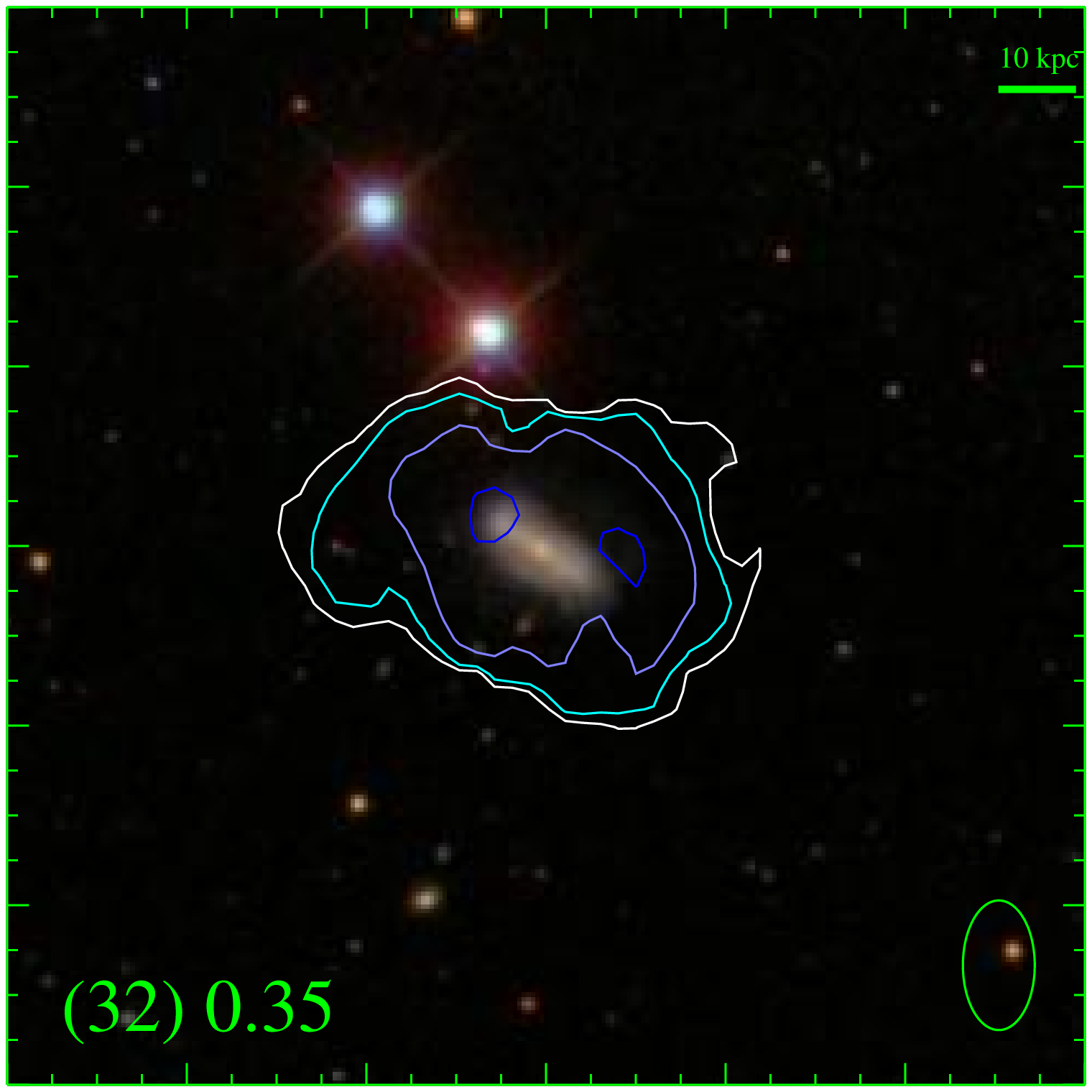}

\includegraphics[width=7cm]{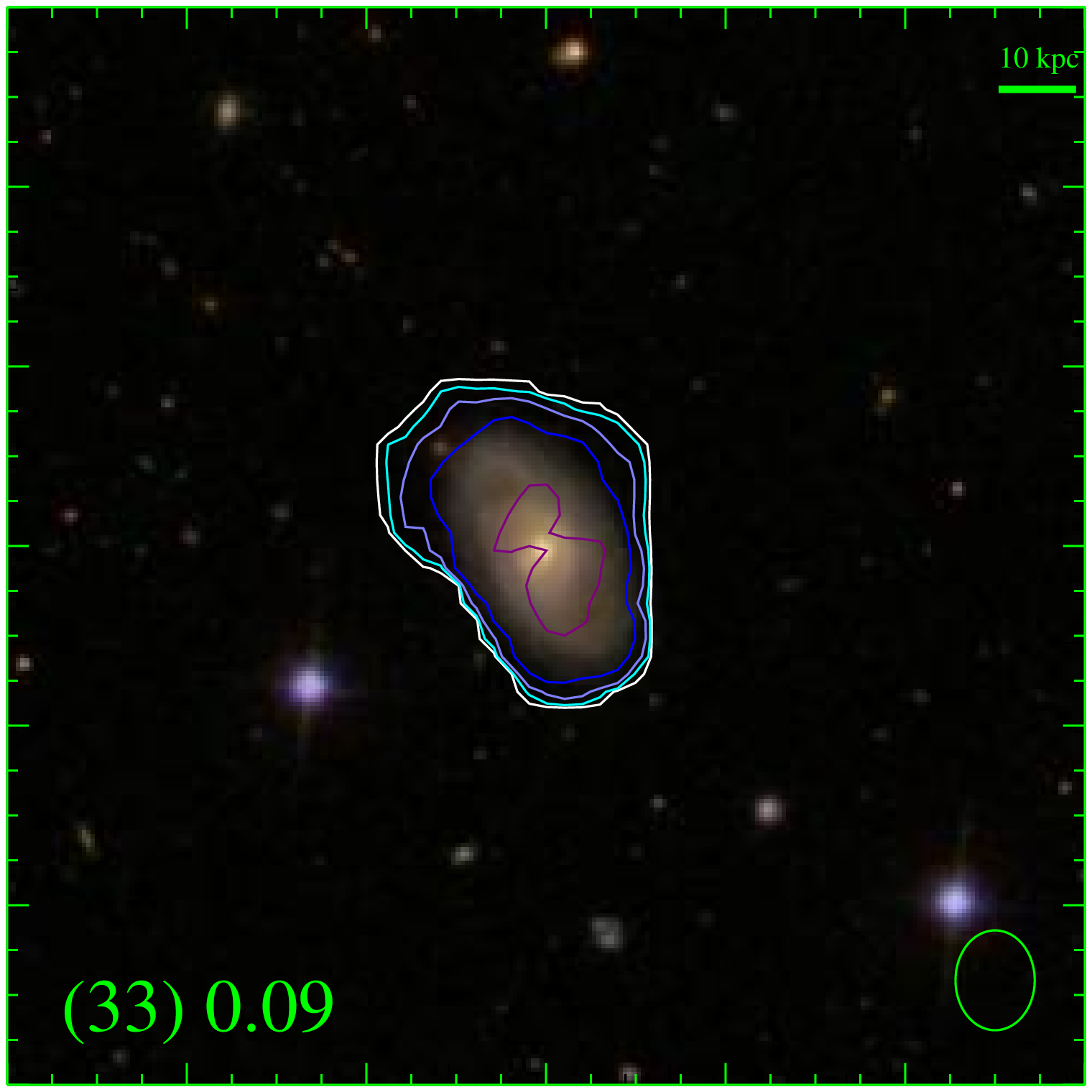}
\includegraphics[width=7cm]{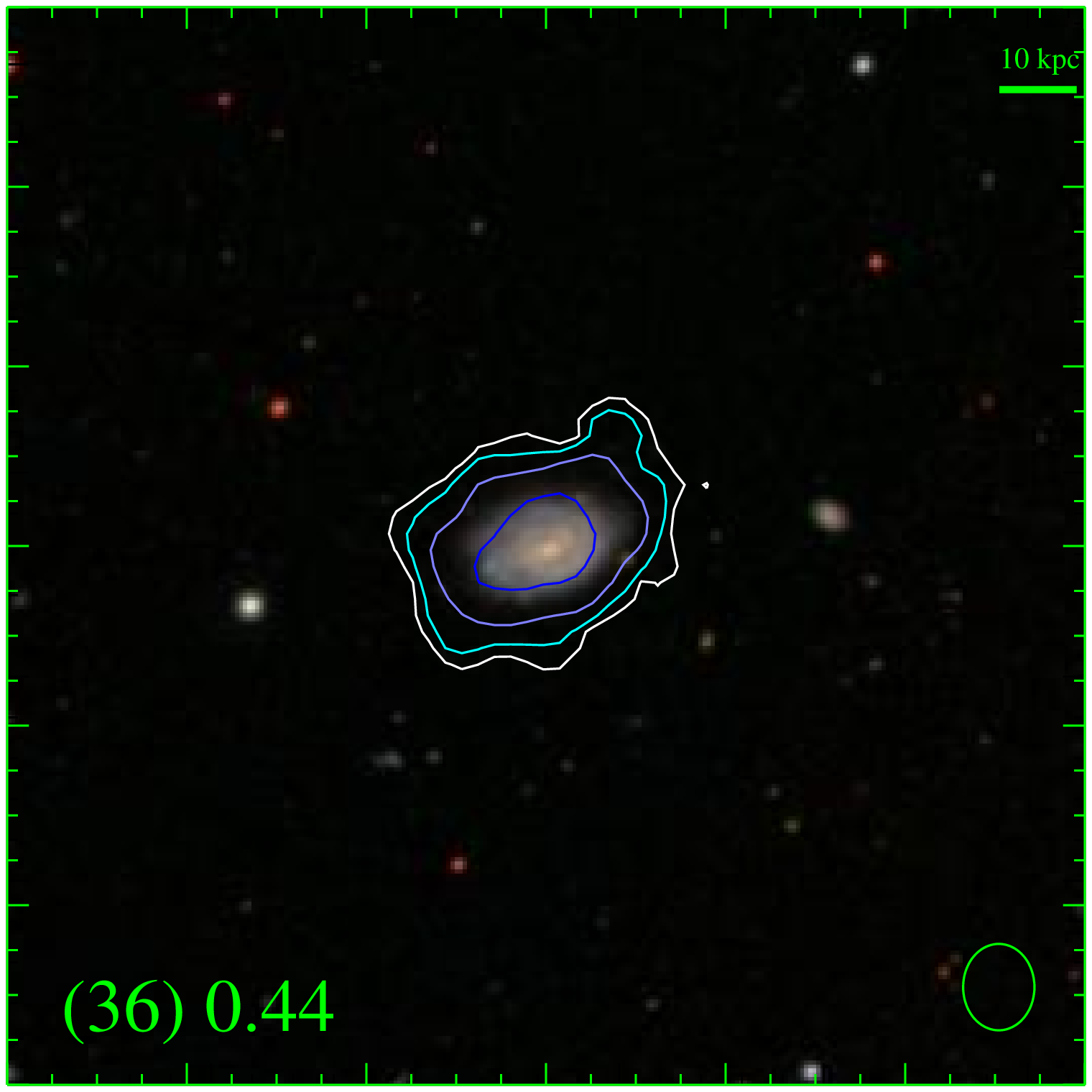}

\includegraphics[width=7cm]{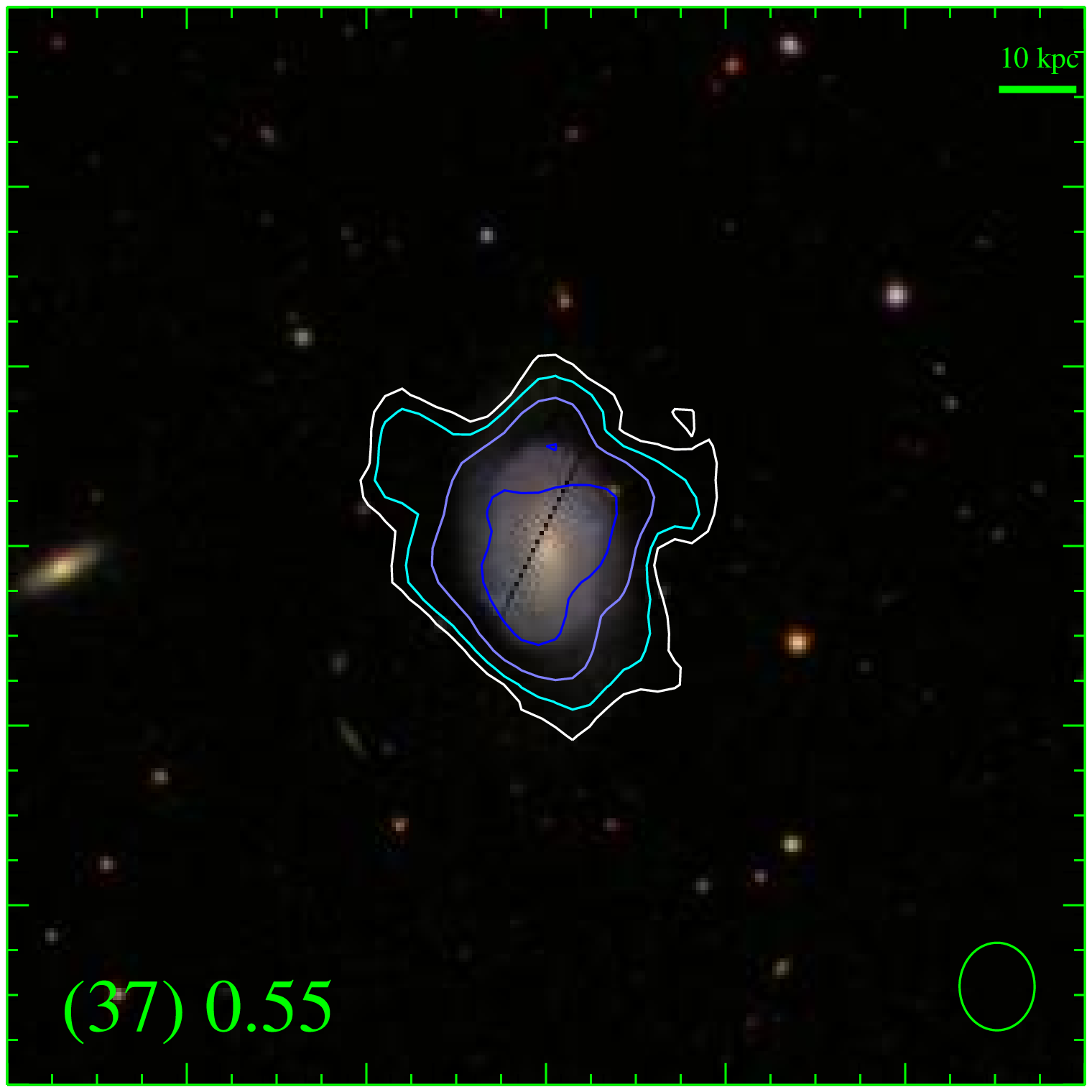}
\includegraphics[width=7cm]{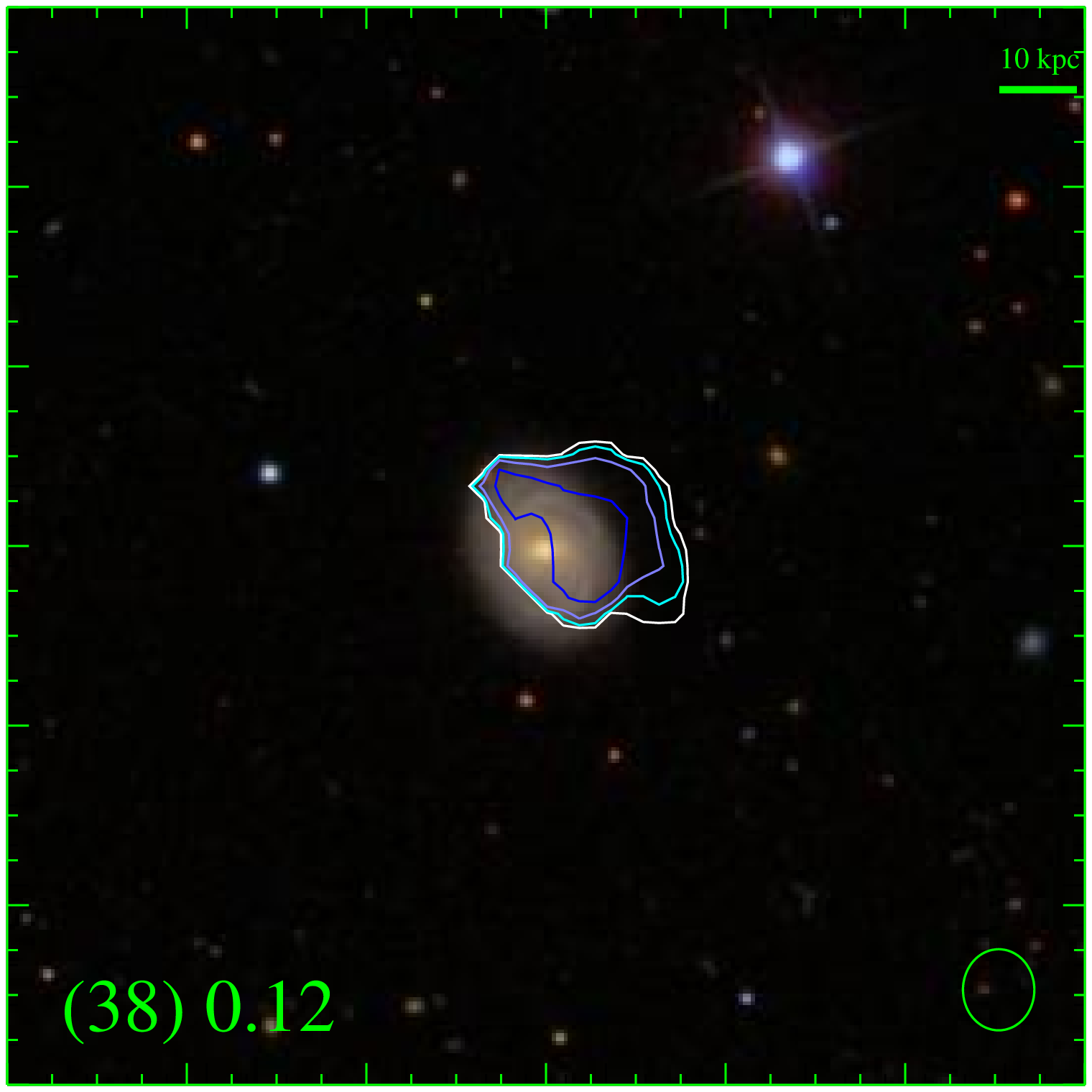}
\caption{To be continued} 
\end{figure*}

\addtocounter{figure}{-1}
\begin{figure*}
\includegraphics[width=7cm]{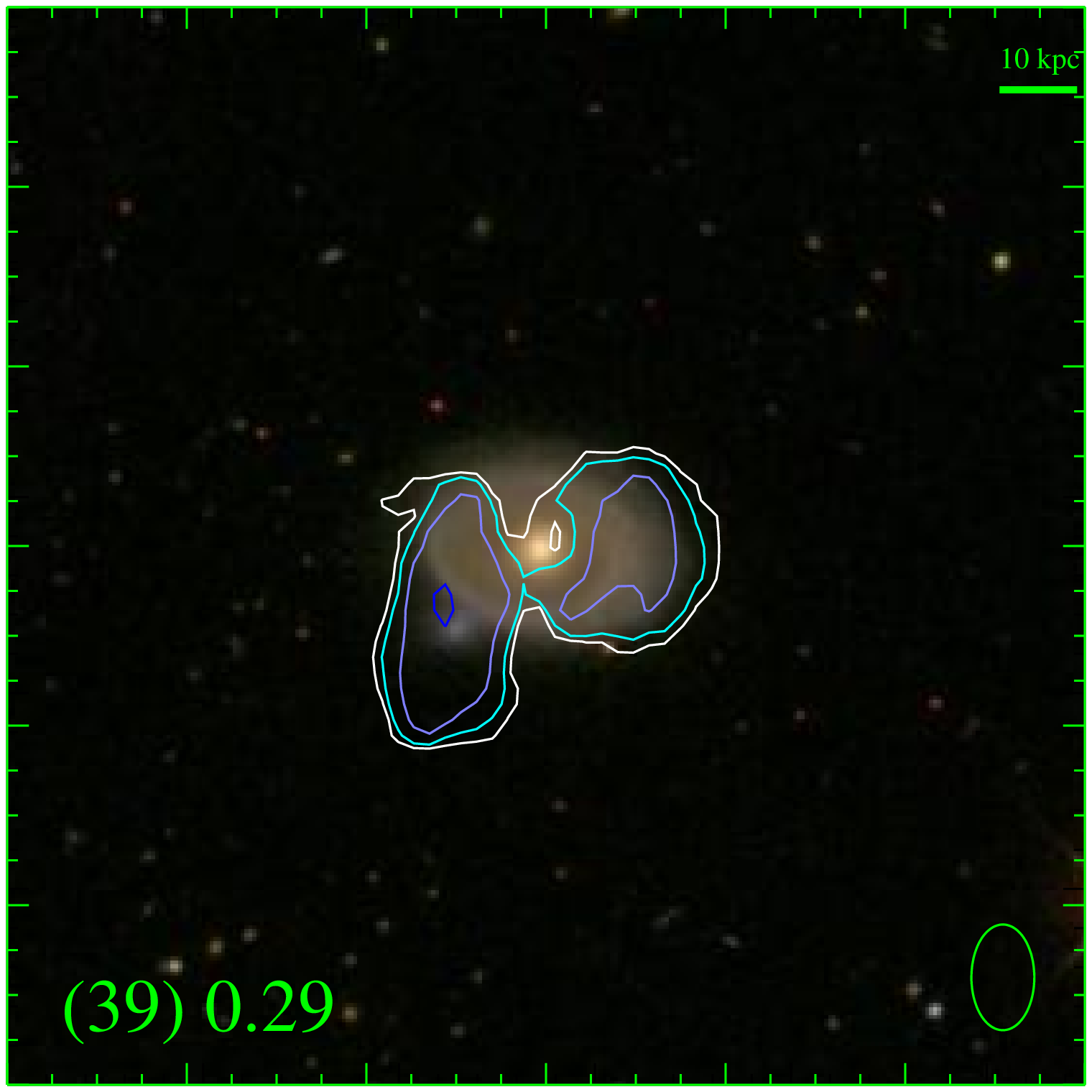}
\includegraphics[width=7cm]{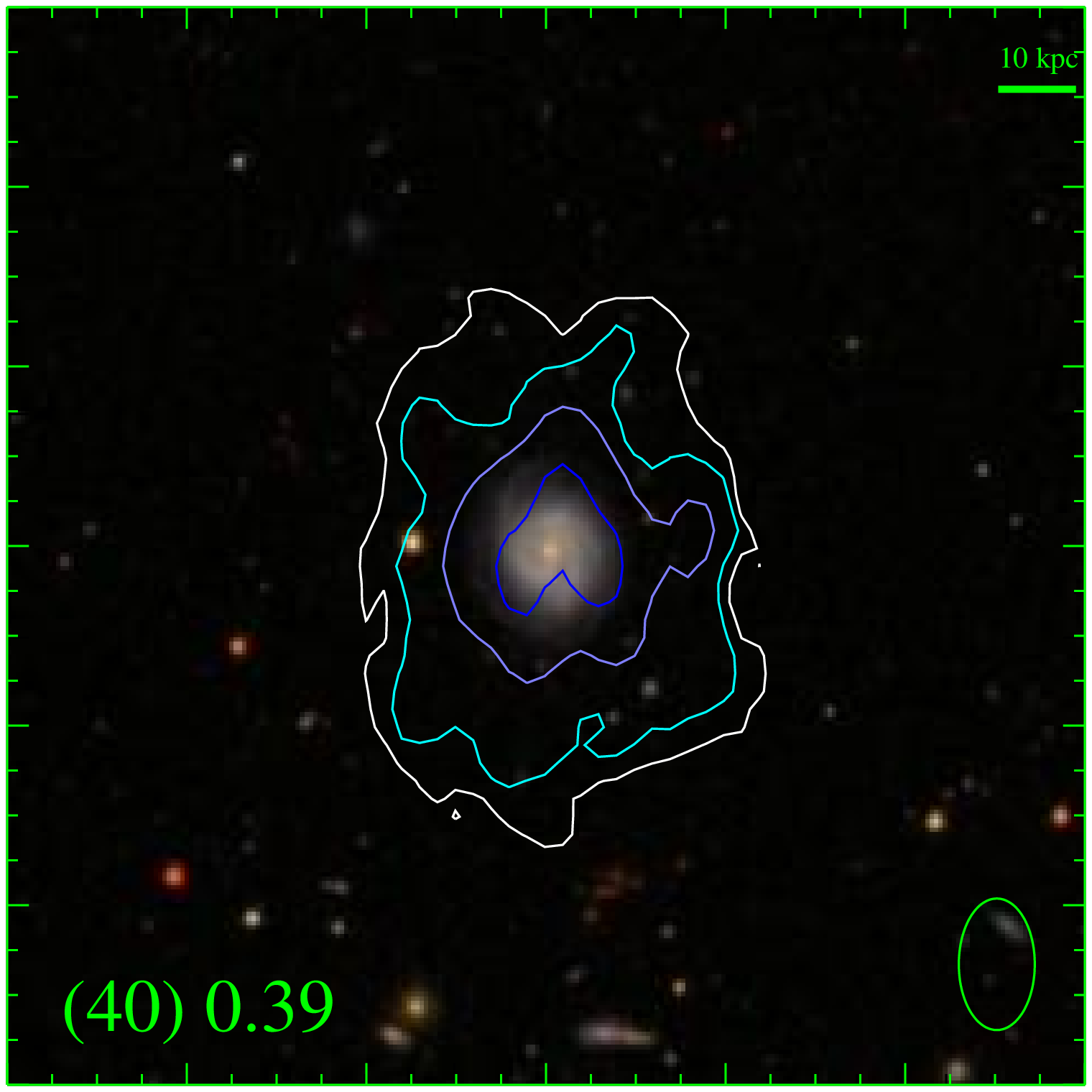}

\includegraphics[width=7cm]{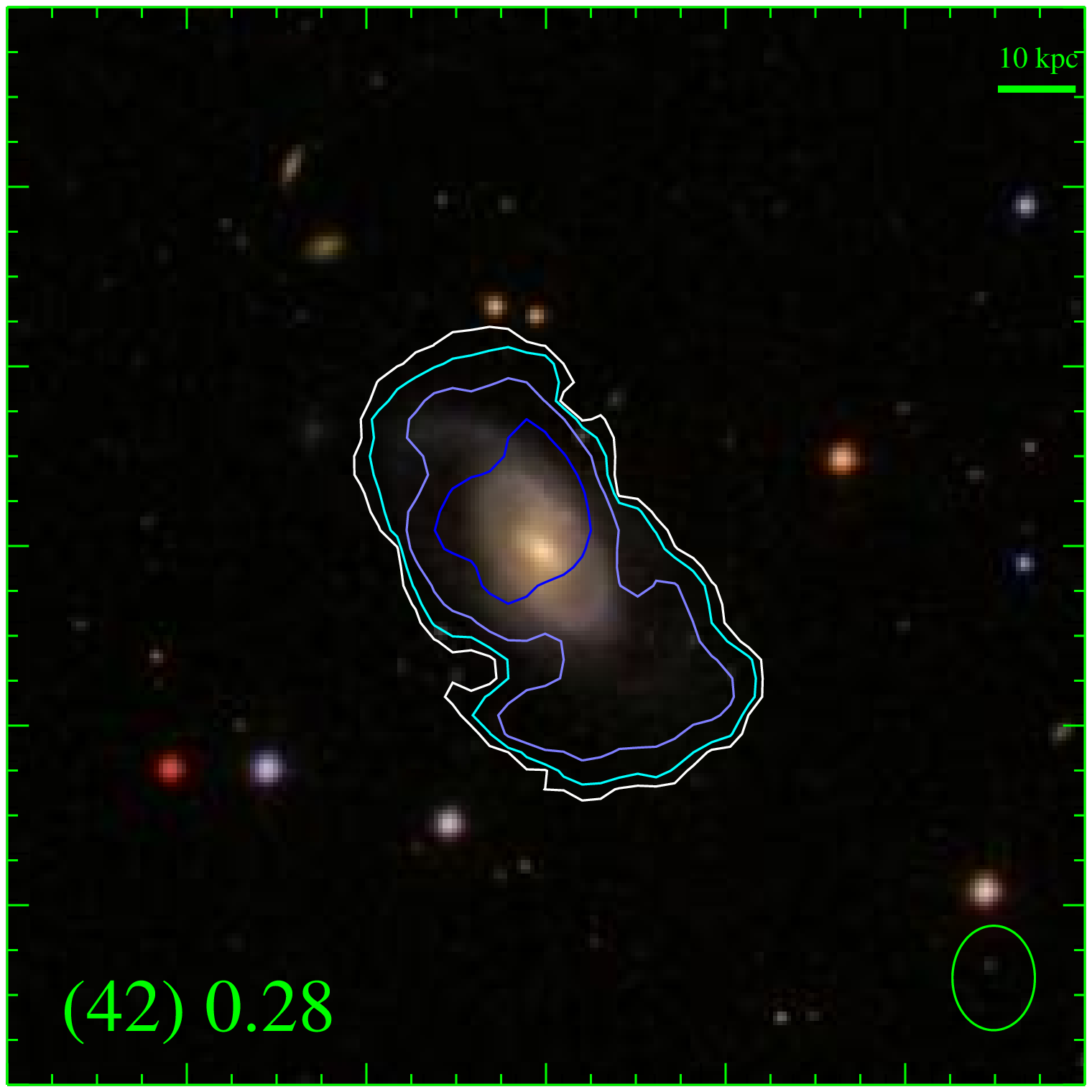}
\includegraphics[width=7cm]{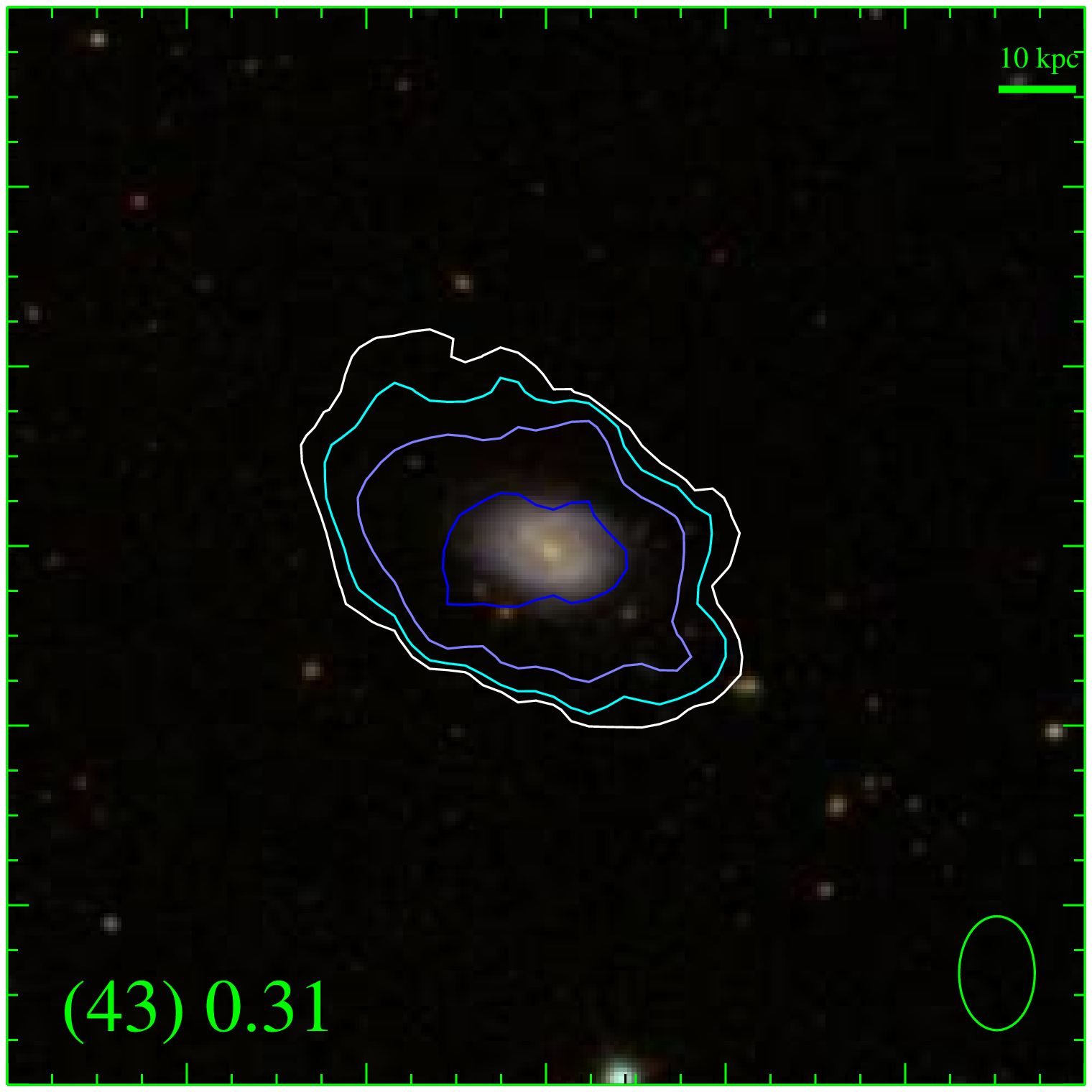}

\includegraphics[width=7cm]{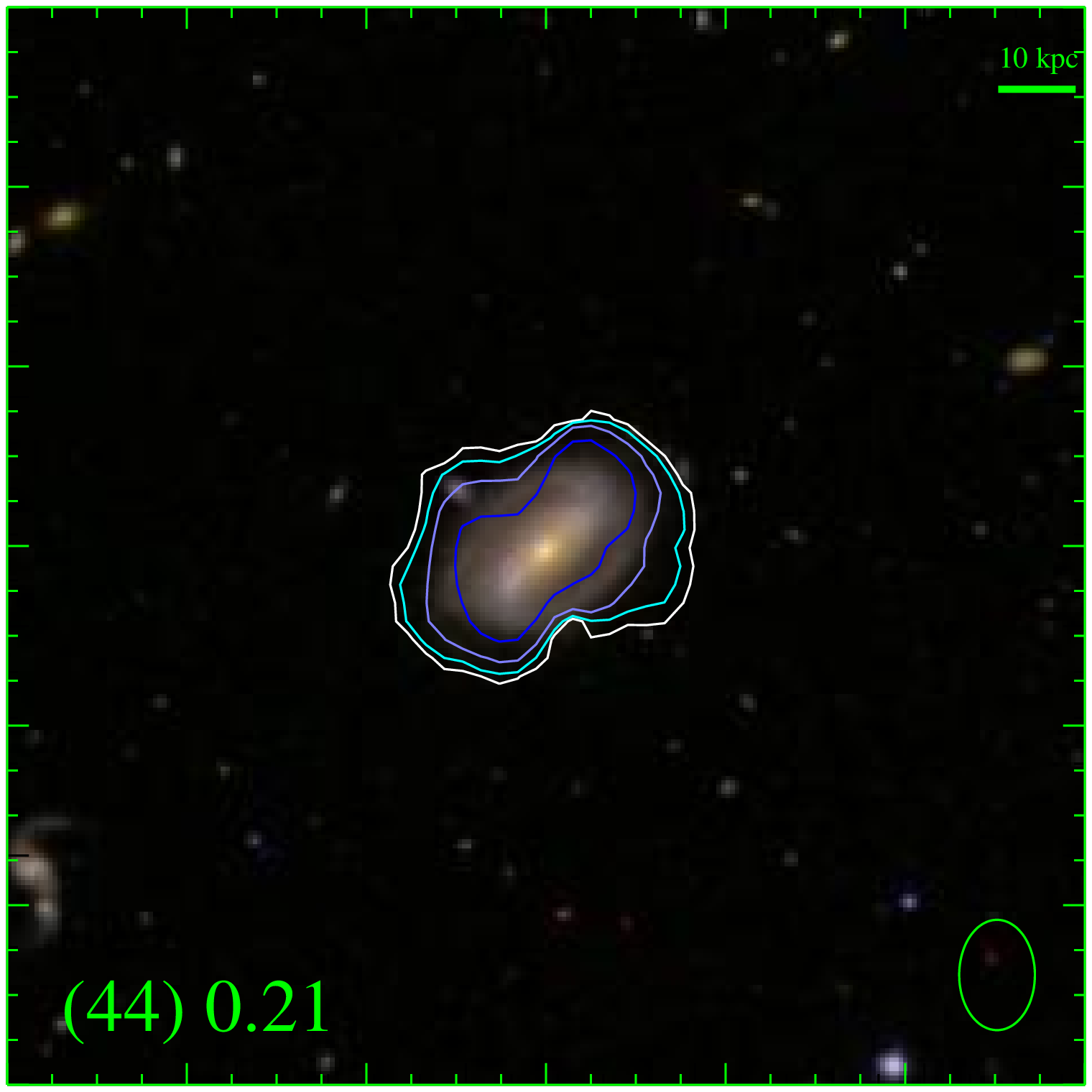}
\includegraphics[width=7cm]{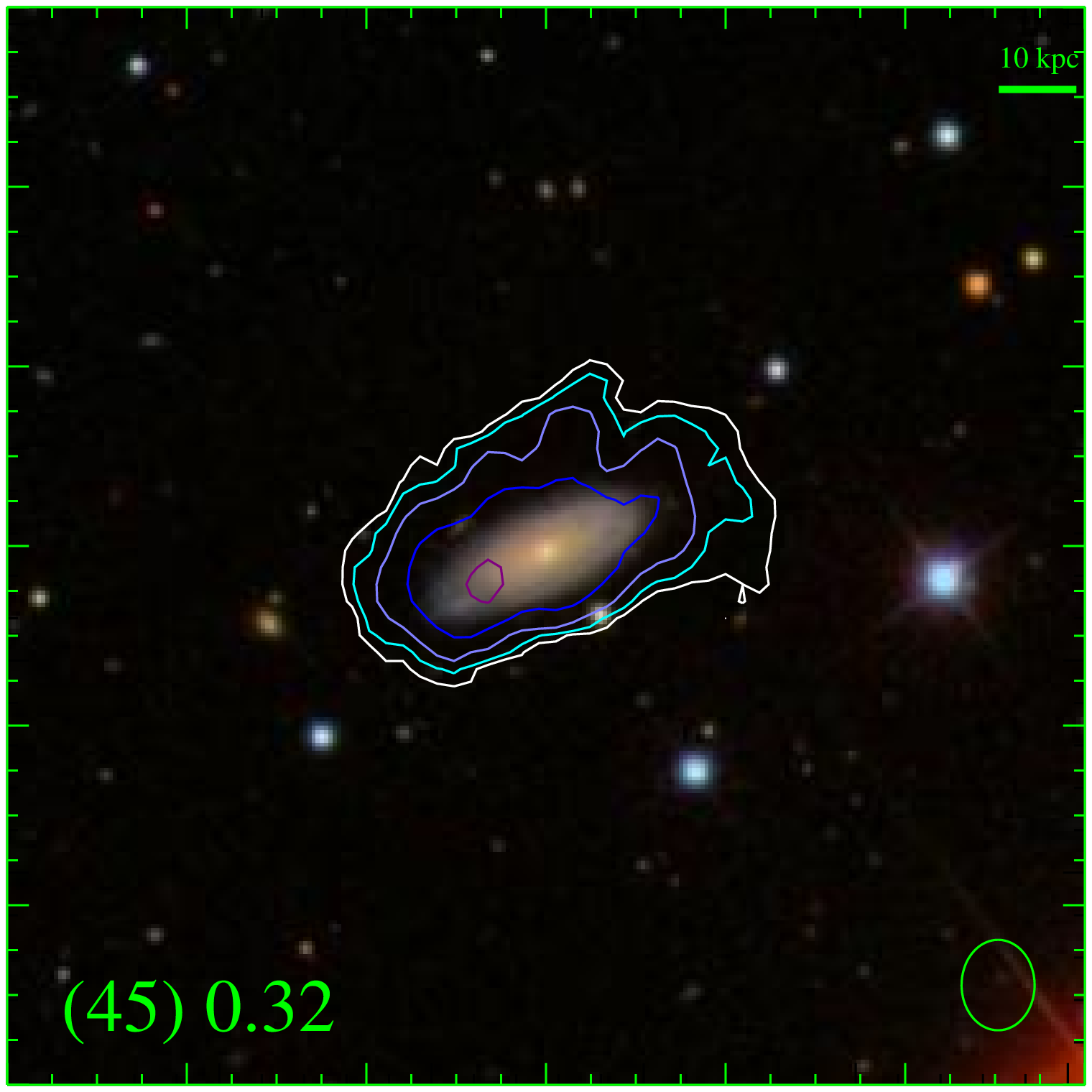}
\caption{To be continued} 
\end{figure*}

\addtocounter{figure}{-1}
\begin{figure*}
\includegraphics[width=7cm]{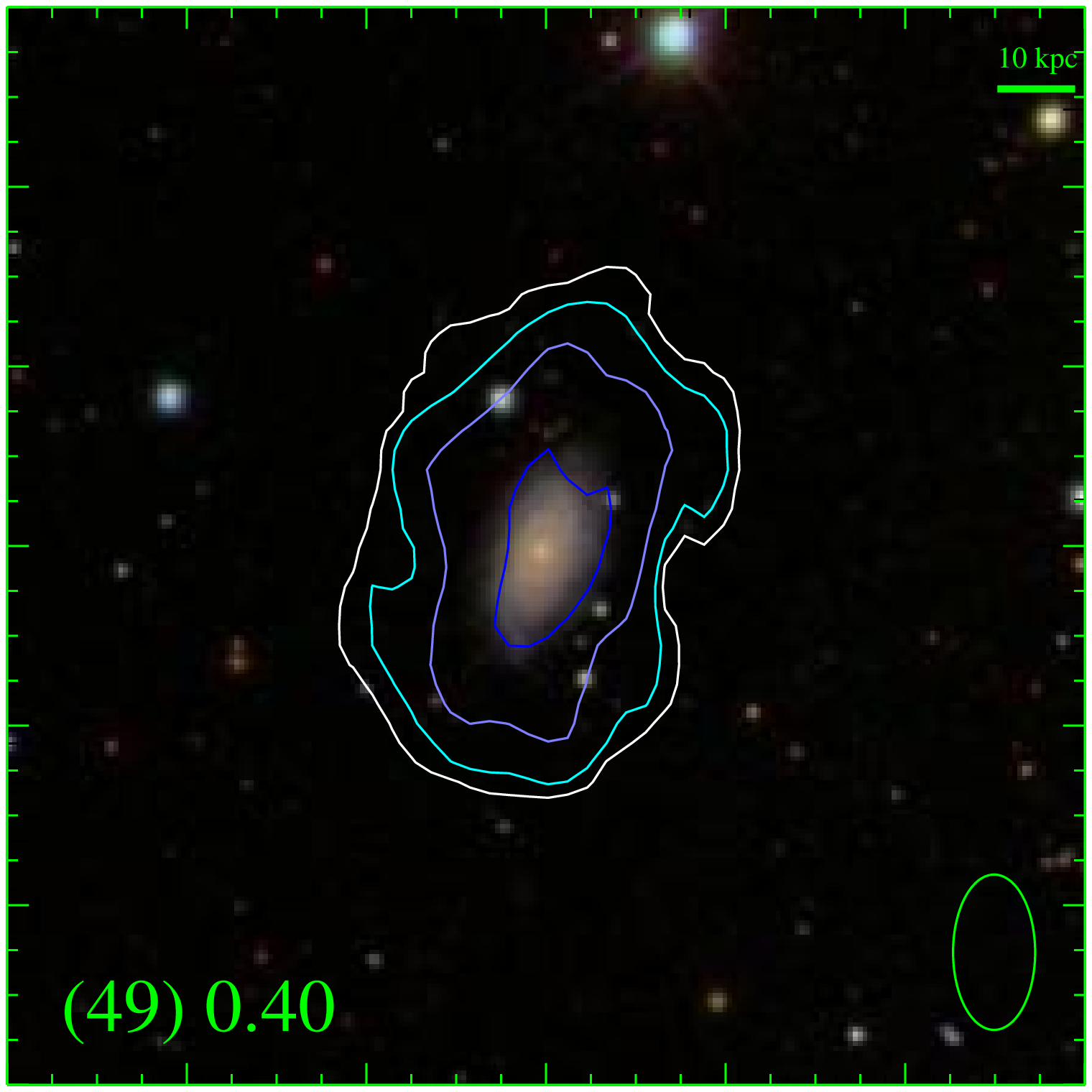}
\hspace{7cm}
\caption{-} 
\end{figure*}

\begin{figure*}
\includegraphics[width=7cm]{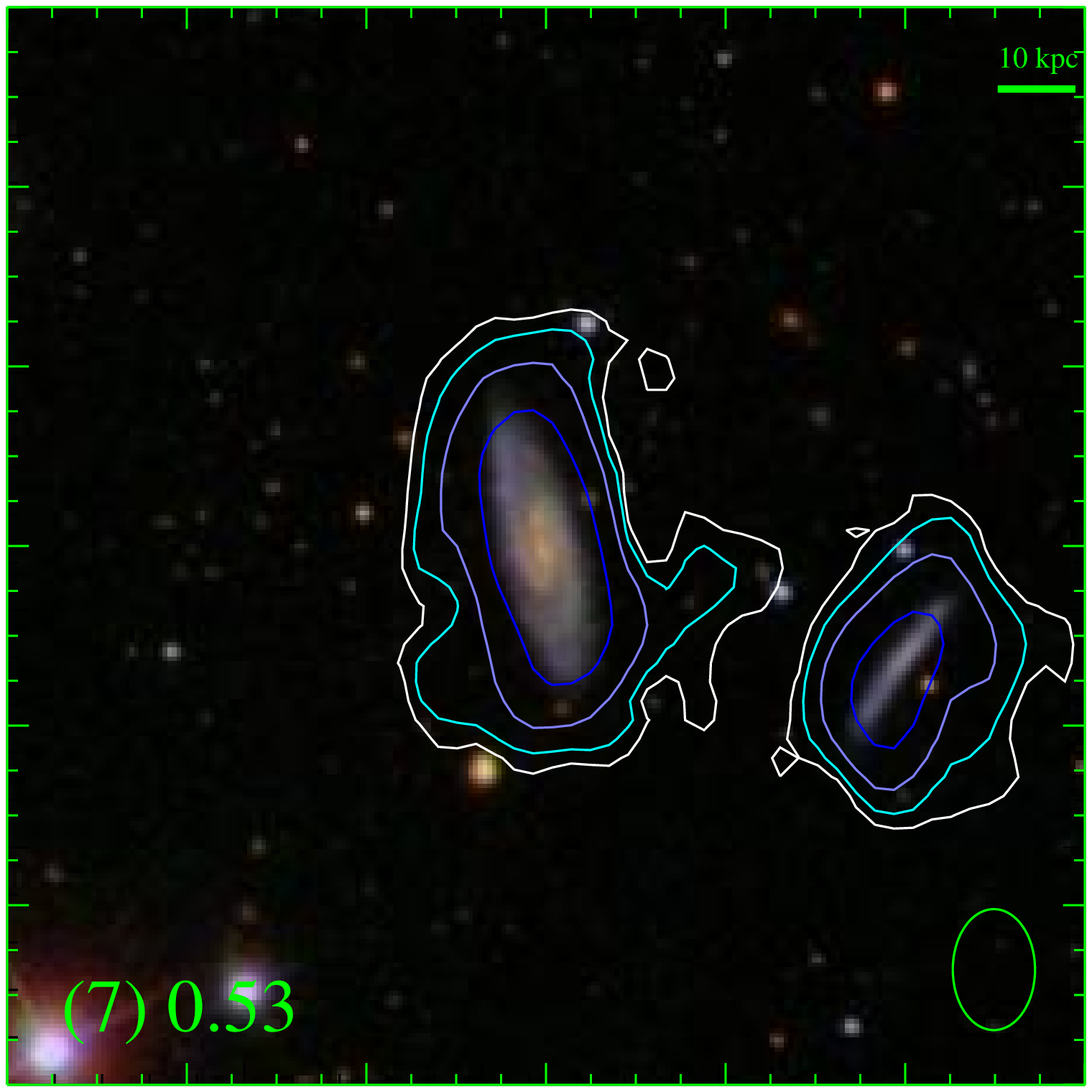}
\includegraphics[width=7cm]{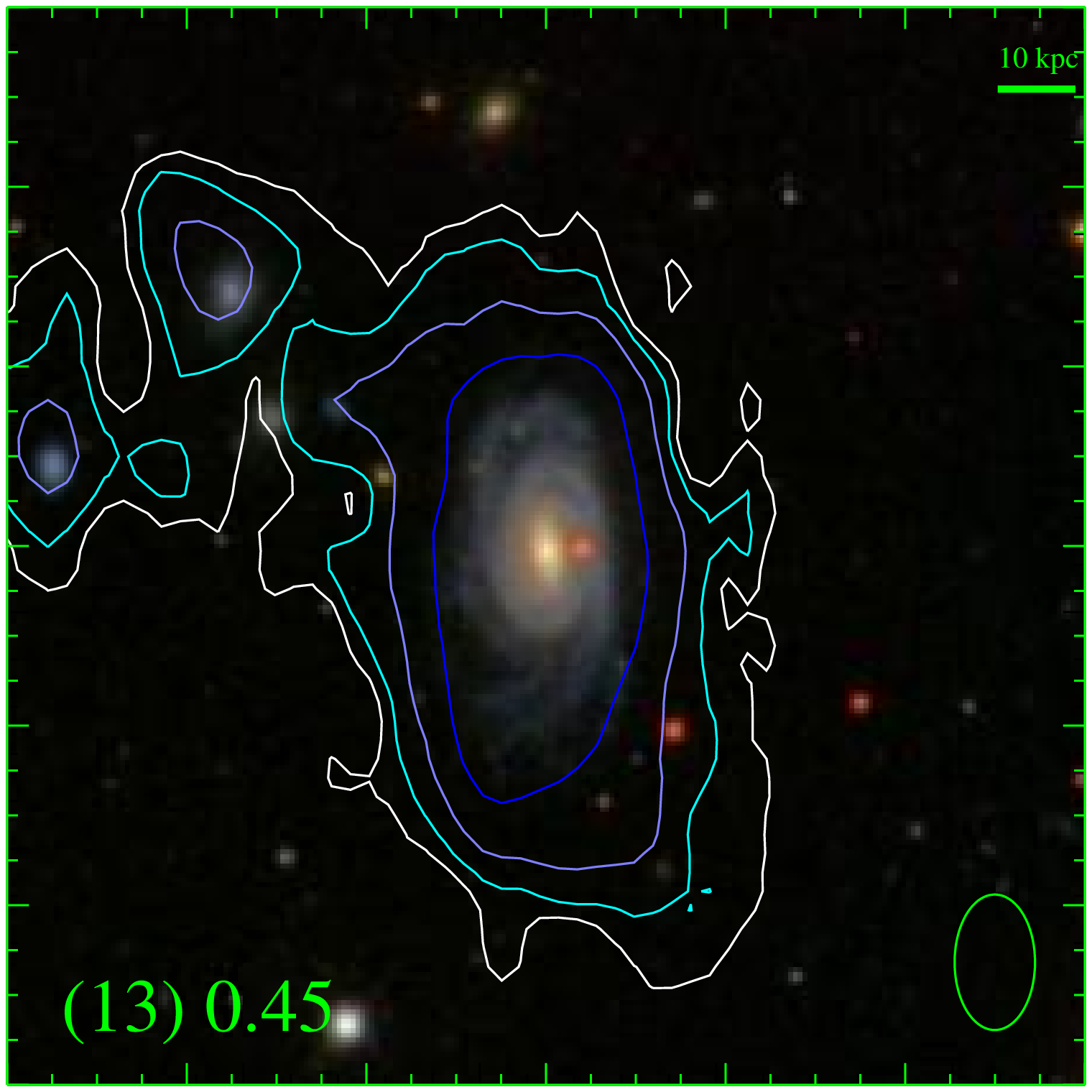}

\includegraphics[width=7cm]{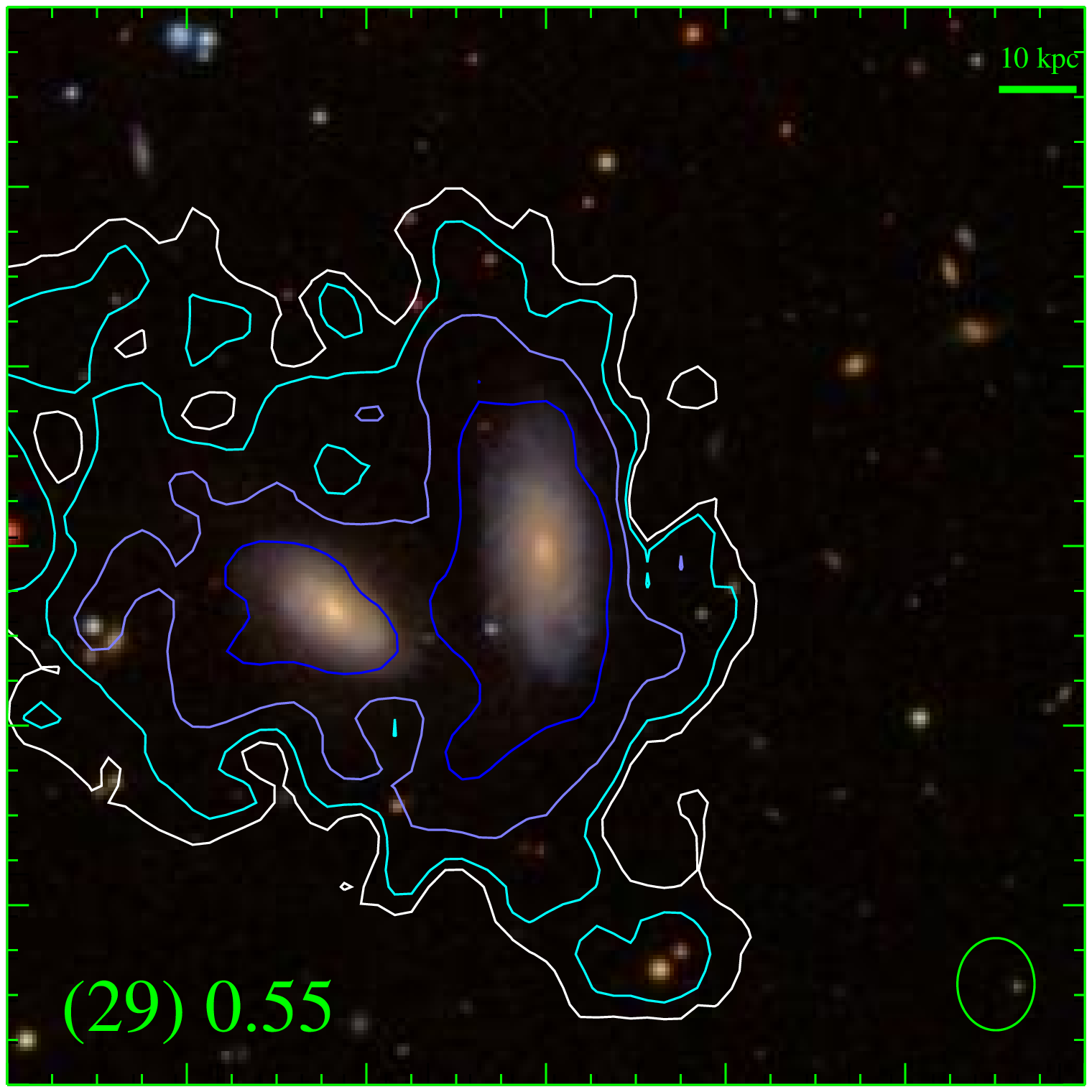}
\includegraphics[width=7cm]{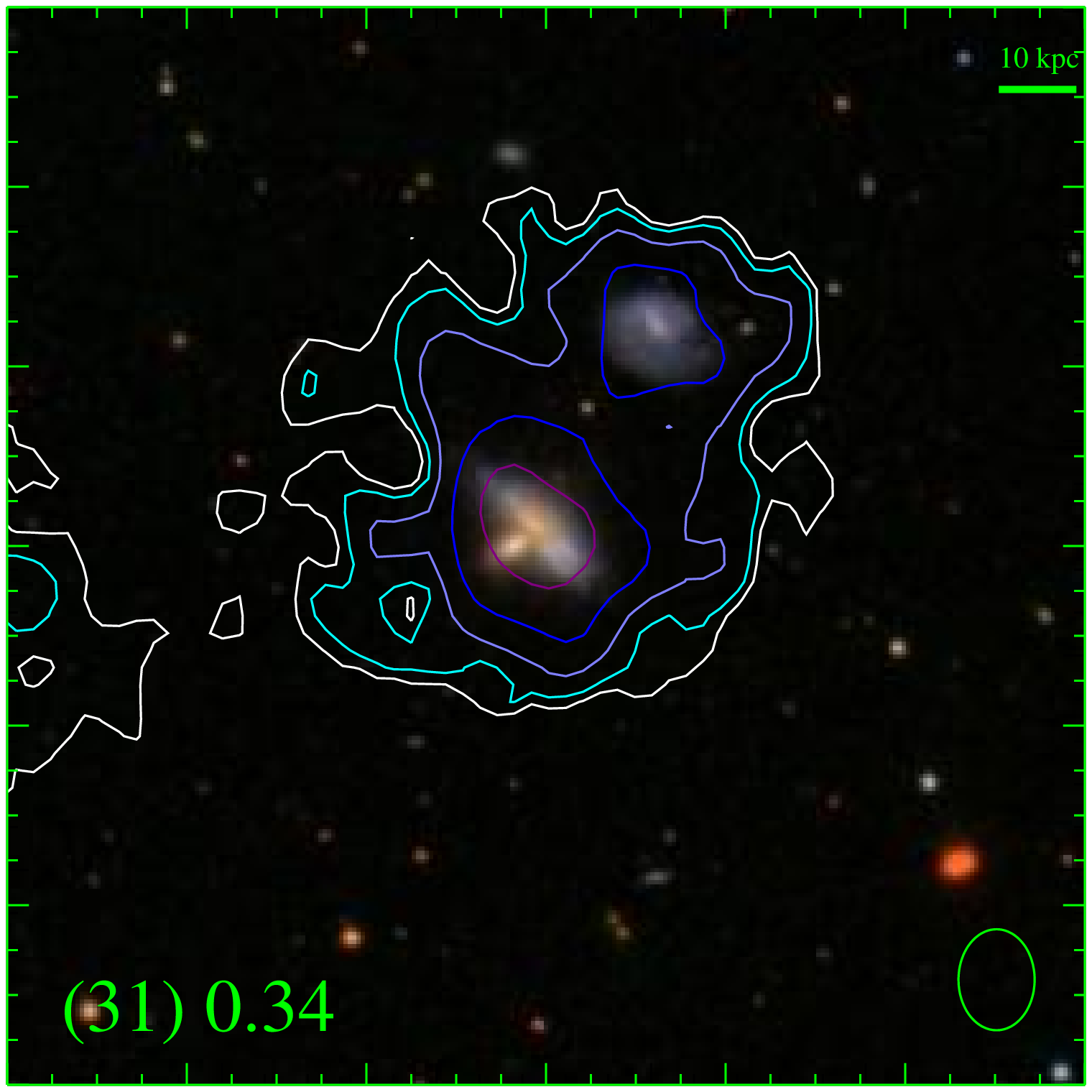}

\includegraphics[width=7cm]{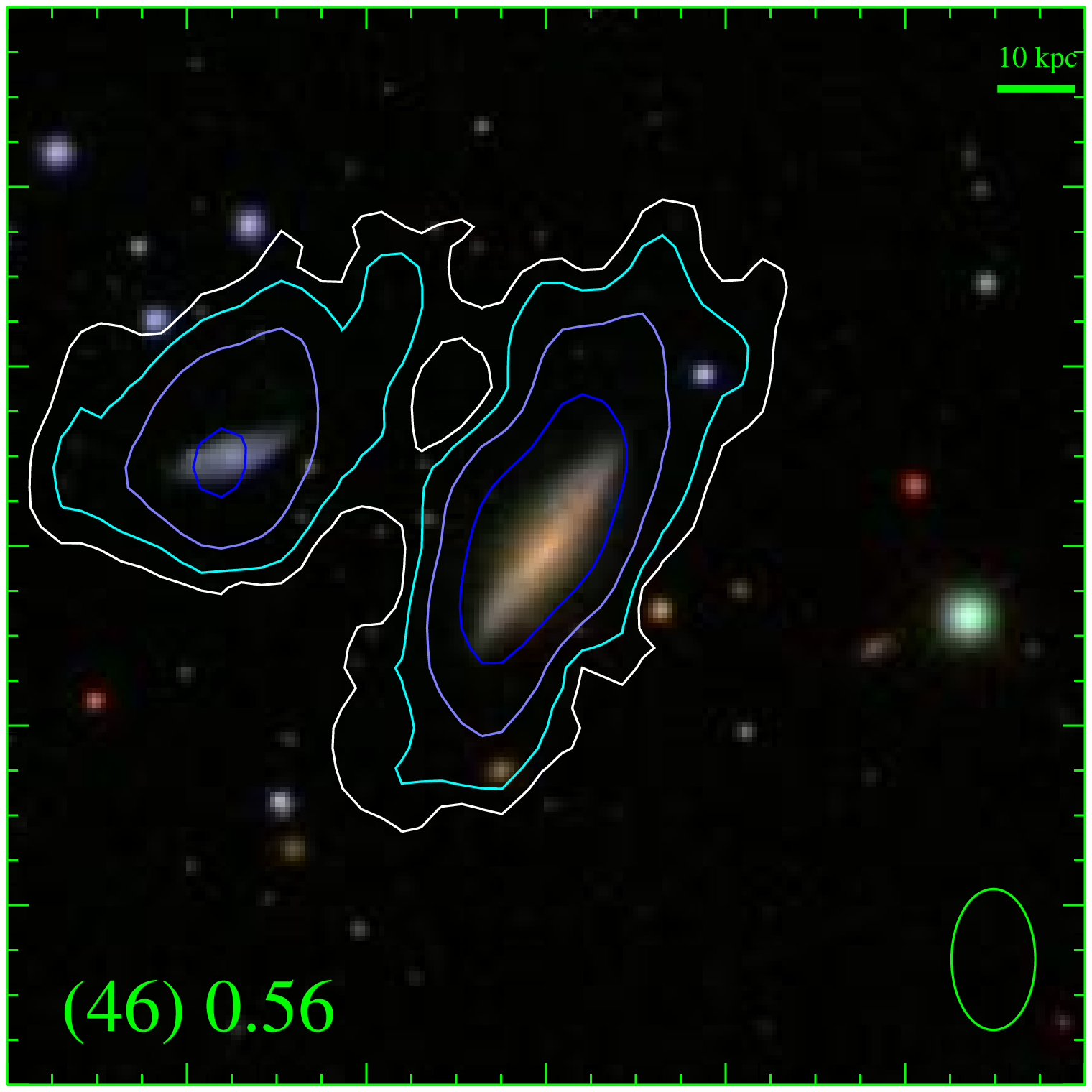}
\includegraphics[width=7cm]{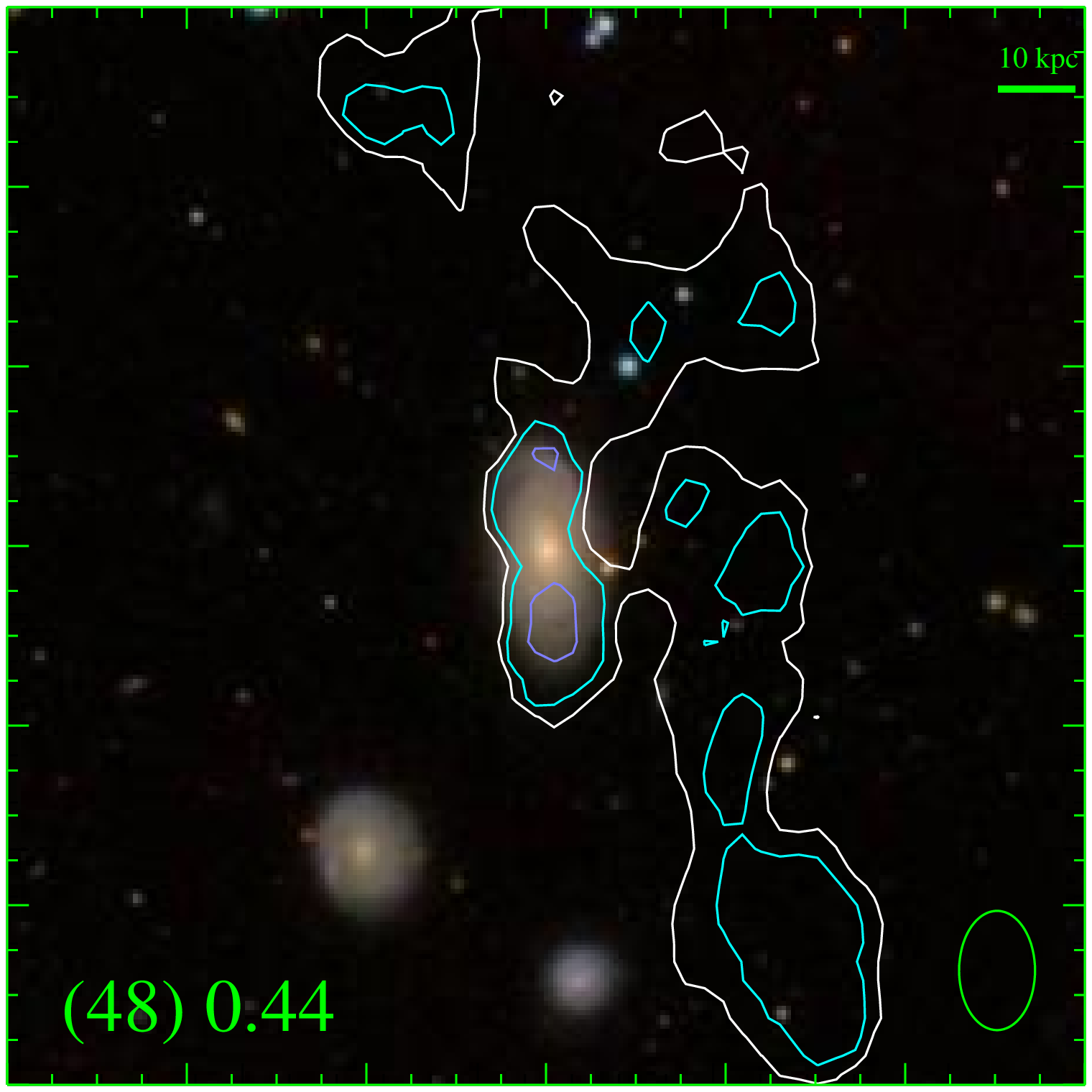}
\caption{HI column density contours on the optical images for the excluded galaxies. See caption of Figure~\ref{fig:G_HIrich}. To be continued.} 
 \label{fig:G_multi}
\end{figure*}

\addtocounter{figure}{-1}
\begin{figure*}
\includegraphics[width=7cm]{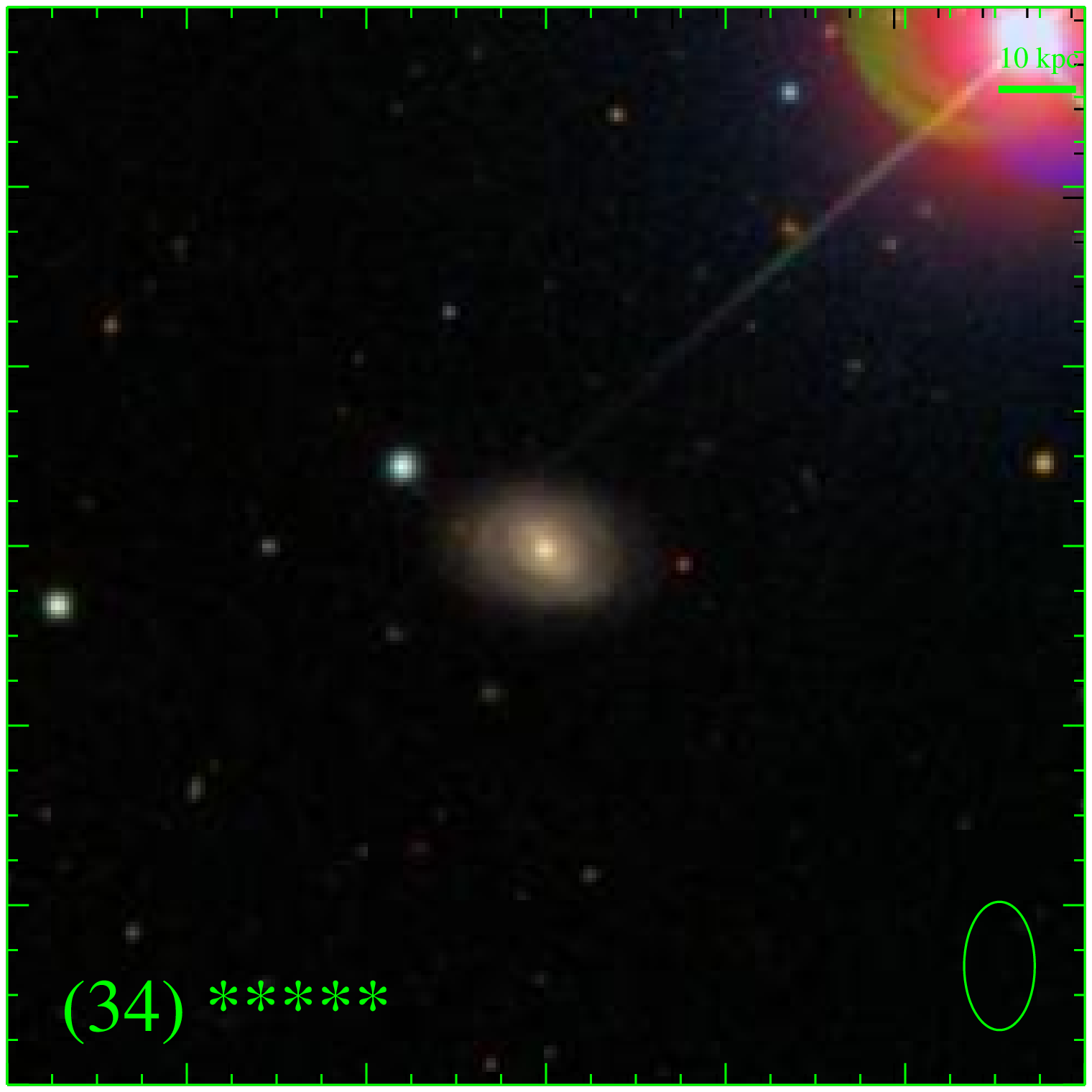}
\hspace{7cm}
\caption{-} 
\end{figure*}

\end{document}